\pdfoutput=1

\RequirePackage{silence}
\documentclass[aps,prx,twocolumn,10pt,nofootinbib]{revtex4-2}

\newif\iftikz
\tikzfalse 

\newif\ifex
\newif\ifexreplace

\extrue

\newcommand{\ex}[1]{%
\ifex{%
  		\ifexreplace\color{gray}\fi%
  		#1
  	}\fi
  }
\newcommand{\exreplace}[1]{\ifexreplace#1\fi}

\newcommand\protchaptitle\textit

\usepackage[utf8]{inputenc}	
\usepackage[english]{babel} 
\usepackage[T1]{fontenc} 

\usepackage{amssymb,amsmath,mathtools,amsthm}
\usepackage{dsfont} 

\usepackage[printonlyused,withpage]{acronym}
\makeatletter
\AtBeginDocument{%
  \renewcommand*{\AC@hyperlink}[2]{%
    \begingroup
      \hypersetup{hidelinks}%
      \hyperlink{#1}{#2}%
    \endgroup
  }%
}
\makeatother

\usepackage{paralist}
\usepackage{enumitem}
\newcommand{\MyItemSep}[1]{\setlength\itemsep{#1 \baselineskip}} 

\newcommand{\func}[1]{{\ensuremath{\mathsf{#1}}}}

\newcommand{\poly}{\func{poly}}
\newcommand{\polylog}{\func{polylog}}

\newcommand{\CNOT}{\func{CNOT}}

\usepackage[table,usenames,dvipsnames]{xcolor}

\usepackage[font=small,labelfont=bf,
   justification=raggedright,singlelinecheck=false,
   format=plain]{caption} 

\usepackage{pgf,tikz}
\definecolor{LightGray}{gray}{0.7}
\definecolor{OliveGreen}{RGB}{0,102,102}

\graphicspath{{./pics/}{./}}
\iftikz
  \usepackage{pgfplots}
  \pgfplotsset{compat=1.16}
  \usetikzlibrary{shadings,shadows}
  \usetikzlibrary{shapes,shapes.arrows}
  \usetikzlibrary{decorations}
  \usetikzlibrary{decorations.text,decorations.markings}
  \usetikzlibrary{arrows.meta}
  \usetikzlibrary{calc}
  \usetikzlibrary{positioning}
  \pgfdeclarelayer{bg}
  \pgfsetlayers{bg,main} 

\tikzset{%
  sbox/.style = {draw, rounded corners = .5ex,%
    minimum height = 1.6em,%
    minimum width = 1.8em},
  blau/.style = {top color=niceblue!12,%
    bottom color=niceblue!90},
  Bbox/.style = {sbox, blau, inner sep = 1pt},
  leg/.style = {rounded corners = .5ex,thick,postaction={decorate},
          decoration = {markings,mark=at position 0.8 with {\arrow{>}} }
          },
  legr/.style = {rounded corners = .5ex,thick,postaction={decorate},
          decoration = {markings,mark=at position 0.8 with {\arrow{<}} }
          },
  Leg/.style = {rounded corners = 1ex,thick,postaction={decorate},
    decoration = {markings,mark=at position 0.2 with {\arrow{>}} }
  },
  dir/.style = {gray,thick},
}%
\else
\fi

\usepackage{xifthen}

\usepackage{graphicx}
\graphicspath{{./pics/}{./}}
\usepackage{subcaption}

\usepackage{scrhack}
\usepackage{algorithm}
\usepackage[noend]{algpseudocode}

\usepackage[colorlinks=true,
			hyperindex, 
		    linkcolor = NavyBlue,
		    anchorcolor = hyperrefblue,
		    citecolor = Green,
		    filecolor = hyperrefblue,
		    urlcolor = NavyBlue,
			breaklinks=true]{hyperref}

\providecommand{\pra}{Phys.\ Rev.\ A}

\providecommand{\prl}{Phys.\ Rev.\ Lett.}

\providecommand{\cmp}{Commun.\ Math.\ Phys.}

\usepackage{tcolorbox} 
\tcbuselibrary{skins,breakable}

\newtcolorbox[auto counter]{theoremblock}[2][]{%
	beamer, 
	breakable,
	colback=LightGray!5,
	colframe=MidnightBlue,
	fonttitle=\bfseries,
	title=Theorem~\thetcbcounter~({#2}):,
	#1,
	}

\newtcolorbox[use counter from=theoremblock]{propositionblock}[2][]{%
	beamer, 
	breakable,
	colback=LightGray!5,
	colframe=MidnightBlue,
	fonttitle=\bfseries,
	title=Proposition~\thetcbcounter~({#2}):,
	#1,
	}

\newtcolorbox[use counter from=theoremblock]{lemmablock}[2][]{%
	beamer, 
	breakable,
	colback=LightGray!5,
	colframe=MidnightBlue,
	fonttitle=\bfseries,
	title=Lemma~\thetcbcounter~({#2}):,
	#1,
	}

\newtcolorbox[use counter from=theoremblock]{corollaryblock}[2][]{%
	beamer, 
	breakable,
	colback=LightGray!5,
	colframe=MidnightBlue,
	fonttitle=\bfseries,
	title=Corollary~\thetcbcounter~({#2}):,
	#1,
	}

\newtcolorbox[use counter from=theoremblock]{definitionblock}[2][]{%
	beamer, 
	breakable,
	colback=LightGray!5,
	colframe=OliveGreen,
	fonttitle=\bfseries,
	title=Definition~\thetcbcounter~({#2}):,
	#1,
	}

\newtcolorbox[use counter from=theoremblock]{protocolblock}[2][]{%
	beamer, 
	breakable,
	colback=LightGray!5,
	colframe=OliveGreen,
	fonttitle=\bfseries,
	title=Protocol~\thetcbcounter~({#2}):,
	#1,
	}

\newtcolorbox[use counter from=theoremblock]{exerciseBlock}[2][]{%
	beamer, 
	breakable,
	colback=LightGray!5,
	colframe=BrickRed!90!black,
	fonttitle=\bfseries,
	title=Exercise~\thetcbcounter~({#2}):,
	#1,
	}

\newenvironment{exerciseblock}[2][]{
		\tcolorbox[beamer,%
		    noparskip,breakable,
		    colback =  LightGray!5,
		    colframe = BrickRed!90!black,%
		    title=Exercise (#2):\phantom{f)}, 
		    #1,
		    ]}%
{\endtcolorbox}

\newenvironment{block}[2][]{
		\tcolorbox[beamer,%
		    noparskip,breakable,
		    colback =  LightGray!5,
		    colframe = MidnightBlue,%
		    title=#2\phantom{f)}, 
		    #1,
		    ]}%
{\endtcolorbox}

\newcommand{\e}{\ensuremath\mathrm{e}}
\renewcommand{\i}{\ensuremath\mathrm{i}}
\newcommand{\rmd}{\ensuremath\mathrm{d}}

\DeclareMathOperator{\LandauO}{\mathrm{O}}

\DeclareMathOperator{\tLandauOmega}{\tilde{\Omega}}

\DeclareMathOperator{\Tr}{Tr}
\renewcommand{\Re}{\operatorname{Re}}
\renewcommand{\Im}{\operatorname{Im}}
\DeclareMathOperator{\Eig}{Eig}

\DeclareMathOperator{\rank}{rank}

\DeclareMathOperator{\id}{id}

\DeclareMathOperator{\diag}{diag}
\newcommand{\fro}{\mathrm{F}}

\DeclareMathOperator{\spec}{spec}

\DeclareMathOperator{\conv}{conv}

\DeclareMathOperator{\dist}{dist}

\newcommand{\avg}{\mathrm{avg}}

\DeclareMathOperator{\sign}{sign}

\renewcommand{\L}{\operatorname{L}}
\DeclareMathOperator{\U}{U}
\DeclareMathOperator{\SU}{SU}

\DeclareMathOperator{\Var}{Var}

\DeclareMathOperator{\Herm}{Herm} 
\DeclareMathOperator{\PSD}{Pos} 
\DeclareMathOperator{\Pos}{Pos} 
\DeclareMathOperator{\DM}{\mathcal{S}} 
\DeclareMathOperator{\M}{\LL} 
\DeclareMathOperator{\CP}{CP} 
\DeclareMathOperator{\CPT}{CPT} 

\newcommand{\CC}{\mathbb{C}}
\newcommand{\RR}{\mathbb{R}}
\newcommand{\QQ}{\mathbb{Q}}
\newcommand{\ZZ}{\mathbb{Z}}

\newcommand{\1}{\mathds{1}}
\newcommand{\EE}{\mathbb{E}}
\newcommand{\PP}{\mathbb{P}}
\newcommand{\FF}{\mathbb{F}}
\newcommand{\LL}{\mathbb{L}}

\newcommand{\R}{\mathbb{R}}
\newcommand{\C}{\mathbb{C}}
\newcommand{\sphere}{\mathbb{S}}
\newcommand{\sphereCd}{\mathbb{S}(\CC^d)}

\newcommand{\mc}[1]{\mathcal{#1}}

\newcommand{\mcP}{\mc{P}}
\newcommand{\mcH}{\mc{H}}
\renewcommand{\H}{\mcH}
\newcommand{\K}{\mc{K}}

\newcommand{\fr}[1]{\mathfrak{#1}}

\newcommand{\frS}{\fr{S}}
\DeclareMathOperator{\choi}{\mathfrak{C}} 
\DeclareMathOperator{\jam}{\mathfrak{J}} 

\newcommand{\argdot}{{\,\cdot\,}}
\renewcommand{\vec}[1]{\boldsymbol{#1}}

\newcommand{\T}{\intercal} 

\newcommand{\myleft}{\mathopen{}\mathclose\bgroup\left}
\newcommand{\myright}{\aftergroup\egroup\right}
\newcommand{\ad}{\dagger}

\newcommand{\norm}[1]{\left\Vert #1 \right\Vert} 
\newcommand{\normb}[1]{\bigl\Vert #1 \bigr\Vert} 
\newcommand{\lpnorm}[2][p]{\norm{#2}_{\ell_{#1}}} 

\newcommand{\pnorm}[2][p]{\norm{#2}_{#1}} 

\newcommand{\iiiNorm}[1]{{\left\vert\kern-0.25ex\left\vert\kern-0.25ex\left\vert #1 
    \right\vert\kern-0.25ex\right\vert\kern-0.25ex\right\vert}}
\newcommand{\tnorm}[1]{\pnorm[1]{#1}} 

\newcommand{\snorm}[1]{\norm{#1}_{\mathrm{op}}} 

\newcommand{\snormb}[1]{\normb{#1}_{\mathrm{op}}}

\newcommand{\fnorm}[1]{\norm{#1}_\fro} 

\newcommand{\dnorm}[1]{\norm{#1}_\diamond} 

\newcommand{\braket}[2]{\left\langle #1 \middle| #2 \right\rangle}

\newcommand{\ket}[1]{\left.\left|{#1}\right.\right\rangle}
\newcommand{\ketn}[1]{| #1 \rangle}

\newcommand{\bra}[1]{\left.\left\langle{#1}\right.\right|}
\newcommand{\bran}[1]{\langle #1 |}

\newcommand{\ketbra}[2]{\ket{#1} \!\! \bra{#2}}

\newcommand{\ketbran}[2]{\ketn{#1} \! \bran{#2}}

\newcommand{\sandwich}[3]
  {\left\langle  #1 \right| #2 \left| #3 \right\rangle}

\renewcommand{\Pr}{\operatorname{\PP}}

\newcommand{\verbat}[1]{\text{\normalfont{\ttfamily{#1}}}}
\newcommand{\vect}[1]{\ket{#1}}

\newcommand{\Jamiolkowski}{Jamio{\l}kowski}

\newcommand{\Cproj}{{\CC\mathbf{P}}}

\newcommand{\MeasurementSymbol}{M} 
\newcommand{\mop}{\MeasurementSymbol} 
\newcommand{\mset}{\mathsf{\MeasurementSymbol}} 

\newcommand{\W}{W} 

\newcommand{\sym}{\mathrm{sym}}

\newcommand{\Sym}{\frS}

\newcommand{\dorder}{k}

\newcommand{\depol}[1][p]{\mc{D}_{#1}}

\newcommand{\mo}{\mc{M}} 

\newcommand{\rhop}{{\tilde{\rho}}}
\newcommand{\Rhop}{{\tilde{\vec{\rho}}}}
\newcommand{\rhot}{\rho}

\newcommand{\np}{{n_\rhop}}
\DeclareMathOperator{\fidelity}{F}
\DeclareMathOperator{\efidelity}{\fidelity_{\!\mathrm{e}}}

\newcommand{\fail}{\verbat{``fail''}}
\newcommand{\pass}{\verbat{``pass''}}
\newcommand{\accept}{\verbat{``accept''}}
\newcommand{\reject}{\verbat{``reject''}}

\DeclareMathOperator{\agf}{F_{\!\avg}} 
\DeclareMathOperator{\hatagf}{\hat{F}_{\!\avg}} 
\DeclareMathOperator{\aer}{r} 
\newcommand{\mcU}{\mc{U}}
\newcommand{\X}{\mc{X}}
\newcommand{\Y}{\mc{Y}}
\newcommand{\mcG}{\mc{G}}

\newcommand{\gset}{\mathsf{G}} 
\renewcommand{\ng}{{n_{\gset}}} 
\DeclareMathOperator{\tw}{tw} 

\newcommand{\gt}{g_T} 
\newcommand{\Gt}{\mcG_T}

\newcommand{\Cl}{\mathrm{Cl}}
\DeclareMathOperator{\dxe}{d_\text{\rm XE}}

\newcommand{\hint}[1]{{\footnotesize\itshape{\sffamily Hint:} #1}}

\makeatletter 
 \hypersetup{pdftitle = {Theory of quantum system certification -- a tutorial},
	     pdfauthor = {Martin Kliesch, Ingo Roth},
	     pdfsubject = {Quantum physics},
	     pdfkeywords = {quantum, state, process, tomography, characterization, 
	     				certification, verification, validation, 
	     				direct, shadow, fidelity, estimation,
	     				interleaved, randomized, benchmarking, 
	     				cross-entropy, XEB, 
	     				concentration, inequality, inequalities, tail, bound, bounds, 
	     				complex, projective, spherical, unitary, design, designs, t-design, t-designs, k-design, k-designs, random, 
						Schur-Weyl duality, Schur's lemma, 
	     				Choi-Jamiolkowski, isomorphism, channel state duality, 
	     				trace, diamond, norm, distance, average gate, entanglement, unitarity}
	    }
\makeatother

\newcommand{\hhu}{%
  Quantum Technology Group, 
  Heinrich Heine University D{\"u}sseldorf, 
  Germany%
}

\newcommand{\fu}{%
  Dahlem Center for Complex Quantum Systems,
  Freie Universit\"{a}t Berlin,
  Germany%
}

\newcommand{\tii}{%
  Quantum Research Centre,
  Technology Innovation Institute, Abu Dhabi,
  UAE
}

\begin{document}

\title{Theory of quantum system certification -- a tutorial}

\author{Martin Kliesch}
\email{info@mkliesch.eu}
\affiliation{\hhu}
\author{Ingo Roth}
\email{i.roth@fu-berlin.de}
\affiliation{\fu}
\affiliation{\tii}

\begin{abstract}
The precise control of complex quantum systems promises numerous technological applications  
including digital quantum computing. The complexity of such devices renders the certification of their correct functioning a challenge. To address this challenge, numerous methods were developed in the last decade. 

In this tutorial, we explain prominent protocols for certifying the physical layer of quantum devices described by quantum states and processes. Such protocols are particularly important in the development of near-term devices. Specifically, we discuss methods of direct quantum state certification, direct fidelity estimation, shadow fidelity estimation, direct quantum process certification, randomized benchmarking and cross-entropy benchmarking. Moreover, we provide an introduction to powerful mathematical methods, which are widely used in quantum information theory, in order to derive theoretical guarantees for 
the protocols. 
\vspace{2cm}
\end{abstract}

\maketitle

\tableofcontents

\newpage

\section{Introduction}
We are witnessing rapid progress in the experimental abilities to manipulate physical systems in their 
inner quantum properties such as state superposition and entanglement. 
Most importantly, we begin to have precise control over complex quantum systems on scales 
that 
are out of reach of simulations on even the most powerful existing classical computing devices. 
Harnessing their computational power promises the development of digital quantum computers that solve important problems 
much faster than any classical computer. 
Envisioned applications also 
include, e.g., the study of complex phases of matter in analogue simulations and cryptographically secure communication \cite{AciBloBuh18}.  
Hence, quantum technology promises highly useful devices with diverse domains of application ranging from  fundamental research
to commercial businesses. 

With the advent of these novel technologies comes the necessity for certifying their correct functioning. 
The certification of quantum devices is a particularly daunting task in the interesting regime of high complexity as 
most straightforward strategies based on classical simulations are bound to fail. 
Indeed, predicting the behaviour of complex quantum devices quickly exhausts the available classical computing power. 
Ironically, it is the same complexity that makes quantum technology powerful that hinders their certification. 
This challenging prospective has already motivated extensive effort in developing certification tools for quantum devices 
in the last decades. 

Intriguingly, numerous fields within the quantum sciences have tackled the 
problem of certification from a variety of different perspectives and have developed a large landscape of different
protocols. 
These protocols operate under very distinct assumptions and resource requirements that are well-motivated by 
the different perspectives. 
For example, certifying the correct function of a small-scale quantum device used in basic research 
allows one to invest sizable efforts. 
Here, one can potentially rely on a precise model of the physics of the device and might aim at a highly discriminative certificate providing plenty of information. 
A very different example is the certification of a server, correctly performing a quantum computation, by a 
remote client with standard desktop hardware. 
Such a protocol should be light-weight on the client-side and not rely on a detailed model of the server.

An attempt at a panoramic overview of the many approaches that all fall within 
the field of quantum certification was recently conducted in Ref.~\cite{Eisert2020QuantumCertificationAnd}. 
Therein, a very general classification framework for quantum certification protocols was proposed that is 
abstract enough to capture their wide range. 
Let us start by sketching the general framework. 
Thereby we can subsequently define the narrower scope of this tutorial. 

\subsection{Anatomy of quantum certification protocols}
\begin{figure*}[t]
  \begin{subfigure}[t]{.65\textwidth}
    \vskip 0pt
    \includegraphics[width=\textwidth]{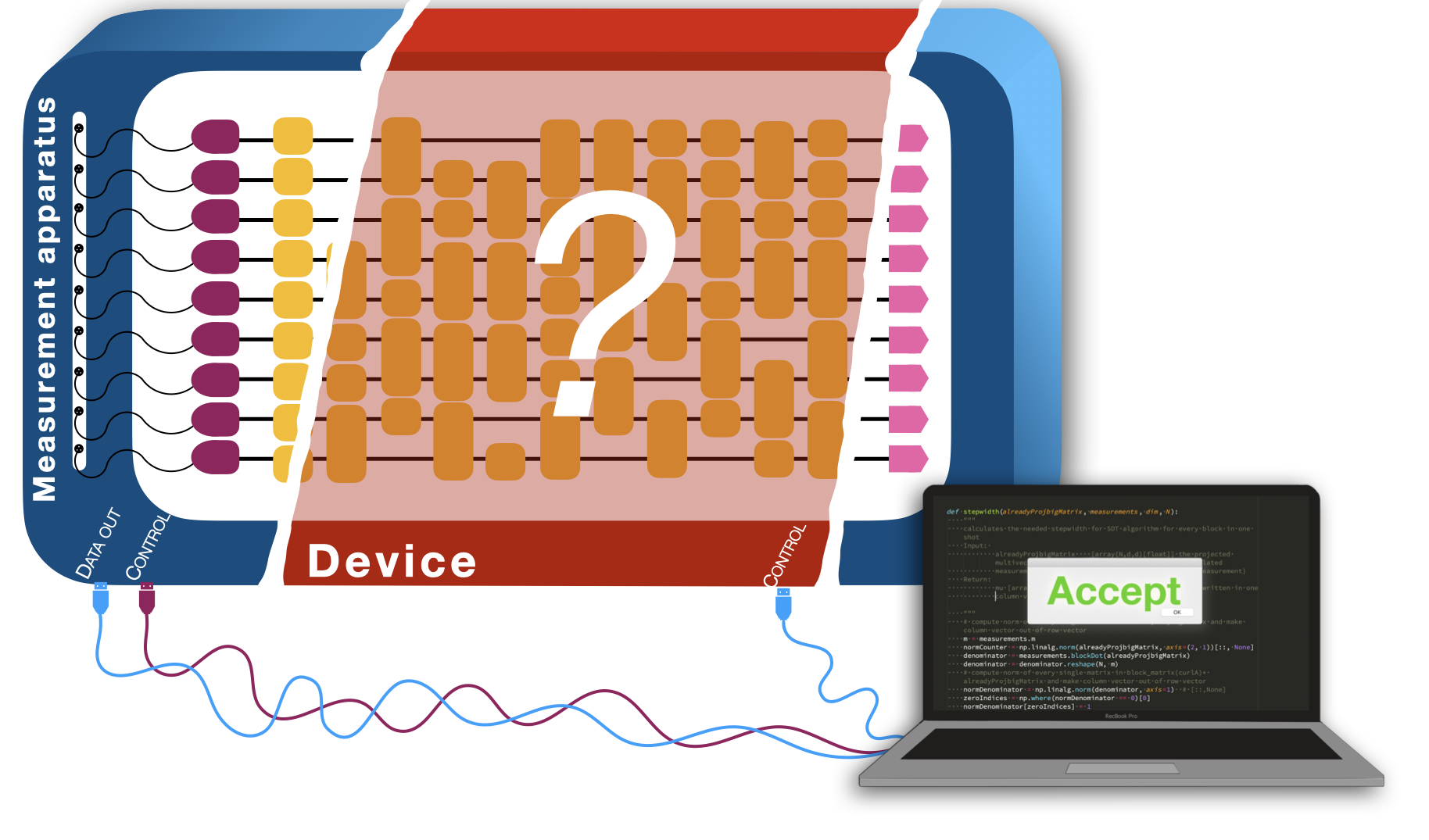}
  \end{subfigure}
  ~
  \begin{subfigure}[t]{.33\textwidth}
    \vskip 25pt
    \hspace{-1.8cm}
    \includegraphics[width=1.1\textwidth]{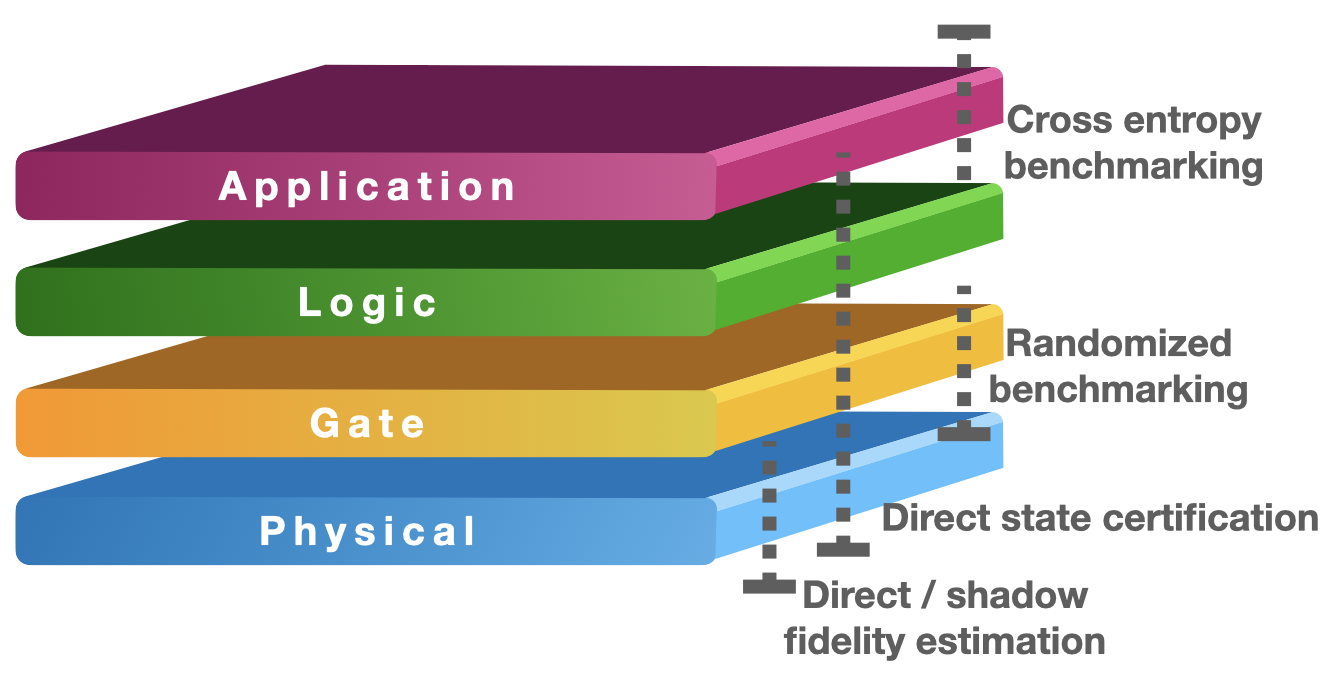}
  \end{subfigure}
  \caption{%
  \emph{Left:} the theoretical description of protocols makes use of the distinction into 
  the device, measurement apparatus and classical processor. 
  \emph{Right:} a complex quantum device 
  comprises multiple abstraction layers. 
  Different protocols aim at certifying the 
  functioning of the device on different layers. 
  NISQ devices are not expected to feature a powerful logical gate layer. 
  Instead, applications 
  are directly tailored to the physical gate layer. 
  \label{fig:anatomy}%
  }
\end{figure*}
A \emph{certification protocol} is a set of instructions that outputs either `accept' or `reject' concerning the 
hypothesis that the device is functioning correctly, with a certain level of confidence. 

The correct functioning of a device is defined in terms of a \emph{measure of quality}. 
Such measures range 
from rigorous worst-case discrimination of `fundamental' physical objects modeling the device, to 
performance benchmarks defined in terms of tasks directly on the application layer. 
Note that in principle a measure of quality can be solely defined in terms of a protocol that 
that can be reproducibly implemented. 
On the other hand, measures of quality that directly aim at the deviation of physical objects modeling 
the function of the device can provide an understanding of the device that is 
highly attractive in the development of the technology. 

In this tutorial, we encounter a couple of such physically motivated measures of quality and study their 
mathematical properties and operational interpretations. 
These measures all map to the real line. 
Certification protocols then provide $\epsilon$-certificates that 
reject the hypothesis of 
the measure of quality being larger than a given $\epsilon$. 
For this reason, most protocols that we present are \emph{estimation protocols} for specific measures of quality that 
can be easily turned into $\epsilon$-certification protocols by a standard method. 

Theoretically, it is convenient to describe the protocol as involving three distinct objects, Fig.~\ref{fig:anatomy} (left): 
First, the \emph{device} that is under scrutiny. 
Ideally, we try to be fairly conservative 
in the model and assumptions describing the device to be on the safe side.  
Second, the protocol employs a \emph{measurement apparatus}. 
The measurement apparatus, also a quantum device, is typically assumed to be much more precisely characterized compared to 
the device itself. 
Note that the device and measurement apparatus are not necessarily physically distinct devices. 
Choosing the split might be ambiguous and yield different formulation of the assumptions 
of the protocol. 
An extreme example are device-independent certification protocols that regard all quantum parts as a single device 
that is not subjected to any assumptions. In particular, they do not involve an anyhow characterized separate quantum measurement apparatus. 
The third object, is the \emph{classical processor}, a classical computing device, that 
might take care of potentially required pre- and post-processing tasks for the device control 
and the processing of the output data to arrive at a certificate or even 
communicates with the device and measurement apparatus in multiple rounds of an interactive protocol. 

The landscape of protocols can be roughly organized according to three `axes'.  
The first axis comprises the set of \emph{assumptions} that are imposed on the device and measurement apparatus to guarantee the functioning of 
the protocol. 

A second axis summarizes the \emph{complexity} of the resources that the protocol consumes. 
Each protocol requires  
a certain number of different measurement settings, its \emph{measurement complexity}, that each require the implementation of 
measurements that involve a certain \emph{quantum measurement complexity}. 
To arrive at a final statistical estimate, a total number of repetitions of device invocations are required, the scaling of which is referred to as the \emph{sample complexity}. 
Furthermore, as we already highlighted at the beginning, a particularly important figure of merit  
for a protocol is that it comes with practically manageable demands in space and time for the classical processing tasks, its \emph{classical processing complexity}. 
For our present scope, the mentioned complexity categories are the most important and are in the focus of our discussion. 
Note, however, that this list is by far not complete, for example, interactive protocols might be compared in terms of challenging demands in the timing of the device's control. 

The third and final axis is the \emph{information gain} of the protocol. 
At a first glimpse this might come as a surprise as a protocol that outputs `accept' or `reject' might 
be regarded as always providing one bit of information. 
But different measures of quality have different discriminatory power among the hypothesis class that models the device 
compatible with the protocol's set of assumptions. 
For example, let us imagine a device preparing quantum states on demand. 
We might require the device to produce a quantum state that is $\epsilon$-close in some distance measure to a specific target state. 
An alternative specification of the device might require it to always output the same quantum state but this quantum state should only be within a specified set of quantum states.
In this situation, we can roughly say that the information gain restricting the device (within its allowed hypothesis class) is higher in the first specification compared to the latter one. 

Concomitant with less information gain, it is conceivable that one can design a protocol for the latter specification with 
significantly less complexity compared to the first specification. 
Analyzing the information gain in performing a certification task often allows one to derive lower bounds on the complexity of any protocol for this task. 
Beside the discriminatory power of the measure of quality, other intermediate steps in the certification protocol can reveal significantly more information about the device than is ultimately reflected in the measure of quality and the final certificate.
For example, a potential certification protocol for our device that prepares quantum states might perform a high-precision, complete tomographic reconstruction of the quantum state and subsequently calculate the measure of quality using the tomographic estimate together with its error bounds. 
Conceptually, this example illustrates that certification is a subtask of the broader task of \emph{quantum system characterization}, that encompasses protocols aiming at different types of information about a quantum system, e.g., identification of a quantum system or testing for a specific property. 
Protocols that perform quantum system identification or property estimation naturally also give rise to certification protocols. 
Note that in practice, the hidden information gain of a certification protocol can provide valuable information to calibrate and improve the device. 

Another related task in \emph{quantum system characterization} is the \emph{benchmarking of quantum devices}. 
Benchmarking aims at comparing the performance of multiple devices. 
This can be done by comparing the achievable $\epsilon$-value of $\epsilon$-certificates of the respective devices. 
Benchmarking especially provides pragmatic impetus towards measures of quality that are not directly 
interpretable on the physical layer. 
Instead, for the benchmarking of quantum devices it suffices to implicitly define a reproducible performance measure 
by specifying a protocol that returns the measure. 
The only required justification is that the measure  
is expected to be correlated with the performance in practically relevant tasks. 

\subsection{Quantum certification for near-term devices -- scalable certification of the physical-layer}
In this tutorial, we focus on protocols that are particularly important for the certification 
of near-term quantum devices. 
These devices are still expected to be fairly noisy 
and of intermediate size, so-called \ac{NISQ} devices \cite{Pre18}. 
However, they are already in a regime of complexity where prominent certification methods that use
 full tomographic characterization become practically infeasible. 
On the other hand, there is still a large technological leap required in order 
to arrive at truly scalable devices, e.g.\ implementing fault-tolerant quantum computing. 
Such a full-fledged quantum device is described using multiple layers of abstraction from the
physical layer over, e.g.\ physical and logical gate layers, to an application layer, see Figure~\ref{fig:anatomy} (right). 
When a device already comes with multiple layers of abstraction one can also 
certify the functioning on the higher levels. 
\ac{NISQ} devices, however, allow only for a bit of abstraction above the physical layer. 
For this reason, near-term quantum devices pose the need for certification techniques 
that aim at the physical layer but are scalable to the intermediate 
system sizes of \ac{NISQ} devices. 
Such \emph{scalable certification methods for the physical layer} are the focus of this tutorial.

In the long term, for complex quantum devices high-level certification on the application level, also referred to as \emph{verification}, 
will become increasingly important. 
With cryptographic techniques quantum computations can be delegated to a remote server without revealing the actual computations. 
The correct execution of such \emph{blind quantum computation} can be verified in different settings without many assumptions \cite{ReichardtUngerVazirani:2013,fitzsimons_unconditionally_2017,mahadev_classical_2018,ColadangeloEtAl:2019:Verifier,GheorghiuVidick:2019}. 
We do not cover these methods in this tutorial. 
Instead, we refer to the review~\cite{GheorghiuEtAl:2019:Verification} of existing approaches for verifying quantum computations on devices that are close to being able to accurately perform a universal set of operations. 
Note that also in the long run, the scalable certification of the physical layer remains 
 important for the diagnostic of the components of more complex quantum devices 
in the development and during run-time. 

We model the physical layer generically in terms of quantum states and processes throughout the tutorial. 
The model is general enough to capture different types of quantum devices used, e.g., in quantum communication networks 
and analogue simulators. 
Nonetheless, we take the certification of digital quantum computing devices as our main 
guiding problem. 
Particularly, the last two methods that we discuss, \acf{RB} and \acf{XEB}, are specifically designed for digital quantum computing devices. 
\ac{RB} aims at estimating the physical noise that compromises a gate layer. \ac{XEB} aims at certifying the generation of samples from a probability distribution encoded in a quantum circuit. 
As such \ac{XEB} can be regarded as a certification for the application layer of 
a digital quantum computing device. But the application is deliberately designed very close to the physical layer. 

In addition, we chose a set of protocols that can be presented 
and analyzed using a common set of mathematical methods. 
This allows us to combine our presentation of the certification protocol with a detailed introduction 
into the mathematical formalism that is required in order to prove rigorous performance guarantees 
for the protocols. 

Lastly, we restrict our focus to certification protocols that employ measures of quality that are 
close to being natural measures of distance on the very fundamental physical description of the devices 
as quantum states and quantum processes. 
Also, important and equally fundamental, but not captured in this tutorial, is the certification of specific properties such as entanglement or non-classicality. 
Certain distinct properties, e.g.\ sufficiently high entanglement, allow for the certification of specific quantum states and processes even  
device-independently. 
This class of so-called \emph{self-testing protocols} is reviewed in Ref.~\cite{Supic2019SelfTesting}. 

One of the most intriguing aspects of the field of quantum certification is definitely 
the impressive stretch over multiple disciplines that come into play. 
Quantum certification is equally a field in applied mathematics, 
theoretical computer science,
applied numerical computer science, experimental physics and quantum hardware and software engineering. 
It comprises proofs of theorems, classical numerical studies of actual implementations, 
and performing the protocol in an actual quantum experiment 
 including a diligent analysis of 
‘real-world’ data. 
Each of the disciplines involved comes with its own methods accustomed to the arising challenges. 
At the same time, looking at certification on different stages from theory to experiment holds 
valuable lessons that go in both directions. 
Having said this,  
we present a practically well-motivated but 
 theoretical formal framework for 
 a set of quantum certification protocols. 
We do not delve into the exciting world of 
numerical and 
experimental implementations of the certification protocols
 that bring our model assumptions to the harsh scrutiny of ‘real-world’ physics. 
Instead, practical considerations and desiderata constantly serve as our motivation 
and inform our discussion. 

\subsection{Overview and structure}
The tutorial is divided into two major subsequent parts: the first part focusing on certification protocols for \emph{quantum states}, Section~\ref{sec:states}, and the second part focusing on certification protocols for \emph{quantum processes}, Section~\ref{sec:processes}. 
Furthermore, the tutorial consists of two different types of chapters: chapters that introduce the \emph{mathematical preliminaries}, and chapters that present and analyze the \emph{certification protocols}. 
We try to bring these two types of chapters in a dialog that goes back and forth between providing the motivation and tools for understanding the mathematical framework and protocols. 
The chapters on certification protocols conclude with suggestions for further reading on variants and extensions of the protocol and its theoretical analysis. 

We would like to highlight that the mathematical methods are core foundations of the broad field of theoretical quantum information and are by far not limited to quantum certification or even quantum characterization in their applications. 
Quite on the contrary, we expect the mathematical introductory chapters to serve as a valuable resource for students and researches working on quantum information in general. 
At the same time experts in quantum information mainly interested in the presented certification methods 
might want to simply skip the mathematical introductory chapters. They can conveniently find the protocol chapters in the table of contents by looking out for chapter titles that are typeset in italic font. 

In more detail, the mathematical methods and certification protocols presented here are the following: 
we start our discussion on quantum states with a brief introduction to the mathematical formalism of quantum mechanics, such as mathematical notions of operators and the modeling of quantum mechanical measurements (Section~\ref{sec:mathQM}). 
This allows us to formally introduce \emph{quantum state certification} as a one-sided statistical test in Section~\ref{sec:StateCertification}. 
Certification protocols rely on quantum mechanical measurements, which are probabilistic in nature. 
Therefore, the confidence of the protocols is controlled using so-called \emph{tail bounds} introduced in Section~\ref{sec:tail_bounds}. 
As an example for an application of tail bounds, we derive the estimation error and the confidence when estimating \emph{expectation values of observables} in Section~\ref{sec:observables_estimation}. 
In order to \emph{quantify} the accuracy of quantum state preparations, we introduce relevant metrics on quantum states in Section~\ref{sec:states:dist}.
A popular metric is given by the \emph{(Uhlmann) fidelity}. 
We provide a certification protocol in terms of the fidelity in Section~\ref{sec:DSC}. 
Stabilizer states are an important class of quantum states that can be certified with particularly few Pauli measurements (Section~\ref{sec:DSCofSTABs}). 
Another approach to certification employs estimation protocols. 
Estimating the fidelity requires more measurements compared to the one-sided certification protocol. 
A tool to reduce the measurement effort is \emph{importance sampling} introduced in Section~\ref{sec:ImportanceSampling}. 
\emph{\Acl{DFE}} uses this method to estimate the fidelity w.r.t.\ pure target states from relatively few state copies, Section~\ref{sec:fidelity_estimation}. 

For the remaining part of the tutorial random quantum states and random unitaries play an important role.  
For this reason, we introduce them in Section~\ref{chap:rep_theory}. 
Certain random unitary operations allow, in general, for an estimation of the fidelity from fewer state copies than \acl{DFE}, which we explain in Section~\ref{sec:shadow_fidelity_estimation} on \emph{\acl{SFE}}. 

We start our discussion of \emph{quantum processes} with some mathematical preliminaries (Section~\ref{sec:quantum_processes_and_measures}), where we introduce the Choi-\Jamiolkowski\ isomorphism (a.k.a.\ channel-state duality), process fidelity measures quantifying average-case error measures and a worst-case error measure, the \emph{diamond norm}. 
Most certification methods for quantum processes use average-case error measures. 
The presented quantum state certification methods can be translated to quantum processes using the Choi-\Jamiolkowski\ isomorphism. 
As an example, Section~\ref{sec:DQPC} presents the resulting protocol for direct quantum process certification.
Such translated protocols, typically require high-quality state preparations and measurements to probe the quantum processes. 
A method tailored to quantum gates that allows the \emph{average gate fidelity} to be extracted without requiring highly accurate state preparations and measurements is \emph{\acl{RB}} (Section~\ref{sec:RB}).
As our last protocol we discuss \emph{\acl{XEB}} in Section~\ref{sec:xeb};
this method has been used by Google to build trust in their recent experiment demonstrating the potential power of quantum computers in the task of generating certain random samples. 

\acresetall

\section{Quantum states}
\label{sec:states}

The first part of the tutorial is devoted to  protocols that aim at certifying that a quantum state generated by a device is the correct one. 
We start by quickly reviewing and introducing the mathematical formalism 
of quantum mechanics. 
We expect that most of the presented material and basic mathematical notions 
are already known to the reader. 
Therefore, we are fairly brief in our presentation and 
aim at quickly setting up the notation that we use throughout the tutorial. 
For sake of completeness, we provide many details on the mathematical formalism. 
However, the main ideas behind the protocols and their theoretical guarantees can also be followed with a more superficial understanding of the mathematical preliminaries. 

\subsection{Mathematical objects of quantum mechanics}
\label{sec:mathQM}
In order to discuss quantum states we set up some mathematical notation. 
We focus on finite-dimensional quantum mechanics in accordance with our emphasis on 
digital quantum computing. Hence, we assume all vector spaces to be finite-dimensional. 
The space of \emph{linear operators} from a vector space $V$ to a vector space $W$ is denoted by $\L(V,W)$, and we set $\L(V) \coloneqq \L(V,V)$. 
A \emph{Hilbert space} is a vector space with an inner product $\langle \cdot, \cdot \rangle$ (w.r.t.\ which it is complete). 
Let $\H$ and $\K$ be complex Hilbert spaces throughout the tutorial. 
We denote the \emph{adjoint} of an operator $X \in \L(\H, \K)$ by $X^\ad$, i.e.\ $\langle k , X h\rangle = \langle X^\ad k, h\rangle$ for all $h \in \mathcal H$ and $k \in \mathcal K$. 

As customary in physics, we use the bra-ket notation (Dirac notation): 
we denote vectors by ket-vectors $\ket\psi \in \H$ and linear functionals on $\H$ by bra-vectors $\bra\psi$, which are elements of the dual space $\H^\ast$. 
Furthermore, we understand ket-vectors and bra-vectors with the same 
label as being related by the canonical isomorphism induced by the inner product. 
In bra-ket notation we frequently drop tensor-product operators to shorten the notation, e.g.\ $\ket\psi\!\ket\phi \coloneqq \ket\psi\otimes \ket\phi \in \K \otimes \H$ or $\ketbra\psi\psi \coloneqq \ket\psi\otimes \bra\psi \in \K \otimes \H' \cong \L(\K, \H)$ for $\ket\psi \in \K$ and $\ket\phi \in \H$. 

To describe the state of a quantum system we require the notion of \emph{density operators}. 
The real subspace of \emph{self-adjoint} operators, $X=X^\ad$, is denoted by $\Herm(\H)\subset \L(\H)$ and the convex cone of \emph{positive semidefinite} operators by $\Pos(\H) \coloneqq \{X \in \Herm(\H) \mid \sandwich \psi X \psi \geq 0 \}$. 
The \emph{trace} of an operator $X \in \L(\H)$ is
$\Tr[X]\coloneqq \sum_i \sandwich iXi$, where $\{\ket{i}\}\subset \H$ is an arbitrary orthonormal basis of $\H$. 
The vector space $\L(\H)$ is itself a Hilbert space endowed with the Hilbert-Schmidt (trace) inner-product 
\begin{equation}\label{eq:HS_inner_prod}
	\langle X, Y\rangle \coloneqq \Tr[X^\ad Y]\, .
\end{equation}
The set of \emph{density operators} is defined as 
$\DM(\H) \coloneqq \{ \rho \in \PSD(\H): \ \Tr[\rho] = 1\}$. 

Outcomes of a quantum measurement are modeled by random variables. 
Abstractly, a \emph{random variable} is defined as a measurable function from a probability space to a measurable space $\mathcal X$. 
Here, we are exclusively concerned with two types of random variables: (i) those that take values in a finite, discrete set 
$\mathcal X \cong [n] \coloneqq \{1, \ldots, n\}$ (understood as the measurable space with its power set as the $\sigma$-algebra) 
and (ii) those that take values in the reals $\mathcal X = \RR$ (with the standard Borel $\sigma$-algebra generated by the open sets). 
In practice, the underlying probability space is often left implicit and one describes a 
random variable $X$ taking values in $\mathcal X$ directly by its probability distribution $\Pr$ that assigns a probability 
to an element of the $\sigma$-algebra of $\mathcal X$. 
For example, for a random variable $X$ taken values in $\RR$ and $I \subset \RR$ an interval, we write $\Pr[X \in I]$ for the probability of $X$ assuming a value in $I$. 
Abstractly speaking, $\Pr$ is the push-forward of the measure of the probability space to $\mathcal X$ induced by the random variable $X$. Thus, $\Pr$ is sufficient to describe $X$. 
The underlying probability space is, however, important to define correlations between multiple random variables which are understood to be defined on the same probability space. 

The probability distribution of a discrete random variable $X$ taking values in a finite set $\mathcal X \cong [n]$ is characterized by its \emph{probability mass function} $p_X: [n] \to [0,1]$, $k \mapsto p_X(k) \coloneqq \Pr[X = k] \coloneqq \Pr(X \in \{k\})$. 
A real random variable $X$ is characterized by its \emph{(cumulative) distribution function} $P_X: \RR \to [0,1]$, $x \mapsto P_X(x) \coloneqq \Pr[X < x] \coloneqq \Pr[X \in (\infty, x)]$ or in case it is absolutely continuous by its \emph{probability density function} $p_X: \RR \to [0,1]$,  $x \mapsto p_X(x) \coloneqq \left.\frac{d}{dt}\right|_x P_X(t)$. 
Note that if a discrete random variable takes values in a discrete subset of $\RR$ we can also assign a non-continuous (cumulative)  distribution function. 

The most general way to define a linear map from density operators $\DM(\H)$ to random variables is by means of a \emph{\ac{POVM}}. 
A \ac{POVM} is a map from (the $\sigma$-algebra) of $\mathcal X$ to $\Pos(\mathcal H)$. 
For a discrete random variable $X$ taking values in $[n]$ a \ac{POVM} is uniquely defined by a set of \emph{effects} $\{E_i \in \Pos(\mathcal H)\}_{i=1}^{n}$ with 
\begin{equation}
	\sum_{i=1}^n E_i = \1_{\H}\, ,
\end{equation}
where $\1_{\H}\in \L(\H)$ denotes the identity operator. 
Strictly speaking the \ac{POVM} is the map on the power set of $[n]$ that extends 
$k \mapsto E_k$ additively. 
It is convenient and common to refer to the set of effects as the \ac{POVM}. 
A \ac{POVM} \ $\mset$ (with effects) $\{E_i \in \Pos(\mathcal H)\}_{i=1}^{n}$ induces a map from $\DM(\H)$ to random variables. 
To this end, we associate to $\rho$ the random variable $\mset_\rho$ with probability mass function $p_{\mset_\rho}(k) \coloneqq \langle \rho, E_k \rangle$.

These are the ingredients to formalize the \emph{static} postulates of quantum theory.
We will only require dynamics in Section~\ref{sec:processes} on quantum process certification. 

\begin{block}[fonttitle=\normalfont]{Postulate (quantum states and measure\-ments):}
\begin{itemize}\MyItemSep{0}
	\item Every quantum system is associated with a (separable) complex Hilbert space $\H$. 

	\item The state of a quantum system, its \emph{quantum state}, is described by a density operator $\rho \in \DM(\H)$

	\item 
	A \emph{measurement} with potential outcomes in a finite, discrete set $O \cong [n]$ is 
	described by a \ac{POVM} $\mset$ with effects $\{E_i\}_{i \in [n]}$.

	\item If a quantum system is in the state $\rho \in \DM(\H)$ and the measurement $\mset$ is performed 
	the observed outcome is a realization of the random variable $\mset_\rho$ associated to $\rho$ by $\mset$.
\end{itemize}
\end{block}

The set $\DM(\H)$ is convex. 
Its extremal points are rank-one operators. 
A quantum state $\rho \in \DM(\H)$ of unit rank is called a \emph{pure} state. 
In particular, there exist a state vector $\ket\psi \in \H$ such that $\rho = \ketbra\psi\psi$. 
The state vector associated to a pure quantum state is only unique up to a phase factor. 
A general quantum state is therefore a convex combination of the form $\sum_i p_i \ketbra{\psi_i}{\psi_i}$, where $p$ is a \emph{probability vector}, i.e., an entry-wise non-negative vector $p\in \RR^d$, $p \geq 0$ that is normalized, i.e., $\sum_i p_i = 1$. 
A quantum state that is not pure is called \emph{mixed}.

Given two 
 quantum systems, their joint system should also be a quantum system. 
This expectation is captured by the following postulate. 

\begin{block}{Postulate (composite quantum systems):}
The Hilbert space of two quantum systems with Hilbert spaces $\mcH_1$ and $\mcH_2$, respectively,  is the tensor product $\mcH_1\otimes \mcH_2$. 
\end{block}

\noindent This construction induces an embedding from $\L(\mcH_1)$ into $\L(\mcH_1\otimes\mcH_2)$ by 
\begin{equation}
	A \mapsto A \otimes \1 \, .
\end{equation}
Dually to that, for any state $\rho\in \DM(\mcH_1\otimes\mcH_2)$, 
\begin{equation}
	\Tr[\rho\, ( A\otimes \1)] = \Tr[\rho_1 A] \, ,
\end{equation}
where $\rho_1$ is $\rho$ \emph{reduced} to system $1$; 
the reduced state captures all information of $\rho$ that can be obtained from measuring system $1$ alone and can be explicitly obtained by the \emph{partial trace} over the second subsystem
\begin{equation}
\begin{aligned}
\Tr_2 : \L(\mcH_1\otimes \mcH_2) &\to \L(\mcH_1) \qquad \text{(linear)}
\\
X\otimes Y &\mapsto \Tr_2[X\otimes Y] \coloneqq X \Tr[Y]
\end{aligned}
\end{equation}
as $\rho_1\coloneqq \Tr_2[\rho]$.

By $\FF\in \L(\mcH\otimes\mcH)$ we denote the \emph{flip operator} (or \emph{swap operator}) that is defined by linearly extending 
\begin{equation}\label{eq:SwapOp}
	\FF \ket\psi\ket\phi \coloneqq \ket\phi\ket\psi  \, .
\end{equation}
In a basis $\{\ket i\}_{i=1}^{\dim(\H)}$ of $\H$, we can express $\ket \psi \in \H \otimes \H$ by a coefficient matrix $A \in \CC^{\dim \H \times \dim \H}$ as $\ket \psi = \sum_{i,j} A_{ij} \ket i \ket j$. 
The coefficient matrix of $\FF\ket \psi$ is given by the matrix transpose $A^{\T}$ of $A$ with entries $(A^{\T})_{i,j} = A_{j, i}$.

\ex{
\begin{exerciseblock}{The swap-trick}
Let $\FF\in \L(\mcH\otimes\mcH)$ be the \emph{flip operator} \eqref{eq:SwapOp}.
Show that 
\begin{equation}\label{eq:swap-trick}
	\Tr[\FF(X\otimes Y)]
	=
	\Tr[XY]
\end{equation}
for any $X \in \L(\H)$. 
\end{exerciseblock}
}
\exreplace{
\begin{lemmablock}{The swap-trick}
Let $\FF\in \L(\mcH\otimes\mcH)$ be the \emph{flip operator} \eqref{eq:SwapOp}.
For any $X \in \L(\H)$ it holds that 
\begin{equation}\ifex\else\label{eq:swap-trick}\fi
	\Tr[\FF(X\otimes Y)]
	=
	\Tr[XY]\, .
\end{equation}
\end{lemmablock}
\begin{proof}
The identity can be checked by direct computation with basis elements or by using tensor network diagrams. 
We leave it as an exercise. 
\end{proof}
}

\subsection{\protchaptitle{A definition of quantum state certification}}
\label{sec:StateCertification}
In this section, we define 
what we mean by a certification test for a quantum state. 
This definition will serve as the blue-print for the specific protocols that we present in the subsequent sections of the chapter. 
A state certification test solves the task of making sure that a quantum state prepared by a device $\rhop$ is a sufficiently good approximation of a target state $\rhot$. 
Due to the statistical nature of quantum measurements, the protocol for a certification test
typically requires multiple copies of the quantum state. 
For this reason, it is appropriate to think of quantum state certification as the certification of a device that 
repeatedly prepares a target state $\rhot$. 

In this tutorial we restrict our attention to single round protocols, 
where a fixed number $\np$ of copies of a target state is prepared and measured subsequently. 
Without further assumptions the output of the device is described by an output state $\Rhop \in \DM((\CC^d)^{\otimes \np})$ on which the measurements are performed. 
Based on the measurement data the classical post-processor then decides to accept or reject the hypothesis that the device 
prepared the target state within a specified accuracy. 

\begin{figure}
  \centering
  \iftikz
  \leavevmode%
  \beginpgfgraphicnamed{fig_cert}%
  \input{./pics/cert.input}
  \endpgfgraphicnamed
  \else
  \includegraphics[width=.65\linewidth]{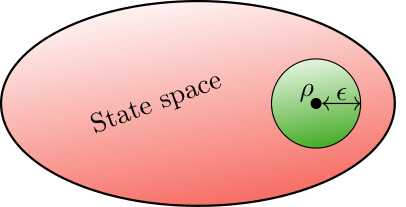}
  \fi
  \caption{%
	The task of quantum state certification is to detect when a state preparation $\rhop$ is not close to a chosen target state $\rhot$, i.e.\ when $\dist(\rhot,\rhop)>\epsilon$.  
  \label{fig:cert}%
  }
\end{figure}

This procedure is formalized by the notion of an $\epsilon$-certification test, illustrated in Fig.~\ref{fig:cert}. 
An $\epsilon$-certification test should output \accept\ if the prepared state is the targeted state in the majority of attempts. 
This requirement is referred to as \emph{completeness}. 
Additionally, one demands an $\epsilon$-certification to likely output \reject{} in case the prepared state deviates from the target state beyond a tolerance. 
The deviation is quantified in terms of a distance measure on $\DM(\CC^d)$ taking values in $\RR_+$, the non-negative reals, and `beyond tolerance' 
means that it exceeds a certain \emph{tolerated error threshold} $\epsilon > 0$. 
We arrive at the following definition for a single-round $\epsilon$-certification test. 

\begin{definitionblock}[{label={def:state_certification},fonttitle=\normalfont}]{Quantum state $\epsilon$-certification test}
Let $\rhot \in \DM(\CC^d)$ be a quantum state, the \emph{target state}, 
$\epsilon > 0$ and $\dist: \DM(\CC^d)  \times \DM(\CC^d) \to \mathbb R_{+}$ 
be a distance measure. 
An \emph{$\epsilon$-certification test for $\rhot$ w.r.t.\ $\dist$} consists 
of a quantum measurement on the device output 
$\Rhop \in \DM((\CC^d)^{\otimes \np})$ 
followed by classical post-processing of the measurement data 
outputting either {\accept} or {\reject} and satisfying the \emph{completeness} condition, 
\begin{equation}\label{eq:right_accept} 
\Rhop = \rhot^{\otimes \np} 
\ \Rightarrow \
\Pr[{\accept}] \geq \frac 23 \, , 
\end{equation}
and the \emph{soundness} condition holds for the reduced states $\rhop_i$ of $\Rhop$, 
\begin{equation}\label{eq:right_reject}
	\dist(\rhot, \rhop_i) > \epsilon \ \forall i \in [\np] 
	\ \Rightarrow \
	\Pr[{\reject}] \geq \frac23 \, .
\end{equation}
\end{definitionblock}

Note that more generally one could also define certification tests with respect to measures directly on the composite space $\DM((\CC^d)^{\otimes \np})$. 

The terms completeness and soundness are inspired by interactive proof systems. 
The role these conditions can be clarified from the 
perspective of statistical hypothesis testing. 
In hypothesis testing one has a \emph{null hypothesis} $H_0$ (often the hypothesis that one hopes to disprove) and an \emph{alternative hypothesis} $H_1$ and one needs to figure out which is true based on statistical data. 
In this setting, there are two types of error, 
\begin{align}
\PP[\,\text{accept $H_1$}\mid H_0]&\quad \text{(type-I error)}
\\
\PP[\,\text{accept $H_0$}\mid H_1]&\quad \text{(type-II error)} \, .
\end{align}
In state certification we choose the null hypothesis $H_0$ to be `$\dist(\rhop, \rhot)>\epsilon$' and `$\rhop=\rhot$' to be the alternative hypothesis $H_1$. 
Then, for the output of the $\epsilon$-state certification test, $\Pr[{\reject} \mid \rhop=\rhot]$ is the type-II error and 
$\Pr[{\accept} \mid \dist(\rhop, \rhot) > \epsilon]$
the type-I error. 
The completeness condition \eqref{eq:right_accept} corresponds to requiring that the type-II error is bounded by $1/3$. 
Analogously, the soundness condition \eqref{eq:right_reject} is the requirement that the type-I error is bounded by $1/3$.

For a test to meet the soundness and completeness condition additional 
assumptions on the prepared state $\Rhop$ can be required. 
A common assumption is that the device prepares a sequence of \emph{independent} states. 
This means that 
\begin{equation}\label{eq:globalRhop}
	\Rhop = \rhop_1 \otimes \rhop_2 \otimes \cdots \otimes \rhop_{n_\rho}
\end{equation}
with $\rhop_i \in \DM(\CC^d)$ for all $i$. 
In principle, it is also conceivable that a device prepares entangled states to maliciously 
trick a certifier working under the independence assumption. 
But in many circumstances minimal control over the device or beliefs about its physically plausible 
limitations justify the independence assumptions. 

An even stronger assumption is that the prepared states are \emph{independent and identically distributed (iid.)}. 
In this case, $\Rhop = \rhop^{n_\rho}$.  
In the experimental practice it can be challenging to fulfill this assumption. For example, drifts 
in environmental parameters of a device can yield to a systematic deviation of the 
state copies that defy the {iid.} assumption. 
Nonetheless, in many instances 
the {iid.} assumption may be justified by a basic understanding of the functioning of the device and valid to a sufficient degree. 
In some situations, the {iid.} assumption can be removed at the cost of a higher measurement effort using \cite{TakMor18} a quantum de Finetti theorem \cite{Li15QuantumDeFinetti} or an improved analysis \cite{Zhu2019VerificationAdversarialPRL,Zhu2019GeneralFrameworkFor}.

The arguably most important measure of complexity for an $\epsilon$-certification test is its sampling complexity. 

\begin{definitionblock}[label={def:certification_sampling_complexity}]{sampling complexity}
The \emph{sample complexity} of a family of 
tests $\{\mc T_{\np}\}$, each consuming $\np$ states, is (the scaling of) $\np$ with $d$ and $\epsilon$. 
\end{definitionblock}

The sampling complexity is the scaling of the number of states that the device needs to prepare for the test with the input parameters. 
In particular, in the context of digital quantum computing the statement that a ``protocol is efficient''  is often understood as 
having sampling complexity in $\LandauO(\polylog(d))$ as this translates into a sampling complexity in $\LandauO(\poly(n))$ for a system of $n$ qubits. 
Most guarantees that we prove for protocols in this tutorial, consist in upper bounds on the sampling complexity of a test. 

Another important measure for the practical feasibility of the protocol is the measurement complexity that quantifies how difficult 
it is to perform the quantum measurements of the protocol. 
In contrast to the precise definition of the sampling complexity, the measurement complexity should be regarded as a collection of different ways to formalize the demands of the measurement. 
For this reason, the discussion of the measurement complexity is of more qualitative nature. 

In the context of state certification, an important aspect of measurement complexity is the number of copies that the \ac{POVM} needs to act on simultaneously. 
The special case that encompasses all the presented protocols are \emph{sequential measurements} where the measurements are only performed on the $\np$ individual state copies separately. 
Therefore, the measurement device does not need to be able to store state copies before performing a measurement significantly lowering its complexity. 

Another relaxation of the measurement complexity of sequential measurements are \emph{non-adaptive measurements} where
the performed measurement on an individual copy does not depend on the previously obtained measurement results. 
Furthermore, the complexity of the implementation of the \ac{POVM} can be quantified, e.g.\ by measures for the complexity of the circuits required for its implementation in terms of local gates.  
The qualitative assessment of the measurement complexity as being experimentally feasible or not can vary widely for different devices and platforms. 

A certification test is only required to accept the target state. 
However, in practice, such test will accept states from some region around the target state with large probability. 
This property of a certification test is called \emph{robustness (against deviations from the target states)}. 
One way of how such a robustness can be guaranteed is by estimating the distance of the targeted state $\rho$ and the prepared state $\rhop$, 
as we see in Section~\ref{sec:fidelity_estimation} on \emph{fidelity estimation}. 
In this way, one obtains more information (a distance) than just {\accept} or {\reject}. 

Clearly, one can also certify through full quantum state tomography. 
However, the number of single sequential measurements in general required for tomography of a state $\rhop \in \DM(\CC^d)$ scales as $\Omega(d\rank(\rho))$ and as $\Omega(d^2\rank(\rho)^2)$ in the case two-outcome Pauli string measurements \cite{FlaGroLiu12}. 
So, for the relevant case of pure $n$-qubit states this number scales at least as $2^{n}$. 
This measurement effort becomes infeasible already for relatively moderate $n$. 

As we will see, fidelity estimation can work with dramatically fewer measurements than full tomography, when the target state has additional structure. 
In many situations, certification can work with even fewer measurements than fidelity estimation thanks to an improved $\epsilon$-dependence in the sample complexity. 

Our definition of a certification test used the somewhat arbitrary confidence value of $2/3$. 
It is not hard to see that as long as the failure probability is bounded away from $1$, the confidence can be amplified 
by repeating the test multiple times. 

\ex{
\begin{exerciseblock}[label={ex:confidence_amplification}]{Confidence amplification}
Let $\mc T_{n_\rho}$ be an $\epsilon$-certification test of a quantum state $\rho$ from $n_\rho$ iid.\ samples with maximum failure probability $\delta = \frac 13$. 
We repeat the certification test $N$ times and obtain a new certification test by performing a majority vote on the outcomes. 
Show that the new test satisfies the \emph{completeness} and \emph{soundness} conditions
\begin{align}
\sigma = \rho 
&\Rightarrow 
\Pr[{\accept}] \geq 1-\delta \, ,
\\
\dist(\rhot,\rhop)>\epsilon 
&\Rightarrow 
\Pr[{\reject}] \geq 1-\delta \, ,
\end{align}
for all $\sigma \in \DM(\CC^d)$, 
where $\delta = \e^{-c\,N}$ and $c>0$ is an absolute constant. 
The parameter $1-\delta$ is also called the \emph{confidence} of the test. 
\end{exerciseblock}
}
\exreplace{
\ifex
\begin{propositionblock}{Confidence amplification}
\else
\begin{propositionblock}[label={ex:confidence_amplification}]{Confidence amplification}
\fi
Let $\mc T_{n_\rho}$ be an $\epsilon$-certification test of a quantum state $\rho$ from $n_\rho$ iid.\ samples with maximum failure probability $\delta = \frac 13$. 
We repeat the certification test $N$ times and obtain a new certification test by performing a majority vote on the outcomes. 
Then the new test satisfies the \emph{completeness} and \emph{soundness} conditions
\begin{align}
\sigma = \rho 
&\Rightarrow 
\Pr[{\accept}] \geq 1-\delta \, ,
\\
\dist(\rhot,\rhop)>\epsilon 
&\Rightarrow 
\Pr[{\reject}] \geq 1-\delta \, ,
\end{align}
for all $\sigma \in \DM(\CC^d)$, 
where $\delta = \e^{-c\,N}$ and $c>0$ is an absolute constant. 
The parameter $1-\delta$ is also called the \emph{confidence} of the test. 
\end{propositionblock}
\begin{proof}
This statement can be checked directly from Definition~\ref{def:state_certification}. 
\end{proof}
}

We remark that the statement of this proposition also holds without the {iid.} assumption. 
Here, only the proof of the soundness condition \eqref{eq:right_reject} changes, since $\Rhop$ might be classically correlated or entangled across the $\np$ subsystems. 
However, one can show (see, e.g.\ \cite[Lemma~14.1]{KitSheVya02} for the argument) that the worst case, given by a $\Rhop$ with minimum rejection probability, corresponds to a product state. 
This statement can be proven by choosing a basis for $(\CC^d)^{\otimes \np}$ for which the local measurements are all diagonal. Then the measurement outcomes only depend on the diagonal entries of $\Rhop$ and, hence, a worst-case $\Rhop$ is a pure product state. 
This means that the worst case corresponds to {iid.} state preparations. 

Finally, we want to mention that, especially in the computer science community, \emph{certification} is often also called \emph{verification}. 
In particular from an epistemological point of view, a physical model or hypothesis can never be fully verified. 
Therefore, we will stick to the term \emph{certification} for the physical-layer where we actually model a device as being in a quantum state. 
This allows one to reserve the term \emph{verification} to certification on higher level of device abstraction such as the application layer.

\subsection{Estimation and tail bounds} 
\label{sec:tail_bounds}
A main technical tool for bounding the sampling complexity of certification protocols are tail bounds. 
The measurement outcomes of a quantum mechanical experiment are random variables. 
Recall that the expected value of a random variable $X$ on a probability space $(\Omega, \Sigma, P)$ is defined as 
\begin{equation}
	\EE[X] = \int_{\Omega} X(\omega) \rmd P(\omega),
\end{equation}
which gives rise to the standard expressions 
\begin{equation}
	\EE[X] = \sum_{k \in [n]} p_X(k) x_k \quad \text{and} \quad \EE[X] = \int_{\RR} x\, p_X(x) \rmd x
\end{equation}
for a discrete finite random variable $X$ taking values in $\{x_k\}_{k\in [n]}$ or a (absolutely continuous) real random variable $X$, with  
$p_X$ being the probability mass function or probability density function, respectively. 

When we want to estimate a measure of quality, such as a distance measure for quantum states, 
we have to construct an estimator for that measure, which is a function of measurement outcomes. 
An estimator $\hat E$ of a quantity $E$ can itself be viewed as a random variable (pushing forward the measure on the probability space). 
It is said to be \emph{unbiased} if $\EE[\hat E] = E$. 
Our estimators are typically families of random variables depending on a number of samples, i.e., the number of quantum states that the protocol consumes. 
In our notation we often leave this dependency implicit. 
We expect that if a protocol provides an estimator $\hat E$ then it reveals $E$ accurately in the limit of infinitely many samples. 
Such an estimator is called \emph{consistent} (if $\hat E$ converges to $E$ in probability). 
To capture the effect of finite statistics, we introduce the notion of an $\epsilon$-accurate estimator. 

\begin{figure}
  \centering
  \iftikz
  \leavevmode%
  \beginpgfgraphicnamed{fig_tail}%
  \input{./pics/tail.input}
  \endpgfgraphicnamed
  \else
  \includegraphics[width=.9\linewidth]{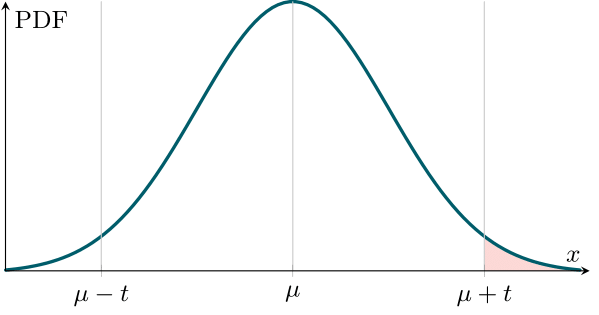}
  \fi
  \caption{%
  The (upper) \emph{tail} of a random variable $X$ is the probability of $X$ being greater than some threshold $t$. 
  This probability is given by the corresponding area under the graph of the probability density function (PDF) of $X$.  
  \label{fig:tail}%
  }
\end{figure}

\begin{definitionblock}{$\epsilon$-accurate estimator}
	Let $E \in \RR$ and $\epsilon, \delta > 0$. 
	A random variable $\hat E$ taking values in $\RR$ is an \emph{$\epsilon$-accurate estimator for $E$ with confidence $1 - \delta$} if 
	\begin{equation}
		\Pr[|\hat E - E| \leq \epsilon] \geq \delta\, .
	\end{equation}
\end{definitionblock}

The (scaling of) number of samples required for a family of estimators to be an $\epsilon$-accurate estimator is its sampling complexity. 
The sampling complexity of estimators 
can be derived using \emph{tail bounds} of random variables. 

Tail bounds for random variables are bounds to the probability that a random variable assumes a value that deviates from the expected value, as visualized by the marked area in Figure~\ref{fig:tail}. 
Indeed, for any non-negative random variable $X$ it is unlikely to assume values that are much larger than the expected value $\EE[X]$, as guaranteed by the following inequality. 

\begin{theoremblock}{{Markov}'s inequality}
Let $X$ be a non-negative random variable and $t > 0$. Then 
\begin{equation}\label{eq:MarkovsIneq}
	P_X(t) = \PP[X\geq t] \leq \frac{\EE[X]}{t} \, .
\end{equation}
\end{theoremblock}

\begin{proof}
Markov's inequality is as elementary as its proof. 
Let $(\Omega, \Sigma, P)$ be the probability space of $X$. 
For the proof we denote 
the indicator function $\vec 1_A$ of a subset $A \subset \Omega$ by
\begin{equation} \label{eq:def:indicator_func}
	\vec 1_A(\omega) \coloneqq 
	\begin{cases} 
	1 & \text{ if } \omega\in A\\
	0 & \text{ otherwise. }
	\end{cases}
\end{equation}
To prove Markov's inequality we set $A\coloneqq \{\omega: \ X(\omega) \geq t\}$ and observe that 
\begin{equation}
	t\, \vec 1_{\{\omega: \ X(\omega) \geq t\}}(\omega') \leq X(\omega')
\end{equation}
for all $\omega' \in \Omega$. 
Taking the expected value of both sides of this inequality finishes the proof. 
\end{proof}

As a consequence of Markov's inequality, the \emph{variance} of a real random variable $X$,
\begin{equation}
	\Var[X] = \EE[X^2] - \EE[X]^2\, ,
\end{equation}
can be used to control its tails: 

\begin{theoremblock}[label={thm:Chebyshev}]{{Chebyshev}'s inequality}
Let $X$ be a random variable, $\EE[X] = 0$, with finite variance $\sigma^2 \coloneqq \EE[X^2]$. 
Then 
\begin{equation}\label{eq:Chebyshev}
	\PP[|X|\geq t ] \leq \frac{\sigma^2}{t^2}
\end{equation}
for all $t\geq 0$. 
\end{theoremblock}

\begin{proof}
The proof follows by simply applying Markov's inequality to the random variable $X^2$. 
\end{proof}

Note that the assumption of mean zero is not really a restriction but only helps to state the theorem more concisely. 
In the case of a random variable $Y$ that does not necessarily have a zero mean, Chebyshev's inequality yields a tail bound by applying it to $X \coloneqq Y - \EE[Y]$; see also Figure~\ref{fig:tail}. 
The same argument can be made for the tail bounds below. 

A random variable $X$ is called \emph{bounded} if it takes values in a bounded subset of the reals almost surely. 
Its empirical mean is $\frac 1n\sum_{i=1}^n X_i$ where $X_i\sim X$ are iid.\ copies of $X$. 
In the case of bounded random variables, the empirical mean concentrates much more than a naive application of Markov's or Chebychev's inequality suggests. 
More precisely, the following inequality holds (see, e.g., \cite[Theorem~7.20]{FouRau13}). 

\begin{theoremblock}[label={thm:Hoeffdings}]{{Hoeffding's} inequality}
Let $X_1, \dots, X_n$ be independent bounded random variables with $a_i \leq X_i \leq b_i$ almost surely for all $i \in [n]$ and denote their sum by $S_n\coloneqq \sum_{i=1}^n X_i$. 
Then for all $t > 0$ it holds that
\begin{align}\label{eq:Hoeffdings}
	&\PP[S_n-\EE[S_n]\geq t] 
	\leq 
	\exp\myleft( -\frac{2\, t^2}{\sum_{i=1}^n (b_i-a_i)^2} \myright)
	\\\notag
	&\text{and}
	\\
	&\PP[|S_n-\EE[S_n]| \geq t] 
	\leq 
	2 \exp\myleft( -\frac{2\, t^2}{\sum_{i=1}^n (b_i-a_i)^2} \myright) . 
	\label{eq:HoeffdingsAbs}
\end{align}
\end{theoremblock}

\begin{proof}
We only sketch the proof and recommend to flesh out the details as an exercise. 
The second statement directly follows from the first one. 
In order to prove the first one, let $s>0$, apply Markov's inequality to 
\begin{equation}
	\PP[S_n - \EE[S_n] \geq t] 
	= 
	\PP\left[\e^{s(S_n - \EE[S_n])} \geq \e^{s\,t}\right]  . 
\end{equation}
The independence of the $X_i$ allows us to factorize the exponential and 
use the bounds on the range of $X_i$ individually. 
Finally, choosing the optimal $s$ yields the theorem's statement. 
\end{proof}

Note that when one can additionally control the variance of bounded random variables then the Bernstein inequality \cite[Corollary~7.31]{FouRau13} can give a better concentration, especially for small values of $t$. 

Another related tail bound is \emph{Azuma's inequality}, which allows for a relaxation on the independence assumption (super-martingales with bounded differences). 

The \emph{median of means estimator} is an estimator that allows for much better tail bounds than the empirical mean for the case of unbounded i.i.d.\ random variables with finite variance. 
The intuition is that taking the median of several empirical means 
is more robust against statistical outliers compared to taking the overall empirical mean. 

\begin{theoremblock}[label={thm:MME}]{Median of means estimator, 
	version of \cite[Theorem 2]{Lugosi2019MeanEstimation}}
Let $\{X_i\}$ be iid.\ random variables with mean $\mu$ and variance 
 $\sigma^2$ and denote by $S_k\coloneqq \frac 1k \sum_{i=1}^k X_i$ the empirical mean from $k$ i.i.d.\ samples. 
Take $l$ empirical means $S_{k,j}$, $j \in [l]$, that are (iid.) copies of $S_k$ and set
\begin{equation}\label{eq:MMestimator}
	\hat \mu \coloneqq \operatorname{median}(S_{k,1}, \dots, S_{k,\ell})\, .
\end{equation}
Then
\begin{equation}
	\PP\bigl[\bigl|\hat \mu - \mu \bigr| > \sigma\sqrt{4/\ell}\bigr] 
	\leq 
	\e^{-k/8} \, .
\end{equation}
In particular, for any $\delta \in (0,1)$, $k=\lceil 8\ln(1/\delta) \rceil$ and $m = k \ell$, 
\begin{equation}
	\bigl|\hat \mu - \mu \bigr|
	\leq \sigma \sqrt{\frac{32\, \ln(1/\delta)}{m}}
\end{equation}
with probability at least $1-\delta$. 
\end{theoremblock}

This theorem can be proven using Chebyshev's inequality for the empirical means $S_{k,j}$ and Hoeffding's inequality for a binomial distribution to obtain the concentration of the median. We refer to Ref.~\cite{Lugosi2019MeanEstimation} for further details.

Finally, it is often required to bound the probability that at least one of several events happens. 
For a series of events $A_1, A_2, \dots$ the \emph{union bound} (Boole's inequality) guarantees that
	\begin{equation}\label{eq:union_bound}
		\PP[A_1 \ \mathrm{or} \ A_2 \ \mathrm{or}\dots ]
		\leq 
		\sum_i \PP[A_i] \, .
	\end{equation}

\subsection{\protchaptitle{Expectation value estimation for observables}}
\label{sec:observables_estimation}

Now we familiarize ourselves with the application of tail bounds for the derivation of sampling complexities and turn our attention to a very basic task in quantum mechanics: 
the estimation of an expectation value of an observable. 

We formulate a general quantum measurement in terms of a \ac{POVM}. 
An important special case of a \ac{POVM} is a \emph{\ac{PVM}} where the effects are orthogonal projectors. 
A measurement described by a \ac{PVM} is also called a \emph{von Neumann / projective measurement}. 

An \emph{observable} quantity is modeled by a self-adjoint operator $A \in \Herm(\H)$. 
A self-adjoint operator has an eigendecomposition $A = \sum_{\alpha = 1}^{n} a_\alpha P_\alpha$ with $a_\alpha \in \mathbb R$ and 
orthogonal projectors $P_i$ onto the eigenspaces. 
The set of outcomes associated to the measurements of $A$ is its real eigenvalue spectrum $\spec(A) = \{a_\alpha\}_{\alpha \in [n]}$ and 
the measurement is described by the \ac{PVM} that has the projectors $P_\alpha$ as effects. 
Thus, associated to an observable $A$ is the map from $\DM(\H)$ to random variables $\rho \mapsto  A_\rho$ taking values in $\spec(A)$ with 
probability mass function $p_{{A}_\rho}(a_\alpha) = \Tr[P_\alpha \rho]$.
This implies that the expectation value of an observable $A\in \Herm(\H)$ in the state $\rho$ is $\langle A \rangle_\rho \coloneqq \EE[ A_\rho] = \Tr[\rho A]$. 

Given a quantum system in some state $\rho\in \DM(\H)$, we wish to estimate $\langle A \rangle_{\rho}$; note that the expectation value itself cannot be observed directly but needs to be estimated from single measurements. 
One protocol for estimating $\langle A \rangle_{\rho}$
 is to perform the projective measurement of the observable multiple times and use the observed 
empirical mean as an estimator for $\langle A \rangle_\rho$.
Let $A^{(i)}_\rho$ be the random variable describing the outcome of the $i$-th measurement of $A$ in state $\rho$. 
The \emph{empirical mean estimator} of $m$ measurements is 
\begin{equation} \label{eq:EmpMean}
	Y^{(m)} \coloneqq \frac 1m \sum_{i=1}^m  A^{(i)}_\rho\, .
\end{equation}
It is easy to see that $Y^{(m)}$ is an unbiased estimator for $\langle A\rangle_{\rho}$. 
So how many copies of $\rho$ does this protocol consume in order to arrive at an $\epsilon$-accurate estimate of $\langle A \rangle_\rho$ with confidence $1 - \delta$? 

If the measurements are independent and the eigenvalue spectrum of $A$ is bounded then Hoeffding's inequality~\eqref{eq:HoeffdingsAbs} yields a bound on the sampling complexity. 

\begin{propositionblock}[label={prop:EstimationExpectationVals}]{Estimation of observables}
Let $\rho \in \DM(\H)$ be a state and $A \in \Herm(\H)$ an observable with $\spec(A) \in [a, b]$. 
Choose $\epsilon > 0$ and $\delta \in (0,1)$. 
The empirical mean estimator \eqref{eq:EmpMean} of the expectation value $\langle A \rangle_\rho$ from 
 measurements of $A$ on $m$ independent copies of $\rho$ satisfies
\begin{equation}
|Y^{(m)} - \langle A \rangle_\rho| \leq \epsilon
\end{equation}
with probability at least $1-\delta$ for all
\begin{equation}\label{eq:ObsEstSampComp}
	m \geq m_0 = \frac{(b-a)^2}{2\epsilon^2} \, \ln \frac2\delta \, .
\end{equation}
\end{propositionblock}

\begin{proof}
Having $m$ independent state copies implies 
that the measurement outcomes are independent random variables. 
We choose $X_1, \ldots, X_m$ as independent copies of the random variable $A_\rho / m$. 
Then, the empirical mean estimator is described by a sum of $m$ independent random variables $Y^{(m)} = \sum_{k = 1}^{m} X_k$ 
with bounded range $X_k \in [a/m , b/m]$ for all $k$. 
Hoeffding's inequality yields 
\begin{equation}
	\PP\left[ \left| Y^{(m)} - \langle A \rangle_\rho\right| \geq \epsilon\right]
	\leq 
	2 \exp\myleft( - \frac{2m\epsilon^2}{(b-a)^2} \myright) 
\end{equation}
for any $\epsilon>0$. 
We wish this probability to be small, i.e., we require that 
\begin{equation}
	2 \exp\myleft( - \frac{2m\epsilon^2}{(b-a)^2} \myright) \leq \delta
\end{equation}
and determine the critical value $m_0$ required for the estimation by solving the inequality for $m=m_0$, which yields \eqref{eq:ObsEstSampComp}.  
\end{proof}

Proposition~\ref{prop:EstimationExpectationVals} guarantees that expectation values of bounded observables can be estimated with a measurement effort that is independent of the Hilbert space dimension. 
The confidence $1-\delta$ can be improved exponentially fast by increasing the measurement effort $m$. 

One can define distance measures on $\DM(\H)$ in terms of expectation values of a set of observables.  
Naturally, the estimation protocol described in this section gives rise to an $\epsilon$-certification test 
w.r.t.\ to such measures. 

\subsubsection*{Further reading}
Using the union bound, one can easily generalize Proposition~\ref{prop:EstimationExpectationVals} to derive the 
sampling complexity of estimating multiple observables. 
The total number of sufficient state copies $\rho$ to estimate $\ell$ different observables then scales as $m_0 \in \LandauO\myleft(\frac{\ell\log(2\ell /\delta)}{\epsilon^2}\myright)$.
In this setting each observable is estimated from a different measurement setting. 
In contrast, \emph{Shadow estimation} \cite{Aaronson2018ShadowTomography,Huang2019PredictingFeatures,Huang2020Predicting} provides a way to estimate multiple observables from a single measurement setting. 
For certain types of observables, the shadow estimation has sampling complexity of  
$m_0\in \LandauO\myleft(\frac{\ln(2\ell/\delta)}{\epsilon^2}\myright)$ \cite{Huang2020Predicting}. 

We further discuss shadow estimation techniques in the context of state certification in Section~\ref{sec:shadow_fidelity_estimation}.

\subsection{Distance measures for quantum states}
\label{sec:states:dist}
In our general definition of an $\epsilon$-certification test, Definition~\ref{def:state_certification}, 
requires a distance measure on $\DM(\H)$. 
In this section we introduce some `natural' measures on quantum states. 

To this end, recall that for any operator $X \in \L(\H,\K)$ between two Hilbert spaces $\H$ and $\K$, 
the operator $X^\ad X$ is positive semidefinite, i.e., in $\Pos(\H)$ (see Section~\ref{sec:mathQM}). 
In consequence, it has a positive semidefinite square root $\left|X\right| \coloneqq \sqrt{X^\ad X} \in \Pos(\H)$. 
Also recall that any \emph{normal} operator $X \in \L(\H)$, i.e., any operator that commutes with its adjoint, $[X,X^\ad] \coloneqq X X^\ad - X^\ad X = 0$, can be written in \emph{spectral composition} $X = \sum_i x_i P_i$, where $x_i\in \CC$ are its eigenvalues and $P_j = P_j^2 \in \Pos(\H)$ the corresponding spectral projectors. This decomposition can be used, for instance, to calculate $\left|X\right|$. 

There are several useful norms of an operator $X \in \L(\H,\K)$. 
The \emph{spectral norm} (a.k.a.\ operator norm) $\snorm{X} \in \RR_{+}$ of $X$ is defined to be the largest eigenvalue of $|X|$. 
The \emph{trace norm} is $\tnorm{X} \coloneqq \Tr[|X|]$ and the \emph{Frobenius norm} $\fnorm{X} \coloneqq \sqrt{\Tr[|X|^2]} = \sqrt{\Tr[X^\ad X]}$. 
These norms can be defined in several equivalent ways: 
the spectral norm coincides with the norm induced by the $\ell_2$-norm on $\H$ via $\snorm{X} = \sup_{\|v\|_{\ell_2} \leq  1}  \|Xv\|_{\ell_2}$, a manifestation of the Rayleigh principle. 
The Frobenius norm is induced by the \emph{Hilbert-Schmidt inner product} \eqref{eq:HS_inner_prod}. 
It can also be expressed in terms of a matrix representation of $X$ as $\fnorm{X} = \sum_{i,j} |X_{ij}|^2$. 
Finally, all three norms are instances of the \emph{Schatten $p$-norms} that are directly defined as $\ell_p$-norms on the singular value spectrum. 
The singular value spectrum $\sigma(X)$ of $X$ is defined as the eigenvalue spectrum of $|X|$ and the $\ell_p$-norms are given by $\norm{x}_{\ell_p} \coloneqq \left(\sum_i |x_i|^p\right)^{1/p}$. 
This gives rise to the unitarily invariant Schatten $p$-norm $\norm{X}_{p} \coloneqq \|\sigma(X)\|_{\ell_p}$ 
and $\snorm{\argdot}$, $\tnorm{\argdot}$,  and $\fnorm{\argdot}$ are the Schatten $p$-norms with $p = \infty, 1, 2$, respectively. 

The Euclidean inner product is bounded by $\ell_p$-norms through the Hölder inequality: for all $x, y \in \CC^d$ and pairs $p, q \in \{1, 2, \ldots, \infty\}$ with $p^{-1} + q^{-1} = 1$ (understanding $1/\infty = 0$) it holds that
\begin{equation}
	|\langle x, y\rangle| \leq \|x\|_{\ell_p} \|x\|_{\ell_q}.
\end{equation}
The Hölder inequality generalizes the Cauchy-Schwarz inequality where $p= q = 2$. 
The Schatten $p$-norms inherit a \emph{matrix Hölder inequality} from the Hölder inequality: 
let $X, Y \in \L(\H, \K)$ and $p, q$ as before, then 
\begin{equation}\label{eq:matrix_Hoelder}
	|\langle X, Y\rangle | \leq \tnorm{X^\ad Y} \leq \|X\|_p \|Y\|_q\, .
\end{equation}
The Hölder inequality directly follows from the von Neumann inequality 
$\Tr[|A B|] \leq \langle \sigma(A), \sigma(B) \rangle$ where the singular value spectra $\sigma(A)$ and $\sigma(B)$ are each in descending \cite{Bha13}. 
Furthermore, the Schatten $p$-norms inherit the ordering of the $\ell_p$-norms, $\|X\|_\infty \leq \ldots \leq \|X\|_2 \leq \ldots \leq\|X\|_1$ for all $X$. 
Norm bounds in reversed order will in general introduce dimensional factors. For low-rank matrices these bounds can be tightened. 
\begin{lemmablock}{Reversed norm bounds}
	For all $X \in \L(\H, \K)$ it holds that 
	\begin{equation}
		\tnorm{X} \leq \sqrt{\rank(X)} \fnorm{X} \leq \rank(X) \snorm{X}\, .
	\end{equation}
\end{lemmablock}
\begin{proof}
	Let $X \in \L(\H, \K)$ and $r = \rank(X)$. 
	We can always write $X = X P_r$ with $P_r$ a rank-$r$ projector onto the orthogonal complement of the kernel of $X$. 
	Now by the matrix Hölder inequality \eqref{eq:matrix_Hoelder}  $\tnorm{X} = \tnorm{X P_r} \leq \fnorm{P_r} \fnorm{X} = \sqrt{r} \fnorm{X}$. 
	For the second inequality, we use again the matrix Hölder inequality to obtain $\left|\Tr[X^\ad X]\right| \leq \tnorm{X^\ad X} \leq \tnorm{P_r} \snorm{X^\ad X} = r \snorm{X}^2$. 
	Taking the square root we conclude that $\fnorm{X} \leq \sqrt{r}\snorm{X}$ from which the second inequality follows. 
\end{proof}

A natural metric on quantum states is the \emph{trace-distance} $\dist_{\Tr}: \DM(\H) \times \DM(\H) \to \R_+$, 
\begin{equation}
	\dist_{\Tr}(\rhot, \rhop) = \frac12 \tnorm{\rhot - \rhop}. 
\end{equation}
We have already seen that compared to the other Schatten $p$-norms the trace norm is the largest one, i.e., provides the most `pessimistic' distance measure. 
Furthermore, the trace norm has an operational interpretation in terms of the distinguishability of quantum states by 
dichotomic measurements.  

\begin{propositionblock}[label={prop:tnorm_operational}]{Operational interpretation of the trace distance}
Let $\rho,\sigma\in \DM(\H)$. 
It holds that 
\begin{equation}
	\dist_{\Tr} (\rho, \sigma)
	=
	\sup_{0 \leq P \leq \1} \Tr[P(\rho-\sigma)] \, .
\end{equation}
Furthermore, the supremum is attained for the orthogonal projector $P^+$ onto the positive part of $\rho-\sigma$. 
\end{propositionblock} 

\begin{proof}
First we show that the supremum is attained for $P^+$. 
The self-adjoint operator difference can be decomposed as 
\begin{equation}
	\rho-\sigma = X^+ - X^-
\end{equation}
into a \emph{positive part} $X^+ \in \Pos(\H)$ and a \emph{negative part} $X^- \in \Pos(\H)$. 
We note that $\snorm{X^\pm} \leq 1$. Since $\Tr[X^+ - X^-]= \Tr[\rho-\sigma]= \Tr[\rho] - \Tr[\sigma] = 0$, we have 
$\Tr[X^+] = \Tr[X^-]$. 
Moreover, $\tnorm{\rho-\sigma} = \Tr[X^+] + \Tr[X^-]$. 
The last two statements together yield that the trace distance between the two states is
\begin{equation}\label{eq:tr_dist_Xplus}
	\frac 12 \tnorm{\rho-\sigma} 
	= 
	\Tr[X^+] 
	= 
	\Tr[P^+ (\rho-\sigma)] \, , 
\end{equation}
where $P^+$ is the orthogonal projector onto the support of $X^+$. 
It can be calculated by means of the singular value decomposition of $\rho - \sigma = U \Sigma V^\ad$ as 
$P^+ = U_+ V_+^\ad$ with $U_+$ and $V_+$ the matrices with singular left and right vectors, respectively, associated to the positive singular values as its columns. 

In order to show that the supremum cannot become larger than the trace distance, we consider some operator $P$ with $0\leq P\leq \1$.
Then, indeed, 
\begin{equation}
\begin{aligned}
\Tr[P (\rho-\sigma)]
&=  
\Tr[P X^+] - \Tr[P X^-]
\leq 
\Tr[P X^+]
\\
&\leq
\tnorm{X^+}
= \frac 12 \tnorm{\rho-\sigma}\, , 
\end{aligned}
\end{equation}
where we use the matrix H\"older inequality \eqref{eq:matrix_Hoelder} and \eqref{eq:tr_dist_Xplus} in the last two steps. 
\end{proof}

Given two quantum states the optimal dichotomic \ac{POVM} measurement $\{P, \1 - P\}$ to distinguish 
the two states is the \ac{POVM} that maximizes the probability of measuring the outcome associated to $P$ in one state and minimizes the 
same probability for the other state. 
Of course exchanging the role of $P$ and $\1 - P$ works equivalently. 
We can think of the achievable differences in probabilities as a measure for the distinguishability of $\rho$ and $\sigma$. 
Proposition~\ref{prop:tnorm_operational} shows that the trace distance of two states coincides with the maximal distinguishability by any dichotomic \ac{POVM} measurements. 
This distinguishability of a single shot measurement can be amplified by measuring multiple iid.\ copies of a quantum state with $\{P, \1-P\}$. 
We turn this insight into an $\epsilon$-certification test for pure states in the next section. 

Before we do this, let us introduce another important distance measure on quantum states. 
The \emph{(squared) fidelity} of two quantum states $\rho,\sigma\in \DM(\H)$ is defined as
\begin{equation}\label{eq:fidelity}
	\fidelity(\rho, \sigma)
	\coloneqq 
	\pnorm[1]{\sqrt \rho \sqrt \sigma}^2 \, .
\end{equation}
Note that 
\begin{equation}\label{eq:sqrt_rho_sigma}
	\pnorm[1]{\sqrt \rho \sqrt \sigma} 
	= 
	\Tr\Bigl[\sqrt{\sqrt{\rho}\, \sigma\sqrt{\rho}}\Bigr] \, .
\end{equation} 
While not any more directly evident from \eqref{eq:sqrt_rho_sigma}, 
the fidelity is symmetric  
as is apparent from \eqref{eq:fidelity}. 

Some authors define the fidelity as $\pnorm[1]{\sqrt \rho \sqrt \sigma}$ without the square. 
For this reason, one might want to refer to the expression of \eqref{eq:fidelity} explicitly as the squared fidelity to avoid 
confusion. 
For brevity, we however call $\fidelity$ simply the fidelity hereinafter. 

The fidelity is more precisely not a measure of `distance' for two quantum states but of ``closeness''. 
In particular, $\fidelity(\rho, \rho) = 1$, which can be seen to be the maximal values of $\fidelity(\rho, \sigma)$ for 
all $\rho, \sigma \in \DM(\H)$. 
Hence, $0 \leq \fidelity(\rho, \sigma) \leq 1$ on $\DM(\H)$. 
Often it is convenient to work with the \emph{infidelity} $1-\fidelity(\rho,\sigma)$ as the complementary measure of `distance'. 

When at least one of the states $\rho$ or $\sigma$ is pure, say $\rho = \ketbra \psi \psi$ then
\begin{equation}\label{eq:pure_state_fidelity}
	\fidelity(\rho,\sigma) 
	= 
	\sandwich \psi \sigma \psi 
	=
	\Tr[\rho \sigma] = \langle \rho, \sigma \rangle
	\, ,
\end{equation}
which can easily be proven using \eqref{eq:sqrt_rho_sigma}. 
Furthermore, for both states being pure we have $\fidelity(\ketbra\psi\psi, \ketbra\phi\phi) = |\!\braket\psi\phi\!|^2$ for all $\ketbra \psi \psi, \ketbra\phi\phi \in \DM(\H)$. 
Thus, for pure states the fidelity is the overlap of the states and can be related to the angle between the state vectors. 
In fact, we mostly encounter the case where at least one of the states is pure and mostly work with \eqref{eq:pure_state_fidelity} instead of \eqref{eq:fidelity}. 

The fidelity is related to the trace distance as follows. 

\begin{propositionblock}{Fuchs-van-de-Graaf inequalities \cite[Theorem~1]{FucGra97}}
 For any states $\rho, \sigma\in\DM(\H)$
\begin{equation}\label{eq:Fuchs-van-de-Graaf}
	1-\sqrt{\fidelity(\rho, \sigma)} 
	\leq 
	\frac 12 \pnorm[1]{\rho-\sigma} 
	\leq 
	\sqrt{1-\fidelity(\rho, \sigma)} \, . 
\end{equation}
\end{propositionblock}

Since the Fuchs-van-de-Graaf inequalities are not explicitly dependent on the Hilbert-space dimension
one can regard the trace-distance and fidelity as equivalent measures of quality in many applications. 
Note however that the square root on the right-hand side can still make a painstaking difference in practice. 
Aiming at a trace-norm distance of $10^{-3}$ can in the worst-case require to ensure 
an infidelity of $10^{-6}$. 
This can be a crucial difference when it comes to the practical feasibility of certification. 
Importantly, the square-root scaling is unavoidable for 
pure states. 
\ex{
\begin{exerciseblock}[label={ex:fvdg_purestates}]{Fuchs-van-de-Graaf inequality for pure states}
The upper bound of the Fuchs-van-de-Graaf inequality for pure states 
$\ketbra\psi\psi, \ketbra\phi\phi \in \DM(\H)$ is tight. 
To show this proof that the following equality holds for 
$p=1$
\begin{equation}\label{eq:pnorm_rank1}
	\pnorm{\ketbra \psi \psi - \ketbra \phi \phi} 
	= 
	2^{1/p}\sqrt{1 - \left|\braket\psi\phi\right|^2} \, .
\end{equation}
Furthermore, show that the equality actually holds for all Schatten-$p$-norms $p \in \{1, 2, \ldots, \infty\}$.
\end{exerciseblock}
}
\exreplace{
\ifex
\begin{lemmablock}{Fuchs-van-de-Graaf inequality for pure states}
\else
\begin{lemmablock}[label={ex:fvdg_purestates}]{Fuchs-van-de-Graaf inequality for pure states}
\fi
The upper bound of the Fuchs-van-de-Graaf inequality for pure states 
$\ketbra\psi\psi, \ketbra\phi\phi \in \DM(\H)$ is tight and 
\begin{equation}\ifex\else\label{eq:pnorm_rank1}\fi
	\pnorm{\ketbra \psi \psi - \ketbra \phi \phi} 
	= 
	2^{1/p}\sqrt{1 - \left|\braket\psi\phi\right|^2} 
\end{equation}
holds for all $p$. 
\end{lemmablock}

\begin{proof}
Denote $X\coloneqq \ketbra \psi \psi - \ketbra \phi \phi$. 
We have $\Tr[X]= 0$ and $\rank(X) \in \{0,2\}$. 
Hence, $X$ has two eigenvalues $\lambda>0$ and $-\lambda<0$. 
This implies that $\lambda^2 = \fnorm{X}^2/2 = 1 - 1\left|\braket\psi\phi\right|^2$, as directly follows by writing $\fnorm{X}^2$ as a Hilbert-Schmidt inner product. 
From the eigenvalues one can calculate $\pnorm{X}$ as Schatten $p$-norm, which yields the result. 
\end{proof}
}

In
\ex{Exercise}%
\exreplace{%
\ifex\else%
Lemma%
\fi%
}%
\ref{ex:fvdg_purestates}, 
we showed that the upper bound of \eqref{eq:Fuchs-van-de-Graaf} is tight for pure states. 
Conversely, one might hope for more mixed states to arrive at an improved scaling closer to the lower bound of Eq.\ 
\eqref{eq:Fuchs-van-de-Graaf}. 
We will review such a bound in the analogous discussion of distance measures of quantum channels, 
Theorem~\ref{thm:RiKbound} in Section~\ref{sec:quantum_processes_and_measures}.

In the next section, we present protocols that aim 
at directly providing an $\epsilon$-certification test for certain states. 
Section~\ref{sec:fidelity_estimation} and \ref{sec:shadow_fidelity_estimation} present 
two protocols that aim at estimating the fidelity: 
\emph{direct fidelity estimation} and \emph{shadow fidelity estimation}.

\subsection{\protchaptitle{Direct quantum state certification}}
\label{sec:DSC}
In this section, we present approaches to certification protocols for quantum states that are direct in that 
they do not use a protocol designed for another task, such as an estimation protocol, as a subroutine. 
Our exposition largely follows the work by Pallister \emph{et al.}~\cite{PalLinMon18}. 
We start with perhaps the most direct attempt building on the insight of Proposition~\ref{prop:tnorm_operational}. 
This proposition illustrates the interpretation of the trace distance as the maximal distinguishability 
by a dichotomic \ac{POVM} and shows that the optimal \ac{POVM} in this regard is given by the projection onto the positive part of 
the state difference. 
This indicates that the best way to distinguish a pure quantum state from all other states 
is to measures the \ac{POVM} that has the state itself as an element. 

We now turn this insight into an $\epsilon$-certification test. 
It can be most easily formulated in terms of the \emph{infidelity} $1-\fidelity$ as the distance measure. 

Given a pure target state $\rhot = \ketbra\psi\psi$ with a state vector $\ket\psi \in \CC^d$, 
we consider the \ac{POVM} $\{\Omega, \1 - \Omega\}$ given by $\Omega = \ketbra\psi\psi$. 
We call the outcome corresponding to $\Omega$ {\pass} and the one of $\1-\Omega$ {\fail}. 
Then, for any $\rhop\in \DM(\CC^d)$ we have 
\begin{equation}
	\Pr[\pass] = \Tr[\Omega \rhop] = \fidelity(\rhot,\rhop) \, ,
\end{equation}
i.e., the probability of the \ac{POVM} returning $\pass$ is the fidelity of the two states. 
This gives us a simple protocol that measures the \ac{POVM} on a single state copy and accepts when the result is $\pass$ and rejects otherwise. 
This protocol is complete but not sound in the sense of Definition~\ref{def:state_certification} as the probability of an acceptance is fixed to be $1-\fidelity(\rhot,\rhop)$, i.e., the probability of a false acceptance in not constantly bounded away from one. 
But using more state copies we can boost the probability to detect deviations of the form $\fidelity(\rhot,\rhop)< 1- \epsilon$ with some targeted confidence $1-\delta$. 

In order to be able to capture a class of large measurement settings we first formulate the protocol
for an arbitrary dichotomic \ac{POVM} measurements. 

\begin{protocolblock}[label={protocol:QSC_unconstrained}]{Naive direct quantum state certification}
Let $\rhot \in \DM(\CC^d)$ be a pure target state and $\Omega\in \Pos(\CC^d)$ with $\snorm{\Omega}\leq 1$. 
Denote by $\{\Omega, \1-\Omega\}$ the binary \ac{POVM} given by $\Omega$, 
call the outcome corresponding to $\Omega$ {\pass} and the one of $\1-\Omega$ {\fail}. 

For state preparations $\rhop_1, \dots, \rhop_{\np}\in \DM(\CC^d)$ the protocol consists of the following steps. 
\begin{algorithmic}[1]
\For{$i \in [\np]$} 
    \State measure $\{\Omega, \1-\Omega\}$ on $\rhop_i$
    \If{the outcome is {\fail}}:
        \State output {\reject} and end protocol
    \EndIf
\EndFor\label{algline:EndFor}
\State output {\accept}
\end{algorithmic}
\end{protocolblock}
As stated, this protocol is adaptive in that it can end early in case of a rejection instance. 
However, one could easily turn into a non-adaptive protocol without changing the number of measurements in the performance guarantee below. 

For a pure state $\rho$ and measurement $\Omega = \rho$ the protocol is a certification protocol w.r.t.\ the infidelity as more precisely summarized by the following proposition. 
\begin{propositionblock}[label={prop:QSCsimple_guarantee}]{Performance guarantee I}
Let $\rhot\in \DM(\CC^d)$ be a \emph{pure} target state and choose $\epsilon, \delta>0$.
Protocol~\ref{protocol:QSC_unconstrained} with $\Omega = \rhot$ is an $\epsilon$-certification test w.r.t.\ the infidelity from $\np$ independent samples for 
\begin{equation}\label{eq:prop:np}
	\np \geq  \frac{\ln(1/\delta)}{\epsilon}
\end{equation}
with confidence at least $1-\delta$.
Moreover, the protocol accepts the target state $\rhot$ with probability~$1$. 
\end{propositionblock}

\begin{proof}
The probability of the measurement outcome {\pass} in step $i\in [\np]$ is 
\begin{equation}
	\PP[{\pass}|\rhop_i]
	= 
	\Tr[\Omega \rhop_i] 
	= 
	\Tr[\rhot\rhop_i]
	= 
	\fidelity(\rhot, \rhop_i) \, .
\end{equation}
Hence, the final probability that the protocol accepts is 
\begin{equation}
	\PP[\accept] 
	= 
	\prod_{i=1}^\np \fidelity(\rhot, \rhop_i) \, .
\end{equation}

Clearly, if $\rhop_i = \rhot$ for all $i \in [\np]$ then the protocol accepts almost surely. 
Now let us consider the case that the fidelity is small, i.e., 
\begin{equation}
	\fidelity(\rhot,\rhop_i) 
	= 
	\sandwich\psi{\rhop_i}\psi \leq 1-\epsilon  \qquad \forall i \in [\np]. 
\end{equation}
Then the probability that the protocol wrongfully accepts is 
\begin{equation}
	\PP[\accept] 
	\leq
	(1-\epsilon)^\np \, .
\end{equation}
Now we wish this probability (type-II error) be bounded by $\delta>0$, i.e., 
\begin{equation}\label{eq:QSC_simple_bound_delta}
	(1-\epsilon)^\np \leq \delta \, .
\end{equation}
This bound on the type-II error satisfied for 
\begin{equation}\label{eq:np}
	\np \geq \frac{\ln\left(\frac 1\delta\right)}{\ln\left( \frac 1{1-\epsilon} \right)} \, .
\end{equation}
We note that for $\epsilon \in [0,a] \subset [0,1)$ the following bounds hold 
\begin{equation}\label{eq:lnfrac_epsilon_bound}
	\epsilon 
	\leq 
	\ln\left( \frac 1{1-\epsilon} \right) 
	\leq 
	\ln\left( \frac 1{1-a} \right) \frac{\epsilon}{a} \, ,
\end{equation}
which can be seen by using the fact that $\epsilon \mapsto \ln\left( \frac 1{1-\epsilon} \right)$ is smooth, has value $0$ at $0$, its first derivative is lower bounded by $1$, and its second derivative is positive. 
Hence, for any $\np \geq  \frac{\ln(1/\delta)}{\epsilon}$ the required bound \eqref{eq:QSC_simple_bound_delta} is satisfied. 
\end{proof}

As a remark, the minimum number of samples in Eq.\ \eqref{eq:np} scales as
\begin{equation}
	\frac{\ln\left(\frac 1\delta\right)}{\ln\left( \frac 1{1-\epsilon} \right)} = \frac{\ln(1/\delta)}{\epsilon} + \LandauO(1/\epsilon^2) \, , 
\end{equation}
so that \eqref{eq:prop:np} captures the leading scaling of \eqref{eq:np}, see also the bounds \eqref{eq:lnfrac_epsilon_bound}. 

Perhaps surprisingly, the sample complexity \eqref{eq:prop:np} of this direct certification protocol does not depend on the physical system size at all. 
It has a zero type~I error and one can control the type~II error via the parameter $\delta$. 
However, for many target states it is not practical to directly implement the required \ac{POVM}. 
This motivates the following more complicated strategies. 
Say, we have access to a set of \ac{POVM} elements 
\begin{equation}\label{eq:mset_subset}
	\mset \subset \{\mop \in \Pos(\CC^d): \ \snorm{\mop} \leq 1 \}\, .
\end{equation}
These encode the measurements that are experimentally feasible. 
As one can only make finitely many measurements, we assume that $|\mset|< \infty$. 
Then for each state preparation we pick a \ac{POVM} element $M \in \mset$ with some probability and consider the corresponding dichotomic \acp{POVM}
$\{M, \1-M\}$, where $M$ has output {\pass} and $\1-M$ has output {\fail}.
We refer to a set $\mset$ of the form \eqref{eq:mset_subset} together with a probability mass $\mu: \mset \to [0,1]$, $\sum_{M \in \mset} \mu(M) = 1$, 
as a \emph{probabilistic measurement strategy}. 
Now we modify Protocol~\ref{protocol:QSC_unconstrained} by including a probabilistic measurement strategy. 

\begin{protocolblock}[label={protocol:QSC}]{Direct state certification}
Let $\rhot \in \DM(\CC^d)$ be a pure target state and $(\mset, \mu)$ be a probabilistic measurement strategy.
For state preparations $\rhop_1, \dots, \rhop_{\np}\in \DM(\CC^d)$ the protocol consists of the following steps. 
\begin{algorithmic}[1]
\For{$i \in [\np]$} 
    \State Draw $M$ from $\mset$ according to $\mu$. 
    \State Measure the \ac{POVM} $\{M, \1-M\}$ on $\rhop_i$.
    \If{the outcome is {\fail}}:
        \State output {\reject} and end protocol. 
    \EndIf
\EndFor
\State output {\accept}
\end{algorithmic}
\end{protocolblock}

Let us assume that the prepared states are iid.\ copies of a state $\rhop$. 
Then 
the overall probability of measuring {\pass} is 
\begin{equation}\label{eq:indirect_measurement_Omega}
	\PP[{\pass}] = \sum_{M\in \mset} \mu(M) \Tr[M \rhop] = \Tr[\Omega \rhop], 
\end{equation}
where
\begin{equation}\label{eq:Omega}
	\Omega\coloneqq \sum_{M \in \mset} \mu(M) M 
\end{equation}
is the so-called \emph{effective measurement operator}. 
Below, we see that it plays a similar role as the measurement operator $\Omega$ in Protocol~\ref{protocol:QSC_unconstrained} when it comes to proving performance guarantees. 
At the same time, it allows capturing more sophisticated measurement strategies. 

However, there is one constraint that allows for a simple analysis of Protocol~\ref{protocol:QSC}: we require that 
\begin{equation}\label{eq:OmegaRhoEqOne}
	\Tr[\Omega\rhot]=1 \, ,
\end{equation}
i.e., that there is no false reject of the target state $\rho$ with probability one. 
In particular, it requires that $\Tr[M\rhot]=1$ for all $M \in \mset$. 
This constraint still allows for optimal measurement strategies, which is guaranteed by the following. 

\begin{propositionblock}{\cite[Proposition~8]{PalLinMon18}}
Let $\rhot=\ketbra\psi\psi$ be a target state.
Let $0\leq \Omega'\leq \1$ be an effective measurement operator \eqref{eq:Omega} with $\Tr[\Omega'\rhot]<1$ so that Protocol~\ref{protocol:QSC} is an $\epsilon$-certification test w.r.t.\ infidelity from $\np'$ iid.\ samples. 
Then there exists an effective measurement operator $0\leq \Omega\leq \1$ with $\Tr[\Omega\rhot]=1$ so that Protocol~\ref{protocol:QSC} is an $\epsilon$-certification test w.r.t.\ infidelity from $\np$ iid.\ samples so that $\np \leq \np'$ holds for sufficiently small $\epsilon$. 
\end{propositionblock}

The proof of this statement is a consequence of the Chernoff-Stein lemma from information theory, which quantifies the asymptotic distinguishability of two distributions in terms of their relative entropy. 

Since the constraint \eqref{eq:OmegaRhoEqOne} implies that there is no \emph{false rejection} the only remaining hypothesis testing error is a \emph{false acceptance}, which is the event where a state $\rhop$ with $\fidelity(\rhot,\rhop)< 1-\epsilon$ is accepted. 
This event has a worst-case probability over all states $\rhop$ in the rejection region that given by the optimization
\begin{equation}\label{eq:false_pass}
	\PP[\pass\mid \text{``$\epsilon$-worst case''}]
	=
	\max_{\substack{\rhop\in \DM(\CC^d):\\ \,\Tr[\rhot\rhop]\leq 1-\epsilon}} \Tr[\Omega\rhop] \, .
\end{equation}
In the following lemma we see that this maximum is determined by the \emph{spectral gap} 
\begin{equation}\label{eq:SpectralGap}
	\nu(\Omega) \coloneqq \lambda_1(\Omega) - \lambda_2(\Omega) \, ,
\end{equation}
of the effective measurement operator $\Omega$, 
where $\lambda_1(\Omega) \geq \lambda_2(\Omega) \geq \ldots \geq \lambda_d(\Omega)$ are the eigenvalues of $\Omega$ in descending order. 

\begin{lemmablock}[{label={lem:QSC-single_measurement_worst_case},fonttitle=\normalfont}]{{\cite{PalLinMon18}}, {\cite[Suppl. material, Section~I]{ZhuHay18}}}
\hfill\\
Let $\rhot\in \DM(\CC^d)$ be a pure state, $0\leq \Omega \leq \1$, $\Tr[\rhot\Omega]=1$, and $\epsilon>0$. 
Then
\begin{equation}\label{eq:worst_case_rhop}
	\max_{\substack{\rhop\in \DM(\CC^d):\\ \,\Tr[\rhot\rhop]\leq 1-\epsilon}} \Tr[\Omega\rhop]
	=
	1 - \nu(\Omega)\epsilon \, . 
\end{equation}
\end{lemmablock}

\begin{proof}
We note that $\Tr[\rhot\Omega] = 1$ means that a state vector $\ket \psi$ with $\rhot = \ketbra\psi\psi$ is an eigenvalue-$1$ eigenvector of $\Omega$. 
Moreover, let us write $\Omega$ in spectral decomposition, 
\begin{equation}
	\Omega = \sum_{j=1}^d \lambda_j P_j
\end{equation}
with $1 = \lambda_1 \geq \lambda_2\geq \dots\geq \lambda_d$ and $P_1 = \rhot$.  
For the case $\lambda_2 = 1$ the choice $\rhop = P_2$ yields a maximum of $1$ in the maximization \eqref{eq:worst_case_rhop}. 
Let us now consider the case $\lambda_2<1$. 
Then for 
\begin{equation}
	\rhop = (1-\epsilon)\rhot + \epsilon P_2
\end{equation}
we have 
\begin{equation}
\begin{aligned}
	\Tr[\Omega\rhop] &= 1-\epsilon \Tr[\Omega\rhot] + \epsilon \Tr[\Omega P_2]\\
	&=
	1-\epsilon + \epsilon \lambda_2 
	= 
	1-(1-\lambda_2)\epsilon \, ,
\end{aligned}
\end{equation}
i.e., 
the claimed maximum in \eqref{eq:worst_case_rhop} is attained for some feasible $\sigma$. 

To show that the claimed maximum is actually the maximum we consider some state $\rhop\in \DM(\CC^d)$ with $\Tr[\rhot\rhop]\leq 1-\epsilon$. 
We write $\rhop$ as convex combination $\rhop = (1-\epsilon') \rhot + \epsilon'\rhot^\perp$ and observe that $\epsilon'\geq \epsilon$. 
Then 
\begin{equation}
\begin{aligned}
\Tr[\Omega\rhop] 
&=
\Tr[\rhot\rhop] + \sum_{j=2}^d \lambda_j \Tr[P_j \rhop]
\\
&\leq 
\Tr[\rhot\rhop] + \lambda_2 \sum_{j=2}^d \Tr[P_j \rhop]
\\
&=
1-\epsilon' + \lambda_2 \epsilon' \Tr\Bigl[ \sum_{j=2}^d P_j \rhot^\perp\Bigr]
\\
&=
1-\epsilon' + \lambda_2 \epsilon' \Tr[\rhot^\perp]
\\
&=
1-\epsilon' + \lambda_2 \epsilon'
= 1-(1-\lambda_2)\epsilon'
\\
&\leq 1 - (1-\lambda_2)\epsilon \, .
\end{aligned}
\end{equation}
\end{proof}

Given a measurement strategy with effective measurement operator $\Omega$ this lemma provides a closed formula for the false acceptance probability \eqref{eq:false_pass}. 
This allows us to state the following guarantee for Protocol~\ref{protocol:QSC}. 

\begin{propositionblock}[{label={prop:QSC_guarantee},fonttitle=\normalfont}]{Performance guarantee II {\cite{PalLinMon18}}}
Let $\rhot\in \DM(\CC^d)$ be a pure target state and $\epsilon,\delta>0$. 
We consider an effective measurement operator \eqref{eq:Omega} satisfying $0\leq \Omega \leq 1$ and $\Tr[\Omega\rhot] = 1$ and having a spectral gap \eqref{eq:SpectralGap} bounded as $\nu(\Omega)>0$. 

Then the certification test from Protocol~\ref{protocol:QSC} 
is an $\epsilon$-certification test w.r.t.\ the infidelity from $\np$ independent samples for 
\begin{equation} \label{eq:SampCompDSC}
	\np \geq  \frac{\ln(1/\delta)}{\nu(\Omega)\, \epsilon}
\end{equation}
with confidence at least $1-\delta$. 
Moreover, the protocol accepts the target state $\rhot$ with probability~$1$. 
\end{propositionblock}

Compared to the sample complexity \eqref{eq:prop:np} of the naive Protocol~\ref{protocol:QSC_unconstrained}, the sample complexity \eqref{eq:SampCompDSC} has an overhead of a factor $1/\nu(\Omega)$, 

\begin{proof}[Proof of Proposition~\ref{prop:QSC_guarantee}]
The proof is mostly analogous to the one of Proposition~\ref{prop:QSCsimple_guarantee}. 

Thanks to Lemma~\ref{lem:QSC-single_measurement_worst_case}, the probability of wrongfully accepting a state $\rhop\in \DM(\CC^d)$ with $\fidelity(\rhot,\rhop_i)\leq 1-\epsilon$ is bounded as
\begin{equation}\label{eq:prob:pass}
\PP[\pass|\rhop_i] \leq 1 - \nu(\Omega)\epsilon \, .
\end{equation} 
Hence, the probability that Protocol~\ref{protocol:QSC_unconstrained} accepts is bounded as
\begin{equation}
	\PP[\accept] \leq \left(1 - \nu(\Omega)\epsilon\right)^\np \, .
\end{equation}
Imposing $\left(1 - \nu(\Omega)\epsilon\right)^\np \leq \delta$ and solving for $\np$ yields 
\begin{equation}
	\np \geq \frac{\ln(1/\delta)}{\ln\!\left( \frac{1}{1-\nu(\Omega)\epsilon} \right)}
\end{equation}
and the bound \eqref{eq:lnfrac_epsilon_bound} finishes the proof. 
\end{proof}

This proposition tells us that as long as $\Omega$ has a constant gap between its largest and second largest eigenvalue the sample complexity of the certification protocol has the same scaling as the one where $\Omega$ is the target state itself. 
Now it depends on the physical situation of what feasible measurement strategies $\Omega$ are. 
Given a set $\mset$ of feasible measurements we can single out an optimal strategy as follows. 

\begin{definitionblock}[label={def:minimax_QSC}]{Minimax optimization}
Let $\rhot$ be a pure state and $\epsilon>0$. 
Moreover, let us assume that we have access to a compact set of binary measurements given by the operators 
$\mset \subset \{ P: \ 0 \leq P \leq \1\, ,\  \Tr[P\rhot]=1\}$. 

Then the best strategy $\Omega$ for the worst-case state preparation $\rhop$ is 
\begin{equation}\label{eq:cert_minimax}
	\min_{\Omega \in \conv(\mset)} \max_{\rhop: \,\Tr[\rhot\rhop]\leq 1-\epsilon} \Tr[\Omega\rhop] \, ,  
\end{equation} 
where $\conv(S)$ denotes the \emph{convex hull} of a set $S$, i.e., the set of all convex combinations of elements in $S$.
This quantity 
is called \emph{minimax value} and a strategy $\Omega$ where the minimum is attained is called \emph{minimax optimal}. 
\end{definitionblock}

Such minimax optimizations are common in game theory and risk analysis. 

If there are no restrictions on the measurements of a pure target state $\rhot$, i.e., 
$\mset = \{ P: \ 0 \leq P \leq \1 \, , \ \Tr[P\rhot]=1\}$, 
then $\Omega = \rhot$ is minimax optimal. 

For a number of settings with physically motivated measurement restrictions the minimax strategy, or at least one that is close to it, has been found. 
For instance for \emph{stabilizer states}, which are ubiquitous in quantum information theory, there are such optimal measurement strategies.  
In the following we introduce stabilizer states and, for two-outcome Pauli measurements, we derive a minimax optimal certification protocol for them.  

\subsubsection{\protchaptitle{Stabilizer states}}
\label{sec:DSCofSTABs}
Now we consider the certification of stabilizer target states by using a particularly suitable measurement strategy in the direct certification Protocol~\ref{protocol:QSC}. 

Let us start with a few preliminaries on stabilizer states. 
An $n$-qubit \emph{Pauli string} is $\sigma_{s_1} \otimes \dots \otimes \sigma_{s_n}$, where $s \in \{0,1,2,3\}^n$ and $\{\sigma_i\}$ are the Pauli matrices
\begin{equation} \label{eq:def:PauliMatrices}
\begin{aligned}
	\sigma_x &\coloneqq \sigma_1 \coloneqq 
		\begin{pmatrix} 0&1\\1&0 \end{pmatrix},
	& 
	\sigma_y &\coloneqq \sigma_2 \coloneqq 
		\begin{pmatrix} 0&-\i\\\i&0 \end{pmatrix},
	\\
	\sigma_z &\coloneqq \sigma_3 \coloneqq 
		\begin{pmatrix} 1&0\\0&-1 \end{pmatrix},
	&
	\sigma_0 &\coloneqq \1_{2\times 2} \, .
\end{aligned}
\end{equation}
Then the \emph{Pauli group} $\mcP_n\subset \U(2^n)$ is the group generated by all $n$-qubit Pauli strings and $\i \1$.  
An $n$-qubit state $\ket \psi$ is a \emph{stabilizer state} if there is an Abelian subgroup $\mc S \subset \mcP_n$, called \emph{stabilizer (subgroup)}, that stabilizes $\ket\psi$ and only $\ket\psi$, i.e., $\ket\psi$ is the unique joint eigenvalue-$1$ eigenstate of all elements in that subgroup. 
Such subgroups are generated by $n$ elements and contain $|\mc S| = 2^n$ elements in total. 
Note that they cannot contain the element $-\1$. 

An example of such a subgroup is the one of all Pauli strings made of $\1$'s and $\sigma_z$'s. 

It is not difficult to show that a general $n$-qubit stabilizer state $\rho$ with stabilizer $\mc S$ is explicitly given as
\begin{equation}\label{eq:STAB_as_sum}
	\rho 
	= 
	\prod_{j=1}^n \tfrac 12 (\1 + G_j)
	= 
	\frac{1}{2^n}\sum_{S \in \mc S} S \, ,
\end{equation}
where $\{G_j\}_{j\in [n]}$ is a set of generators of $\mc S$.

The measurement strategy for our direct certification of stabilizer states essentially consists in measuring stabilizer observables that are drawn uniformly at random from the stabilizer group of the target state. 
We accept exactly when the measurement outcome corresponds to the stabilized eigenspaces of eigenvalue $+1$. 
This strategy is minimax optimal (Definition \ref{def:minimax_QSC}) among all strategies based on measuring Pauli observables, i.e.\ two-outcome Pauli measurements. 

\begin{theoremblock}[label={thm:QST:stabilizer}]{Minimax optimal $2$-outcome Pauli measurements for STABs {\cite{PalLinMon18}}}
Let $\ket\psi$ we an $n$-qubit stabilizer state with stabilizer group $\mc S \subset \mc P_n$ with elements 
$\mc S = \{\1 = S_0, S_1, \dots, S_{2^n-1}\}$.
For $i \in [2^n-1]$ denote by 
$P_i \coloneqq \frac 12\left( \1 + S_i \right)$ 
the projector onto the positive eigenspace of $S_i$. 

Then the minimax optimal measurement strategy for having Pauli observables $\mc P_n$ as accessible measurements (see Definition~\ref{def:minimax_QSC})
is given by measuring $S_i$ with probability $\frac{1}{2^n-1}$. 
The resulting effective measurement operator $\Omega = \frac{1}{2^n-1} \sum_{i=1}^{2^n-1} P_i$ satisfies 
$\Omega\ket\psi=\ket\psi$ and has the second largest eigenvalue 
\begin{equation}
	\lambda_2(\Omega) = \frac{2^{n-1}-1}{2^n-1}\, . 
\end{equation}
\end{theoremblock}

\begin{proof}
By Lemma~\ref{lem:QSC-single_measurement_worst_case}, the minimax optimum is
\begin{equation}
\begin{aligned}
	\min_{\Omega\in \mc X} 
	\max_{\rhop: \,\Tr[\rhot\rhop]\leq 1-\epsilon} \Tr[\Omega\rhop]
	&=
	\min_{\Omega \in \mc X} \left( 1-\nu(\Omega) \epsilon \right)
	\\
	&=
	1- \epsilon \, \max_{\Omega \in \mc X}  \nu(\Omega)  \, ,
\end{aligned}
\end{equation}
where 
\begin{equation}
\begin{aligned}
 	\mc X &\coloneqq \{\Omega \in \conv(\mc P_n):\ \Omega\ket\psi = \ket\psi\} 
 	= \conv(\mc S)\, . 
\end{aligned}
\end{equation} 
We argue that the minimization over $\conv(\mc S)$ can be replaced by a minimization over $\conv(\mc S')$ with $\mc S'\coloneqq \mc S \setminus\{\1\}$. 
To see this, 
observe that if $\Omega = (1-\alpha) \Omega' + \alpha \1$ for $\alpha \in [0,1]$ then $\nu(\Omega)\leq \nu(\Omega')$. 
Then minimax optimal measurement strategies are hence of the form
\begin{equation}\label{eq:Omega:opt_strategy_form}
	\Omega = \sum_{i=1}^{2^n-1} \mu_j P_i
\end{equation}
for a probability vector $\mu$. 
We note that 
\begin{equation}
	\Tr[\Omega] = 2^{n-1}
\end{equation}
since $\Tr[P_i] = 2^{n-1}$. 

Next, since $\ket\psi$ is an eigenvalue-$1$ eigenvector of $\Omega$, we can write  
\begin{equation}
	\Omega = 1\oplus \tilde \Omega 
\end{equation}
with
\begin{equation}
	\lambda_2(\Omega) = \snormb{\tilde \Omega}. 
\end{equation}
Moreover, $\Tr[\tilde\Omega] = 2^{n-1}-1$. 
The operator $\tilde\Omega$ with the minimal norm $\snormb{\tilde \Omega}$ under this constraint is of the form $\tilde \Omega = a \1$ for $a>0$. 
Taking the trace of that equality, solving for $a$ and denoting the orthogonal projector of $\ketbra\psi\psi$ by 
$\ketbra\psi\psi^\perp\coloneqq \1-\ketbra\psi\psi$ 
yields 
\begin{equation}\label{eq:Omega_suggestion}
	\Omega = \ketbra\psi\psi + \frac{2^{n-1}-1}{2^n-1} \, \ketbra\psi\psi^\perp 
\end{equation}
with
\begin{equation}
	\lambda_2(\Omega) 
	= 
	\frac{2^{n-1}-1}{2^n-1} 
	\, .
\end{equation}
In order to finish the proof we show that $\Omega\in \conv(\mc S)$, i.e., that this choice of $\Omega$ is indeed compatible with \eqref{eq:Omega:opt_strategy_form}. 

We write the stabilizer state $\ketbra\psi\psi$ as combination of the stabilizers (see~\eqref{eq:STAB_as_sum}) and use that $S_j = 2P_j-\1$, 
\begin{equation}
\begin{aligned}
	\ketbra\psi\psi 
	&= \frac{1}{2^n}\left( \1 + \sum_{j=1}^{2^n-1} S_j \right) 
	\\
	&=
	\frac{1}{2^n}\left( \1 + 2 \sum_{j=1}^{2^n-1} P_j - (2^n-1) \1 \right) 
	\\
	&= \left( \frac{1}{2^{n-1}}-1 \right) \1 + \frac{1}{2^{n-1}} \sum_{j=1}^{2^n-1} P_j \, .
\end{aligned}
\end{equation}
With $\1 = \ketbra\psi\psi + \ketbra\psi\psi^\perp$ this implies that
\begin{equation}
\begin{aligned}
	 \sum_{j=1}^{2^n-1} P_j 
	&=
	 (2^n-1)\ketbra\psi\psi + (2^{n-1}-1) \ketbra\psi\psi^\perp 
\end{aligned}
\end{equation}
and, hence, 
\begin{equation}
\begin{aligned}
	 \frac{1}{2^n-1} \sum_{j=1}^{2^n-1} P_j 
	&=
	 \ketbra\psi\psi + \frac{2^{n-1}-1}{2^n-1} \ketbra\psi\psi^\perp \, ,
\end{aligned}
\end{equation}
which is the $\Omega$ from \eqref{eq:Omega_suggestion} and also the measurement strategy from the theorem statement. 
\end{proof}

\begin{corollaryblock}{Sampling complexity {\cite{PalLinMon18}}}
Let us call the outcome corresponding to $P_i$ {\pass} and the one corresponding to $1-P_i$ \fail. 
Then Protocol~\ref{protocol:QSC} is an $\epsilon$-certification test of $\rhot$ w.r.t.\ infidelity from $\np$ independent samples for 
\begin{equation}
	\np \geq 2\, \frac {\ln(1/\delta)}{\epsilon}
\end{equation}
with confidence $1-\delta$.  
Moreover, $\rhot$ is accepted with probability $1$. 
\end{corollaryblock}

\begin{proof}
According to Proposition~\ref{prop:QSC_guarantee} a number of measurements 
\begin{equation}
	\np \geq  \frac{\ln(1/\delta)}{\epsilon\, \nu(\Omega)}
\end{equation}
is sufficient, where 
\begin{equation}
\begin{aligned}
	\nu(\Omega) &= 1-\lambda_2(\Omega) 
	\\
	&= 
	1 -\frac{2^{n-1}-1}{2^n-1}
	\\
	&=
	\frac{2^{n-1}}{2^n-1} \, .
\end{aligned}
\end{equation}
This results in 
\begin{equation}
	\np \geq \frac{2^n-1}{2^{n-1}} \frac{\ln(1/\delta)}{\epsilon} \, .
\end{equation}
\end{proof}

So, restricting from all measurements to Pauli measurements results in at most a constant overhead of $2$, cmp.\ Proposition~\ref{prop:QSCsimple_guarantee}. 
We note that only very few of the $2^n-1$ non-trivial stabilizers of $\rhot$ are actually measured. 
More precisely, the measurements are the ones of randomly subsampled stabilizer observables.

\subsubsection{\protchaptitle{Extension towards fidelity estimation}}
Direct certification provides minimum information to solve the certification task by  just giving an accept/reject answer. 
Often it is also desirable to actually know explicitly what the distance or fidelity of a quantum state implementation $\rhop$ to its target $\rhot$ is. 

The direct quantum state certification protocol~\eqref{protocol:QSC} with effective measurement operator $\Omega$ (see \eqref{eq:Omega}) can be turned into a estimation protocol for the fidelity $\fidelity(\rhop,\rhot)$ if $\Omega$ is \emph{homogeneous}, i.e.\ of the form 
\begin{equation}
	\Omega = \ketbra\psi\psi + \tau \, \ketbra\psi\psi^\perp 
\end{equation}
for some $\tau>0$ \cite{LiHanZhu19,Zhu2019GeneralFrameworkFor};
for instance, for stabilizer states \eqref{eq:Omega_suggestion} we have $\tau = \frac{2^{n-1} - 1}{2^n-1}$. 
In this case, we have 
\begin{equation}
	\Tr[\rhop\,\Omega] = \fidelity(\rhop,\rhot) + \tau (\Tr[\rhop] - \fidelity(\rhop,\rhot))
\end{equation}
and, hence, 
\begin{equation}
	\fidelity(\rhop,\rhot)
	= 
	\frac{\Tr[\rhop\,\Omega] - \tau}{1-\tau}
	=
	\frac{\Tr[\rhop\,\Omega] - \lambda_2(\Omega)}{\nu(\Omega)} \, .
\end{equation}
Therefore, an estimate on the expectation value of $\Omega$ yields an estimate of the fidelity $\fidelity(\rhop,\rhot)$. 

We note that the number of measurements required for estimating the expectation value up to an additive error $\varepsilon$ scales as $1/\varepsilon^2$. 
We also remember that in the case where we can measure $\Omega$ as observable the estimation task can be solved with a number of measurements $m\geq  \frac{1}{2\varepsilon^2}\ln(2/\delta)$ with confidence $1-\delta$, see Proposition~\ref{prop:EstimationExpectationVals}. 
However, in Protocol \ref{protocol:QSC} we only assume access to measurements 
$M\in \mset$ that on some average give the expectation value of $\Omega$. 
In general, the number of measurements $\left|\mset\right|$ can be exponentially large, as is the case for stabilizer states, cp.\ Theorem~\ref{thm:QST:stabilizer}. 
Here, one could use ides of Monte Carlo estimation and importance sampling in order to perform this estimation efficiently; a method that we introduce in Section~\ref{sec:ImportanceSampling}. 
Subsequently, we discuss \acl{DFE}, which relies on this idea.

\subsubsection*{Further reading}
The direct certification of maximally entangled states was studied by Hayashi \emph{et al.}
\cite{HayashiMatsumotoTsuda:2006,Hayashi:2009}. 
Building on these earlier works and the discussed framework of Pallister \emph{et al.}\ \cite{PalLinMon18}, direct certification protocols were then developed for other classes of quantum states featuring a (close to) optimal sampling complexity. 
They include the following settings:
\begin{itemize}
	\item Stabilizer states and two-qubit states with single-qubit measurements \cite{PalLinMon18}. 

 	\item Bipartite states \cite{LiHanZhu19,YuShaGuh19}, qubit case in an LOCC setting \cite{WanHay19}. 

 	\item Hypergraph states \cite{TakMor18} with improvements in efficiency by Zhu and Hayashi \cite{ZhuHay18} and a generalization to weighted graph states \cite{HayashiTakeuchi:2019}.

 	\item Dicke states \cite{LiuEtAl:2019, LiEtAl:2020}.

 	\item A general adversarial scenario without the assumptions of the state preparations being identical and independent \cite{Zhu2019VerificationAdversarialPRL,Zhu2019GeneralFrameworkFor}. 
\end{itemize} 
Protocols for the efficient verification of graph states, which are certain stabilizer states, were developed in the context of measurement-only blind quantum computation \cite{Hayashi15VerifiableMeasurementOnly} and interactive proof systems \cite{McKague2016InteractiveProofsFor,Morimae16QuantumProofsCan}. 

Efficient certification protocols for ground states of locally interacting Hamiltonians were developed by Cramer \emph{et al.}\ \cite{CraPleFla10} 
and extended by Hangleiter \emph{et al.}\ \cite{HanKliSch17} to ground states enabling universal quantum computation. 
In this line of research, fidelity witnesses \cite{AolGogKli15,HanKliSch17,GluKliEis18} can be used to  estimate fidelity lower bounds from simple measurements. 
Also in the context of interactive proof systems efficient ground state certification schemes have been developed \cite{Morimae16QuantumProofsCan,Ji15ClassicalVerificationOf}. 

Kalev \emph{et al.}~\cite{KalKyrLin19} have extended arguments from direct fidelity estimation \cite{FlaLiu11} (see Section~\ref{sec:fidelity_estimation}) and ground state certification \cite{HanKliSch17} to the certification of stabilizer states. 
They also use Bernstein's inequality to give a quadratically improved $\epsilon$-scaling for large $\epsilon$. 

The work \cite{CraPleFla10} solves the certification problem by efficiently reconstructing the state 
assuming it to be of matrix product form. 
Similar ideas based on ansatz state tomography also work for permutationally invariant states 
\cite{Toth2010PermutationallyInvariant,Moroder2012PermutationallyInvariant,Schwemmer2014EfficientTomographic}. 

Takeuchi and Morimae \cite{TakMor18} provide efficient results on the certification of ground states of locally interacting Hamiltonians, and hypergraph states, where the iid.\ assumptions on the state preparations is removed using a quantum de Finetti theorem \cite{Li15QuantumDeFinetti}. 
Hypergraph states include quantum states that are generated by so-called IQP circuits designed for demonstrating quantum supremacy \cite{BreMonShe16}. 

Global von Neumann measurements on multiple iid.\ copies of the prepared quantum state have been considered \cite{BadDonWri17} (even with mixed target states), 
which leads to a sample complexity scaling as $\np \in \LandauO(d/\epsilon)$ a version of $\epsilon$-certification of quantum states in $\DM(\CC^d)$. 

For a very helpful survey on \emph{quantum property testing} we refer to Ref.~\cite{MonWol14}, where several methods and notions of certification are reviewed.

\subsection{Importance sampling}
\label{sec:ImportanceSampling}
In the next section, we study \acl{DFE}, where the fidelity between a target state and a state preparation is estimated from measurements that are drawn randomly from a certain distribution depending on the target state. 
The idea is to perform the measurements more often that are particularly relevant to the fidelity estimation. 

This idea is formalized by a Monte Carlo integration technique called \emph{importance sampling}. 
Monte Carlo integration aims at computing an integral $F$ that is written as an expected value 
of some function $f$ 
over a probability distribution with density function $p$:
\begin{equation}\label{eq:EE_of_f}
 	F \coloneqq \EE_{X \sim p} [f(X)] = \int f(x) p(x) \, \rmd x \, .
\end{equation} 
The general idea is to draw iid.\ samples $X^{(1)}, \dots, X^{(m)} \sim p$ and take the empirical average 
\begin{equation} \label{eq:MCintegration}
	\hat F \coloneqq 
	\frac 1m \sum_{i=1}^m f(X^{(i)}) 
\end{equation}
as estimator for $F$. 
It is not difficult to see that $\hat F$ is unbiased. 
If $\Var[f(X)] < \infty$ then $\hat F$ can be proven to be \emph{consistent}, i.e., $\hat F$ converges to $F$ for $m \to \infty$ in an appropriate sense. 
Moreover,  
\begin{equation}\label{eq:Variance_of_sum}
	\Var[\hat F] = \frac{\Var[f(X)]}m \, .
\end{equation}
Thereby the empirical variance also gives an estimate of the estimation error. 
The estimation error can be controlled by increasing the number of samples $m$. 

Now, the integration \eqref{eq:MCintegration} relies on the ability to sample from $p$. 
A popular way to make such sampling efficient is \emph{importance sampling}. 
The main idea of importance sampling is to rewrite the integrand $f\, p$ in the expectation value \eqref{eq:EE_of_f} as 
\begin{equation}
	f p = \frac{fp}qq
\end{equation}
for some probability distribution with density function $q$. 
Then we can apply the Monte Carlo sampling idea \eqref{eq:MCintegration} w.r.t.\ $q$ and draw $X^{(1)}, \dots, X^{(m)} \sim q$ iid.\ to obtain the estimator
\begin{equation}
	\hat F_q 
	\coloneqq 
	\frac 1m \sum_{i=1}^m f(X^{(i)}) \frac{p(X^{(i)})}{q(X^{(i)})} \, .
\end{equation}
It holds that $\EE_q[\hat F_q] = F$ and 
\begin{equation}
	\Var_q[\hat F_q] = \frac 1m \, \Var_q[fp/q] 
= 
\frac 1m \left[\int \frac{f^2 p^2}{q} - F^2 \right] .
\end{equation}

One can show that the minimal variance 
is achieved by choosing $q$ as
\begin{equation}
	q^\ast \coloneqq \frac{p\, |f|}{Z}
\end{equation}
with a normalization factor $Z$ such that $q^\ast$ is a probability density. 
Note that for $f\geq 0$ we have 
$\EE_{q^\ast}[(f p /q^\ast)^2] = \EE_p[f]^2= \EE_{q^\ast}[f p /q^\ast]^2$ and, thus,  
$\Var_{q^\ast}[\hat F_{q^\ast}] = 0$. 
So, if $f$ does not change its sign then a single sample from $q^\ast$ is sufficient for the exact estimation. 
This might seem miraculous at first sight. 
But its is important to notice that in order to determine the optimal $q^\ast$ 
one needs to know the value of normalization $Z$ and
calculating $Z$ is equivalent to solving the integration problem. 
However, finding non-optimal but good choices for $q$ can already speed up the integration, as we will see in the case of direct fidelity estimation.

\subsection{\protchaptitle{Direct fidelity estimation}}
\label{sec:fidelity_estimation}
We assume to be given access to state preparations $\rhop\in \DM(\CC^d)$ of some target state $\rhot\in\DM(\CC^d)$. 
\emph{\Ac{DFE}} \cite{FlaLiu11,SilLanPou11} is a protocol to estimate the fidelity $\Tr[\rhop\rhot]$ for the case where the $\rhot$ is a pure state, i.e.\ of the form $\rhot = \ketbra \psi\psi$. 
In order to do so, the target states is expanded into products of Pauli matrices 
\eqref{eq:def:PauliMatrices} of the form 
$\sigma_{s_1} \otimes \dots \otimes \sigma_{s_n}$ 
with $s_i\in \{0,\dots, 4\}$ and $d= 2^n$ being the Hilbert space dimension. 
For sake of readability we denote these Pauli products by $\W_1, \dots, \W_{d^2}$ in some order and note that they are an orthogonal basis for the space of Hermitian operators $\Herm(\CC^d)$ w.r.t.\ the Hilbert-Schmidt inner product \eqref{eq:HS_inner_prod}:  
\begin{equation}\label{eq:W_orthogonal}
	\frac 1 d \Tr[\W_k \W_{k'}] = \delta_{k,k'}
\end{equation}
for all $k,k'\in[d^2]$. 

Given any operator $\sigma\in \Herm\myleft(\CC^d\myright)$ we define its 
\emph{characteristic function} (or \emph{quasi-probability distribution}) $\W_\sigma: [d^2] \to \RR$ by 
\begin{equation}\label{eq:dfe:charFunc}
	\chi_\sigma(k) 
	\coloneqq 
	\Tr\Bigl[\sigma \frac{W_k}{\sqrt d} \Bigr] \, . 
\end{equation} 
Thanks to the orthogonality relation \eqref{eq:W_orthogonal} we have 
\begin{equation}
	\sigma = \sum_{k=1}^{d^2} \chi_\sigma(k) \frac{W_k}{\sqrt d}
\end{equation}
and hence
\begin{equation}\label{eq:rhosigmaFrameExpansion}
\begin{aligned}
	\Tr[ \rho \sigma ] 
	&=
	\sum_{k=1}^{d^2}  \chi_\rho(k) \chi_\sigma(k) 
\end{aligned}
\end{equation} 
for any $\rho,\sigma\in\Herm(\CC^d)$. 

Now, we use \emph{importance sampling} (Section~\ref{sec:ImportanceSampling}) to estimate the sum~\eqref{eq:rhosigmaFrameExpansion} for a pure target state $\rhot\in \DM(\CC^d)$ and its preparation $\rhop\in \DM(\CC^d)$. 
For this purpose we rewrite the overlap \eqref{eq:rhosigmaFrameExpansion} as
\begin{align}
	\Tr[ \rhot \rhop ] 
	= 
	\sum_{k=1}^{d^2} 
	\frac{\chi_\rhop(k)}{\chi_\rhot(k)}\chi_\rhot(k)^2
\end{align}
and define 
\begin{equation}\label{eq:def:q}
	q_k \coloneqq \chi_\rhot(k)^2 \, . 
\end{equation}
We choose $q$ as the probability mass function of the importance sampling distribution on the sampling space $[d^2]$. 
The purity of $\rhot$ can be written as 
\begin{equation}
\Tr[\rhot^2]
=
\sum_{k=1}^{d^2} \Bigl|\Bigl\langle \frac{\W_k}{\sqrt d}, \rhot\Bigr\rangle\Bigr|^2
=
\sum_{k=1}^{d^2}  \chi_\rhot(k)^2 
\end{equation} 
and equals $1$ for any pure state $\rhot$.
Thus, $q$ is indeed a normalized probability vector. 

We define a random variable 
\begin{equation}\label{eq:Xlambda}
	X_k \coloneqq \frac{\chi_\rhop(k)}{\chi_\rhot(k)} 
\end{equation}
with $k \sim q$ and find that $X_k$ is an unbiased estimator of the fidelity: 
\begin{equation} \label{eq:XunbiasedEstimator}
\EE_{k \sim q} [X_k] 
= 
\sum_{k=1}^{d^2} \frac{\chi_\rhop(k)}{\chi_\rhot(k)} q_k
= 
\sum_{k=1}^{d^2} \chi_\rhot(k) \chi_\rhop(k)
=
\Tr[\rhot \rhop]\, ,
\end{equation}
where the last identity is again \eqref{eq:rhosigmaFrameExpansion}. 

In order to estimate the random variable $X_k$, we need to know the value of the characteristic function $\chi_{\tilde \rho}(k)$. 
By definition \eqref{eq:dfe:charFunc}, $\chi_{\tilde \rho}(k)$ can be estimated as the expectation value 
from repeated measurements of the observable $W_k$ in the prepared state $\tilde \rho$. 
Thus, we end up with an estimation procedure of $\Tr[\rhot \rhop]$ that involves two sources of randomness and correspondingly 
proceeds in two steps. 
(i) We classically sample $k$ from $[d^2]$ according to the importance sampling distribution \eqref{eq:def:q} defined by the target state $\rhot$. 
(ii) For the randomly drawn $k$, we estimate $X_k$ from repeated probabilistic measurements of $W_k$. 
Combining the estimates of the $X_k$ we arrive at an estimate for $\Tr[\rhot \rhop]$. 

The following protocol summarizes these steps. 
\begin{protocolblock}[label={protocol:DFE}]{\ac{DFE} \cite{FlaLiu11}}
Let $\rhot \in \DM(\CC^d)$ be a pure target state and 
 $\{\W_k\}$ a set of observables $\{\W_k\}$. 
Let $\epsilon>0$ and $\delta>0$ be the parameters for the desired estimation accuracy and maximum failure probability. 

The protocol consists of the following steps requiring $n_\rhop$ state preparations in total: 
\begin{compactenum}[(i)]
\item \label{item:sampling}
Draw iid.\ samples $k_1, \dots, k_\ell \sim q$ from the importance sampling distribution \eqref{eq:def:q}, where $\ell \coloneqq \left\lceil \frac{1}{\epsilon^2\delta}\right\rceil$
 (or as \eqref{eq:sample_complexity_DFE_wc} for well-conditioned states). 

\item \label{item:mi}
Measure each observable $\W_{k_i}$ a number of $m_i$ times for $i \in [\ell]$ with $m_i$ chosen as 
\begin{equation} \label{eq:mi_choice}
	m_i \coloneqq 
	\left\lceil \frac{ 2 }{d \chi_\rhot(k_i)^2 \ell \epsilon^2} \ln(2/\delta)  \right\rceil
\end{equation}
(or as $m_i= 1$ for well-conditioned states). 

\item \label{item:Westimation}
For each $i \in [\ell]$ calculate empirical estimate of the expectation value $\langle \W_{k_i}\rangle_{\rhop}$ from the 
measurement outcomes. 
From these estimates calculate the empricial estimator $\hat X_{k_i}$ of $X_{k_i} \coloneqq \frac{\chi_\rhop(k_i)}{\chi_\rhot(k_i)} =\frac{\langle \W_{k_i}\rangle_{\rhot}}{\sqrt{d}\,  \chi_\rho(k_i)}$.

\item \label{item:Yestimation}
Calculate $\hat Y \coloneqq \frac 1\ell \sum_{i=1}^\ell \hat X_{k_i}$. 

\item Output $\hat Y$ as a fidelity estimator. 
\end{compactenum}
\end{protocolblock}

To derive a guarantee for \ac{DFE} we have to control the error made 
in the two estimation steps. 
To this end, we consider the steps in reversed order:
we consider $Y \coloneqq \frac 1\ell \sum_{i=1}^\ell X_{k_i}$ with $\ell$ iid.\ samples $k_i\sim q$
 assuming perfect estimates $X_{k_i}$ for the moment. 
The accuracy of $Y$ as an estimator of $\Tr[\rhot\rhop]$ can be controlled by increasing $\ell$. 
Subsequently, we have to analyze the accuracy of the estimator $\hat Y$ of $Y$ that uses the finitely many measurement outcomes. 
Altogether we arrive at the following guarantee: 

\begin{theoremblock}[label={thm:DFE}]{Guarantee for \ac{DFE} \cite{FlaLiu11}}
Let $\rhot\in \DM(\CC^d)$ be a pure target state. 
The number of expected state preparations in Protocol~\ref{protocol:DFE} is 
\begin{equation}\label{eq:DFEsampleComplexity}
	\EE[n_\rhop] = \EE\, \sum_{i=1}^\ell m_i
	\leq   
	1 + \frac{1}{\epsilon^2 \delta} + \frac{ 2\, d }{\epsilon^2 } \ln(2/\delta) \, .
\end{equation}
If the state preparations are iid.\ given by $\rhop\in \DM(\CC^d)$ 
then the fidelity estimate $\hat Y$ is an $2\epsilon$-accurate unbiased estimator of $\fidelity(\rhot,\rhop)$ with confidence $1-2\delta$. 
\end{theoremblock}

Note that the sample complexity scales linearly in the Hilbert space dimension. 
In contrast, the number of Pauli measurements required for state tomography scales as $\tLandauOmega(d^2\rank(\rhop)^2)$ \cite{FlaGroLiu12}. 

\begin{proof}[Proof of Theorem~\ref{thm:DFE}]
We start with bounding the estimation error arising by taking the empirical average in step~(\ref{item:Yestimation}) of Protocol~\ref{protocol:DFE}. 
We note that $X_k$ defined in\eqref{eq:Xlambda} is an unbounded random variable in general, as 
$\chi_\rhot(k)$ can be arbitrarily small. 
Hence, we will use Chebyshev's inequality~\eqref{eq:Chebyshev} to derive a tail bound for $Y$. 
Using the definitions \eqref{eq:def:q} and \eqref{eq:Xlambda} of $q$ and $X$ and that $X$ is the unbiased estimator \eqref{eq:XunbiasedEstimator},
the variance of $X$ becomes 
\begin{equation}
\begin{aligned}
\Var_{k \sim q} [ X_k ]
&=
\EE_{k \sim q}[ X_k^2 ] - \Tr[\rho\sigma]^2
\\
&= \sum_{k=1}^{d^2}  \frac{\chi_\rhop(k)^2}{\chi_\rhot(k)^2} \chi_\rhot(k)^2
	- 
	 \Tr[\rho\sigma]^2
\\
&= \chi_\rhop(k)^2 
	- 
	 \Tr[\rho\sigma]^2 
\\
&=
\Tr[\rhop^2]-\Tr[\rhot \rhop]^2 \, . 
\end{aligned}
\end{equation}
Hence, 
\begin{equation}
	\Var_{k \sim q} [ X_k ]
	\leq 
	\Tr[\rhop^2]
	\leq 
	1 \, .
\end{equation}
Using the basic insight of Monte Carlo estimation \eqref{eq:Variance_of_sum},
we obtain
\begin{equation}
	\Var_q[Y] = \EE_q[(Y -\Tr[\rhot\rhop])^2] \leq 1/\ell \, .
\end{equation}
As $Y$ is an unbiased estimator of $\Tr[\rhot\rhop]$, i.e., 
$\EE_q[Y -\Tr[\rhot\rhop]]=0$, we can directly apply Chebyshev's inequality \eqref{eq:Chebyshev} to arrive at
\begin{equation}\label{eq:appCheb}
\PP\myleft[\bigl|Y - \Tr[\rhot\rhop]\bigr|\geq \epsilon \myright] 
\leq 
\frac{1}{\epsilon^2\ell}
\end{equation}
for any $\epsilon>0$. 
Hence, for any $\delta>0$ and
\begin{equation} \label{eq:ell_bound}
	\ell \geq \frac1 {\epsilon^2 \delta}
\end{equation}
the failure probability is bounded by $\delta$, 
\begin{equation} \label{eq:Ytail}
  \PP\myleft[\myleft|Y - \Tr[\rhot\rhop]\myright|\geq \epsilon\myright] \leq \delta\, .
\end{equation}

Now we bound the statistical error that arises from the estimation of $X_{k_i}$ from the measurement setup $i\in [\ell]$ in step~(\ref{item:Westimation}) of Protocol~\ref{protocol:DFE}. 
For this purpose we write for each $k$ the eigendecomposition of $\W_k$ as
\begin{equation}\label{eq:MlambdaEig}
	\W_k = \sum_\alpha a_{k,\alpha} P_{k,\alpha}
\end{equation}
with $\{P_{k,\alpha}\}$ being the projector onto the eigenspaces and 
$\{a_{k,\alpha}\} \subseteq \{-1,1\}$ 
the eigenvalues of the Pauli string $\W_k$. 
We note that the expected measurement outcome is
\begin{equation}
	\EE[a_{k,\alpha}] = \Tr[\W_k \rhop] = \sqrt d\,  \chi_\rhop(k) \, .
\end{equation}
We denote by $a_{k_j, \alpha_j}$ the measurement outcome for measurement $j\in[m_i]$ 
and consider the following corresponding empirical estimate of $X_{k_i}$ 
(see \eqref{eq:Xlambda}) 
\begin{equation}
	\hat X_{k_i} 
	\coloneqq 
	\frac{1}{m_i \sqrt d \chi_\rhot(k_i)} \sum_{j=1}^{m_i} a_{k_i, \alpha_j} \, .
\end{equation}
Then we consider the sum 
\begin{equation}
\begin{aligned}
	\ell \hat Y 
	&= 
	\sum_{i=1}^\ell \hat X_{k_i} 
	\\
	&=
	\sum_{i=1}^\ell \sum_{j=1}^{m_i}
		\frac{1}{ m_i \sqrt d\,  \chi_\rhot(k_i)}   
			a_{k_i, \alpha_j} \,  .
\end{aligned}
\end{equation}
As $\EE[\ell \hat Y] = \ell Y$, using Hoeffding's inequality \eqref{eq:HoeffdingsAbs} on the double sum with 
$t = \epsilon \ell$ and bounds
\begin{equation}
	b_{i} = -a_{i}
	= \frac{1}{m_i \sqrt d\, \chi_\rhot(k_i)}   \, ,			
\end{equation}
we find that 
(w.l.o.g.\ we assume that there are no $i$ with $\chi_\rhot(k_i)=0$)
\begin{equation}\label{eq:hatY_Y_tail_bound_explicit}
\begin{aligned}
	\PP[ |\hat Y -Y| \geq \epsilon ] &= \PP[ |\ell \hat Y -\ell Y| \geq \ell \epsilon ] 
	\\
	&\leq
	2 \exp\left( 
		\frac{- 2\epsilon^2 \ell^2 }{\sum_{i=1}^\ell \sum_{j=1}^{m_i}
		 \frac{ 2^2 }{ m_i^2 d\,  \chi_\rhot(k_i)^2 }} 
		\right) 
	\\
	&=
	2 \exp\left( 
		\frac{- \epsilon^2 \ell^2 }{\sum_{i=1}^\ell
		 \frac{ 2 }{ m_i d\,  \chi_\rhot(k_i)^2 }}
		\right) 
\end{aligned}
\end{equation}
We wish that the tail bound 
\begin{equation}\label{eq:hatYtail} 
	\PP\myleft[\bigl| \hat Y - Y \bigr| \geq \epsilon \myright] \leq \delta
\end{equation}
holds. 
Therefore, we impose the right-hand-side of \eqref{eq:hatY_Y_tail_bound_explicit} to be bounded 
by $\delta$, which is equivalent to 
\begin{equation}\label{eq:mi_cond}
\ln(2/\delta) 
\leq 
\frac{ \epsilon^2 \ell^2 }{\sum_{i=1}^\ell 
		 \frac{ 2 }{ m_i d\, \chi_\rhot(k_i)^2 }} \, .
\end{equation}
The choice of $m_i$ as in \eqref{eq:mi_choice} guarantees that this bound it always satisfied and, thus, \eqref{eq:hatYtail} holds. 
Then combination of the tails bounds \eqref{eq:Ytail} and \eqref{eq:hatYtail} with the union bound \eqref{eq:union_bound} proves the confidence statement, 
\begin{equation}\label{eq:DFE_final_confidence}
	\PP[|\hat Y - \fidelity(\rhot,\rhop)| \leq 2\epsilon] 
	\geq 1 - 2\delta \, . 
\end{equation}

In order to obtain the final sample complexity~\eqref{eq:DFEsampleComplexity} note that $m_i$ is a random variable itself, since $k_i$ and hence $\chi_{\rhot}(k_i)$ is randomly chosen. 
By the definition of the sampling \eqref{eq:def:q}, for fixed $i$ we have 
\begin{equation}
\begin{aligned}
	\EE[m_i] &= \sum_{k_i=1}^{d^2} m_i q_{k_i}
	\\
	&\leq  
	1+ \frac{ 2 d }{\ell \epsilon^2 } \ln(2/\delta)   \, ,
\end{aligned}
\end{equation}
where the $+1$ comes from the ceiling in \eqref{eq:mi_choice}. 
Using the bound \eqref{eq:ell_bound} on $\ell$, the expected total number of measurements is 
\begin{equation}
	\EE\, \sum_{i=1}^\ell m_i
	\leq   
	1 + \frac{1}{\epsilon^2 \delta} + \frac{ 2d }{ \epsilon^2 } \ln(2/\delta) \, .
\end{equation}
\end{proof}

We remark that \ac{DFE} estimation can be extended to sets of observables that are arbitrary orthonormal bases of $\Herm(\CC^d)$. 
However, in this case the operator norm used to bound the eigenvalues $a_{k,\alpha}$ and hence the sampling complexity can be larger. 
One can generalize \ac{DFE} further to frames that include over-complete bases, see Ref.~\cite{Kli19}. 

The main contribution to the number of measurements in the derivation of the sample complexity above can be traced back to the application of Chebyshev's inequality in \eqref{eq:appCheb}. 
This step can, however, be improved for the following class of states. 

\begin{definitionblock}{Well-conditioned states}
We call an operator $\rho \in \Herm(\CC^d)$ \emph{well-conditioned with parameter $\alpha>0$} if for each $k\in [d^2]$
either $\left|\Tr[\W_k \rho]\right|\geq \alpha$ or $\Tr[\W_k\rho] = 0$. 
\end{definitionblock}

A prominent example for well-conditioned states are stabilizer states \eqref{eq:STAB_as_sum}. 
It is easy to show that every stabilizer state $\rhot$ on $n$ qubits with stabilizer $\mc S$, \eqref{eq:STAB_as_sum}, is well-conditioned with parameter $\alpha=1$:
\begin{equation}
\begin{aligned}
	\Tr[\W_k \rhot] 
	&= 
	\frac 1 d \sum_{S \in \mc S} \Tr[\W_k S] \in \{-1,0,1\}\, ,
\end{aligned}
\end{equation}
where the sum evaluates to $1$ if $\W_k \in \mc S$, to $-1$ if $-\W_k \in \mc S$ and to $0$ otherwise. 
For such well-conditioned states the sample complexity can be improved as follows. 

\begin{theoremblock}[{label={thm:DFE_well-cond},fonttitle=\normalfont}]{\ac{DFE}, well-conditioned states \cite{FlaLiu11}}
Let $\rhot\in \DM(\CC^d)$ be a pure target state that is well-conditioned with parameter $\alpha>0$. 
Consider the estimator $\hat Y$ from Protocol~\ref{protocol:DFE} modified by setting $m_i=1$ for all $i \in [\ell]$ in step~(\ref{item:mi}) and 
\begin{equation}\label{eq:sample_complexity_DFE_wc}
	\ell \coloneqq \left\lceil \frac{2}{\alpha^2\epsilon^2} \, \ln(2/\delta) \right\rceil 
\end{equation}
in step~(\ref{item:sampling}). 
If the state preparations are iid.\ given by $\rhop\in \DM(\CC^d)$ 
then the fidelity estimate $\hat Y$ is an $\epsilon$-accurate unbiased estimator of $\fidelity(\rhot,\rhop)$ with confidence $1-\delta$. 
\end{theoremblock}
\begin{proof}
With probability $1$ we have $\sqrt d \chi_\rhot(k_i) \geq \alpha$ for all $i\in [\ell]$. 
Moreover, $|\sqrt d \hat \chi_\rhop(k_i)| \leq 1$. 
The estimator from step~(\ref{item:Westimation}) of Protocol~\ref{protocol:DFE} is hence bounded as 
\begin{equation}
	|X_{k_i}| \leq \frac{1}{\alpha}
\end{equation}
with probability $1$. 
The estimator $\hat Y$ is, thus, bounded as $|\hat Y| \leq \frac{ 1 }{\alpha}$ almost surely. 
Hoeffding's inequality~\eqref{eq:HoeffdingsAbs} with $t=\epsilon \ell$ yields
\begin{equation}\label{eq:appHoeff_wc}
\PP\myleft[\bigl|\hat Y - \Tr[\rhot\rhop]\bigr|\geq \epsilon \myright] 
\leq 
2\, \exp\myleft( - \frac{\ell\, \alpha^2 \epsilon^2 }{2} \myright) .
\end{equation}
Imposing 
\begin{equation}
	2\, \exp\myleft( - \frac{\ell\, \alpha^2 \epsilon^2 }{2} \myright) \leq \delta
\end{equation}
and solving for $\ell$ yields that
\begin{equation}\label{eq:DFE_final_confidence_well_cond}
	\PP[|\hat Y - \fidelity(\rhot,\rhop)| \leq \epsilon] 
	\geq 1 - \delta 
\end{equation}
for $\ell$ chosen as in Eq.~\eqref{eq:sample_complexity_DFE_wc}. 
\end{proof}

Theorem~\ref{thm:DFE_well-cond} tells us that 
for well-conditioned states \ac{DFE} has a sampling complexity independent of the system size. 
Ref.~\cite{FlaLiu11} also investigates the idea of removing ``bad events'' which are those that violate the well-conditioning condition. 
Moreover, a two-step estimation procedure as in Theorem~\ref{thm:DFE} is considered also for well-conditioned states.  

Finally, we look at how to turn \ac{DFE} into a certification protocol with respect to the trace distance. 
\ex{%
The details are worked out in the following exercise.
\begin{exerciseblock}[label={ex:estimation_to_certification}]{Certification w.r.t.\ the trace distance via \ac{DFE}}
Fix parameters $\tilde \epsilon, \epsilon,\delta>0$ with $\tilde \epsilon \leq \frac 12 \epsilon^2$. 
Let $\hat Y$ be the direct fidelity estimator of the fidelity $\fidelity(\rho,\sigma)$ so that 
$|\hat Y -\fidelity(\rho,\sigma)|\leq \tilde \epsilon$ with confidence $1-\delta$. 
We consider the protocol that \emph{accepts} if 
$\hat Y \geq 1 - \tilde \epsilon$ and 
\emph{rejects} otherwise. 
We choose the trace distance 
$\dist_{\Tr}(\rho,\sigma) \coloneqq \frac 12 \tnorm{\rho-\sigma}$ as the distance measure. 

\begin{itemize}
    \item Show that this protocol is an $\epsilon$-certification test w.r.t.\ the trace distance in the sense of Exercise~\ref{ex:confidence_amplification}, i.e., that the completeness and soundness conditions are satisfied with confidence $1-\delta$. 

	\item What is the resulting sampling complexity of \ac{DFE} fore well-conditioned states? 

	\item Let $\epsilon'<\epsilon$. 
	Turn this protocol into a \emph{robust} $(\epsilon,\epsilon')$-certification test, i.e., into an $\epsilon$-certification test that is guaranteed to accept all states within an $\epsilon'$-trace norm ball around $\rhot$ with confidence $1-\delta$.
\end{itemize}
\end{exerciseblock}
}
\exreplace{
We summarize the result in the following proposition.
\ifex
\begin{propositionblock}{Certification w.r.t.\ the trace distance via \ac{DFE}}
\else
\begin{propositionblock}[label={ex:estimation_to_certification}]{Certification w.r.t.\ the trace distance via \ac{DFE}}
\fi
Fix parameters $\tilde \epsilon, \epsilon,\delta>0$ with $\tilde \epsilon \leq \frac 12 \epsilon^2$. 
Let $\hat Y$ be the direct fidelity estimator of the fidelity $\fidelity(\rho,\sigma)$ so that 
$|\hat Y -\fidelity(\rho,\sigma)|\leq \tilde \epsilon$ with confidence $1-\delta$. 
We consider the protocol that \emph{accepts} if 
$\hat Y \geq 1 - \tilde \epsilon$ and 
\emph{rejects} otherwise. 
We choose the trace distance 
$\dist_{\Tr}(\rho,\sigma) \coloneqq \frac 12 \tnorm{\rho-\sigma}$ as the distance measure. 

\begin{itemize}
    \item This protocol is an $\epsilon$-certification test w.r.t.\ the trace distance in the sense of Proposition~\ref{ex:confidence_amplification}, i.e., the completeness and soundness conditions are satisfied with confidence $1-\delta$. 

	\item The resulting sampling complexity of \ac{DFE} for well conditioned states scales as $1/\epsilon^4$. 

	\item Let $\epsilon'<\epsilon$. 
	This protocol can be turned into a \emph{robust} $(\epsilon,\epsilon')$-certification test, i.e., into an $\epsilon$-certification test that is guaranteed to accept all states within an $\epsilon'$-trace norm ball around $\rhot$ with confidence $1-\delta$.
\end{itemize}
\end{propositionblock}
\begin{proof}
The proof follows from Definition~\ref{def:state_certification} by direct calculations. 
We leave filling in the details as an exercise. 
\end{proof}
}

\subsection{Random states and unitaries}
\label{chap:rep_theory}
Random ensembles of quantum states and unitary matrices find ubiquitous applications in quantum information processing and, 
in particular, in certification and estimation protocols. 
Roughly speaking, random unitary operations together with a fixed quantum measurement allow information about the entire state space to be gained quickly. 
Arguably the simplest probability distribution on the unitary group $\U(d)$ is given by the \emph{Haar measure} $\mu_{\U(d)}$. 
In general, for a compact Lie group the Haar measure is the unique left and right invariant probability measure, which generalizes the notion of a uniform measure. 
In applications one is often interested in random variables that are polynomials in matrix elements of a Haar-random unitary $U$ and its complex-conjugate $U^\ad$. 
In this case, also all moments of the random variable are the expected value of such polynomials. 
In this section we introduce the mathematical theory required to explicitly calculate such moments. 
To this end, we observe that any polynomial $p_t(U, U^\ad)$ of degree $\dorder$ can be written as the contraction with two matrices $A, B\in \CC^{d\dorder \times d\dorder}$ 
\begin{equation}\label{eq:poly_as_contraction}
	p_\dorder(U, U^\ad) = \Tr[B U^{\otimes \dorder} A (U^\ad)^{\otimes \dorder}]\, .
\end{equation}
This motivates the definition of the \emph{$\dorder$-th moment operator} 
of a probability measure $\mu$ on $\U(d)$ as 
$\mo^{(\dorder)}_{\mu}: \CC^{d\dorder \times d\dorder} \to \CC^{d\dorder \times d\dorder}$,  
\begin{equation}\label{eq:moment_op}
\begin{split}
	\mo^{(\dorder)}_{\mu}(A) 
	&= 
	\EE_{U\sim \mu}\bigl[ U^{\otimes \dorder} A (U^\ad)^{\otimes \dorder} \bigr]\\
		&= \int_{\U(d)} U^{\otimes \dorder} A (U^\ad)^{\otimes \dorder} \rmd\mu(U). 		
\end{split}
\end{equation}
If we have an expression for the $\dorder$-th moment operator for the Haar measure $\mu_{\U(d)}$, we can calculate the expectation value of arbitrary polynomials $p_\dorder(U, U^\ad)$ over $U\sim \mu_{\U(d)}$ by a linear contraction \eqref{eq:poly_as_contraction}. 

The crucial property that characterizes the $\dorder$-th moment operator of $\mu_{\U(d)}$ is the following: 
consider a fixed unitary $U \in \U(d)$ then a short calculation exploiting the unitary invariance of the Haar measure reveals that
\begin{equation}\label{eq:mop:commutes}
	U^{\otimes \dorder} \mo^{(\dorder)}_{\mu_{\U(d)}}(A) 
	= 
	\mo^{(\dorder)}_{\mu_{\U(d)}}(A) U^{\otimes \dorder}.
\end{equation}
We find that $\mo^{(\dorder)}_{\mu_{\U(d)}}(A)$ commutes with every unitary $U$ raised to the $\dorder$-th tensor power. 

For a set of endomorphisms $\mathcal A \subset \L(W)$ on a vector space $W$ one calls the set 
\begin{equation}
	\operatorname{comm}(\mathcal A) 
	= 
	\{ B \in \L(W) \mid BA = AB \quad \forall A \in \mathcal A \}
\end{equation}
of all endomorphisms that commute with all elements of $\mathcal A$ the \emph{commutant of $\mathcal A$}. 
The following lemma establishes that not only does $\mo^{(\dorder)}_{\mu_{\U(d)}}(A)$ commute with every unitary 
of the form $U^{\otimes \dorder}$ but it is in fact the orthogonal projector onto the commutant of $\mc A = \{U^{\otimes \dorder} | U\in \U(d)\}$, 
where orthogonality is understood with respect to the Hilbert-Schmidt inner product \eqref{eq:HS_inner_prod}. 
As becomes motivated shortly, we refer to 
\begin{equation}
	\Delta^{(\dorder)}: \U(d) \to \U(d^\dorder)
	\, , \qquad 
	U \mapsto U^{\otimes \dorder}
\end{equation}
as the \emph{diagonal representation} of $\U(d)$. 

\begin{lemmablock}[label={lem:momentoperatorIsProjector}]{$\dorder$-th moment operator}
	The $\dorder$-th moment operator $\mo^{(\dorder)}_{\mu_{\U(d)}}$ is the orthogonal projector onto $\operatorname{comm}(\Delta^\dorder[\U(d)])$, the commutant of the $\dorder$-order diagonal representation of $\U(d)$. 
\end{lemmablock}

\begin{proof}
	With \eqref{eq:mop:commutes} we established that the range of $\mo^{(\dorder)}_{\mu_{\U(d)}}$ is in $\operatorname{comm}(\Delta^\dorder(\U(d)))$. 
	The converse also holds since for $A \in \operatorname{comm}(\Delta^\dorder(\U(d)))$ we calculate that $\mo^{(\dorder)}_{\mu_{\U(d)}}(A) = A \mo^{(\dorder)}_{\mu_{\U(d)}}(\1) = A$. 
	Thus, it remains to check the orthogonality 
	$
		\mo^{(\dorder)\,\ad}_{\mu_{\U(d)}} = \mo^{(\dorder)}_{\mu_{\U(d)}}.
	$
	The orthogonality requirement follows in very few lines of calculation using linearity and cyclicity of the trace.
\end{proof}

Note that the argument of the proof applies more generally and yields the analogous result for arbitrary groups equipped with a Haar measure, e.g.\ the uniform measure on a finite group. 

The commutant of the diagonal representation of the unitary group can be 
characterized using a powerful result from representation theory: 
Schur-Weyl duality. 
To set the stage for explaining the result we start by reviewing some basic definitions and results from representation theory. 

\subsubsection{Representation theory}
Let us start with the most basic definitions. 
For a proper introduction we refer to Simon's book~\cite{Sim96} and to Goodman and Wallach's book \cite{GooWal09} for the representation theory of the standard matrix groups. 

Let $G$ and $H$ be groups. 
\begin{itemize}
	\item $f:G\to H$ is a \emph{(group) homomorphism} if $f(g_1 g_2) = f(g_1)f(g_2)$ for all $g_1,g_2\in G$. 
	Note that this condition implies that $f(e_G) = e_H$ and $f\myleft(g^{-1}\myright) = f(g)^{-1}$ for all $g\in G$. 

	\item Let $V$ be a vector space. 
	By $\operatorname{GL}(V)$ we denote the \emph{general linear group} over $V$, i.e.\ the group of invertible operators. 
	A homomorphism $R: G \to \operatorname{GL}(V)$ is called a \emph{linear (group) representations}. 
	$R$ is a \emph{unitary representation} if $R: G \to \U(\H)$ is a homomorphism to a unitary group $\U(\H) \subset \L(\H)$ on a Hilbert space $\H$. 
	Here, we are only concerned with such unitary representations and, hence, often omit the word ``unitary''. 

	\item A subspace $V\subset \H$ is said to be \emph{invariant} if $R(g) V \subseteq V$ for all $g\in G$. 
	$R$ is called \emph{irreducible} if the only invariant subspaces are $\{0\}$ and $\H$ itself. 
	\emph{Irreducible representations} are also called \emph{irreps} for short. 

	\item Two representations $R : G \to \U(\H)$ and $\tilde R : G \to \U(\tilde \H)$ are said to be \emph{unitarily equivalent} if there is a unitary operator $W: \H \to \tilde \H$ such that $\tilde R(g) = W  R(g) W^\ad$ for all $g\in G$. 
\end{itemize}

If $R_i:G \to \H_i$ for $i=1,2$ are two representations of $G$ then 
$(R_1\oplus R_2)(g) \coloneqq R_1(g)\oplus R_2(g)$ defines another representation $R_1\oplus R_2: G \to \H_1\oplus \H_2$. 
This representation has $\H_1$ and $\H_2$ as invariant subspaces. 
Conversely, if a representation $R$ has a non-trivial invariant subspace $V$ then it can be decomposed as $R = R|_V \oplus R_{V^\perp}$. 
By iterating this insight, 
we have the following statement (see e.g.\ \cite[Theorem~II.2.3]{Sim96}). 

\begin{propositionblock}[label={prop:irreps_decomp}]{Decomposition into irreps}
Let $R: G \to \L(\H)$ be a unitary representation of a 
group $G$ on a finite-dimensional Hilbert space $\H$. 
Then $(R,\H)$ can be decomposed into a direct sum of irreps $(R_i,\H_i)$ of $G$ as  
\begin{equation}\label{eq:irrep_decomp}
	\H = \bigoplus_i \H_i  
	\quad \text{and} \quad
	R(g) = \bigoplus_i R_i(g)\, .
\end{equation}
\end{propositionblock}

Several irreps $R_{i_1}, \dots, R_{i_m}$ in the decomposition \eqref{eq:irrep_decomp} might be unitarily equivalent to each other. 
The maximum number $m$ is called the \emph{multiplicity} of that irrep. 
The space $\CC^m$ in the resulting identification
\begin{equation}
	\bigoplus_{j=1}^m R_{i_j} (g)
	\cong
	R_{i_1}(g) \otimes \1_m  \quad \in \L(\H_1 \otimes \CC^m)
\end{equation}
is called the \emph{multiplicity space} of $R_{i_1}$. 
The decomposition \eqref{eq:irrep_decomp} is called \emph{multiplicity-free} if all irreps $R_i$ are inequivalent, i.e., not isomorphic. 

\begin{theoremblock}[label={thm:Schurs_lemma}]{Schur's lemma}
Let $R : G \to \U(\H)$ be an irrep of $G$ on $\H$. 
If $A \in \L(\H)$ satisfies 
\begin{equation}\label{eq:SchursLemmaCond}
	A R(g) = R(g) A \qquad \forall g \in G
\end{equation}
then $A= c\, \1$ for some $c\in \CC$. 
\end{theoremblock}

\begin{proof}
The condition \eqref{eq:SchursLemmaCond} implies that $R(h) A^\ad = A^\ad R(h)$ for all $h = g^{-1} \in G$. 
Hence, this condition also holds for $\Re(A) \coloneqq \frac 12 (A+A^\ad)$ and $\Im(A) \coloneqq \frac 1{2\i} (A-A^\ad)$ and $A$ is a constant if they both are. 
Hence, it is sufficient to prove the theorem for $A\in \Herm(\H)$. 

Let $\ket \psi$ be an eigenvector with $A \ket \psi = \lambda \ket \psi$ and 
$\Eig_\lambda(A) \coloneqq \{\ket \psi: \  A\ket \psi = \lambda\psi\}$ the full eigenspace. 
Then $R(g)\ket \psi \in \Eig_\lambda(A)$ for all $g\in G$ because $A R(g) \ket \psi = R(g) A \ket \psi = \lambda R(g) \ket\psi$. 
So, $\Eig_\lambda(A)$ is an invariant subspace. 
Since $\Eig_\lambda(A)\neq \{0\}$ and $R$ is an irrep, $\Eig_\lambda(A) = \H$ follows. 
\end{proof}

\begin{corollaryblock}[label={cor:irreps_abelian}]{Irreps of Abelian groups}
If $G$ is Abelian then every irrep has dimension $1$. 
\end{corollaryblock}

\begin{proof}
	Let $R$ be an irrep of of an Abelian group $G$ on $\H$. 
	Theorem~\ref{thm:Schurs_lemma} implies that each $g\in G$ has representation $R(g) = c\, \1$ for some constant $c$. 
	Hence, every subspace of $\H$ is invariant under $R$. 
	Since $R$ is an irrep this is only possible if $\dim(\H) = 1$. 
\end{proof}

There is also a slightly more general version of Schur's lemma: 

\begin{theoremblock}[label={thm:Schurs_lemmaII}]{Schur's lemma II}
Let $R : G \to \U(\H)$ and $\tilde R : G \to \U(\tilde \H)$ be two irreps of $G$ on finite-dimensional Hilbert spaces $\H$ and $\tilde \H$. 
If $A \in \L(\H, \tilde \H)$ satisfies 
\begin{equation}\label{eq:SchursLemmaCondII}
	A R(g) = \tilde R(g) A \qquad \forall g \in G
\end{equation}
then either $A = 0$ or $R_1$ and $R_2$ are unitarily equivalent up to a constant factor. 
\end{theoremblock}

\begin{proof}
The condition~\eqref{eq:SchursLemmaCondII} implies that for all $g\in G$
\begin{equation}
	R(g) A^\ad  
	= 
	A^\ad \tilde R(g) 
\end{equation}
and, hence, 
\begin{align}
	R(g) A^\ad A &= A^\ad A R(g)
	\\
	\tilde R(g) AA^\ad &= AA^\ad \tilde R(g) \, . 
\end{align}
Schur's lemma (Theorem~\ref{thm:Schurs_lemma}) implies that 
$A^\ad A = c \, \1$ and 
$AA^\ad =  \tilde c \, \1$ for constants $c, \tilde c$. 
Since the singular values of both operators have to coincide we find that $c=\tilde c$. 
It follows that either $c=0$ so that $A=0$ or that $W = A/\sqrt{c}$ is a unitary. 
In the latter case 
\begin{equation}\label{eq:intertwiningW}
	W R(g) = \tilde R(g) W
\end{equation}
for all $g\in G$, i.e., $R$ and $\tilde R$ are unitarily equivalent. 
\end{proof}

A unitary $W$ relating two representations $R$ and $\tilde R$ as in Eq.~\eqref{eq:intertwiningW} is called an \emph{intertwining} unitary of $R$ and $\tilde R$. 

\subsubsection{Schur-Weyl duality and the commutant of the diagonal action}
We wish to calculate the moments of random variables depending on Haar-random unitaries. 
Therefore, we are interested in understanding the commutant of the diagonal representation of the unitary group. 
Formally, we define the \emph{diagonal representation} of $\U(d)$ on $(\CC^d)^{\otimes \dorder}$ as 
\begin{equation}
	\Delta^\dorder_d : \U(d) \to \U\myleft( (\CC^d)^{\otimes \dorder} \myright) 
\end{equation}
by linearly extending the action 
\begin{equation}\label{eq:Ud-action}
	\Delta^{\dorder}_d (U) (\ket{\psi_1} \otimes \cdots \ket{\psi_\dorder}) \coloneqq (U \ket\psi_1) \otimes \cdots (U\ket{\psi_\dorder})\, .
\end{equation}

The representation $\Delta^\dorder_d$ has a duality relation with another well-known representation on $\mathbb {C^d}^\dorder$: the  representation $\pi_\dorder$ of the \emph{symmetric group} $\Sym_\dorder$ permuting the $\dorder$ tensor components: 
\begin{equation}\begin{split}\label{eq:Sk-action}
	\pi_\dorder &: \Sym_k \to \U\myleft( (\CC^d)^{\otimes \dorder} \myright) \, , \\
	\pi_\dorder&(\sigma)\myleft(\ket{\psi_1}\otimes \dots \otimes\ket{\psi_\dorder}\myright)
	\coloneqq
	\ket{\psi_{\sigma^{-1}(1)}}\otimes \dots \otimes\ket{\psi_{\sigma^{-1}(\dorder)}} \, .
\end{split}\end{equation}
We note that $\pi_\dorder(\sigma)$ and $\Delta_d(U)$ commute for any $\sigma\in \Sym_\dorder$ and $U\in \U(d)$. 

Let us consider the following two irreducible representations of the symmetric group which appear in the decomposition \eqref{eq:irrep_decomp} of $\pi_\dorder$ for any $\dorder$. 
We call $\ket\Psi \in (\CC^d)^{\otimes \dorder}$ \emph{symmetric} if $\pi_\dorder(\sigma)\ket\Psi = \ket\Psi$ for all $\sigma \in \Sym_\dorder$ and \emph{anti-symmetric} if $\pi_k(\sigma)\ket\Psi = \sign(\sigma)\ket\Psi$  for all $\sigma \in \Sym_\dorder$. 
The \emph{symmetric subspace} $\H_{\sym^\dorder}$ 
 and \emph{anti-symmetric subspace} $\H_{\wedge^\dorder}$ of $(\CC^d)^{\otimes \dorder}$ are the subspaces consisting of all symmetric and all anti-symmetric vectors, respectively. 
By $P_{\sym^\dorder}$ and $P_{\wedge^\dorder}$ we denote the orthogonal projectors onto these two subspaces. 

\ex{
\begin{exerciseblock}{Symmetric subspace}
\begin{itemize}
	\item Calculate $P_{\sym^\dorder} \ket \psi$ for a product state $\ket \psi = \ket{\psi_1}\otimes \dots \otimes \ket{\psi_\dorder}$. 
	\item Show that the dimension of the symmetric subspace $P_{\sym^\dorder} (\CC^d)^{\otimes \dorder}$ is 
	\begin{equation}\label{eq:dimSymSpace}
	\Tr[P_{\sym^\dorder}] = \binom{\dorder+d-1}{d-1} \, .
	\end{equation}
	\hint{Argue first that this is the number of ways to distribute $\dorder$ indistinguishable particles (bosons) into $d$ boxes (modes).}
	\item Show that 
	\begin{equation}\label{eq:psym_expansion}
	\begin{split}
		P_{\sym^\dorder} &= \frac1{k!} \sum_{\sigma \in \Sym_\dorder} \pi_\dorder(\sigma) \\
	 	\text{and\quad} P_{\wedge^\dorder} &= \frac1{k!}\sum_{\sigma \in \Sym_\dorder} \sign(\sigma)\pi_\dorder(\sigma) \, .
	 \end{split}
	 \end{equation}
\end{itemize}
\end{exerciseblock}
}
\exreplace{
\begin{lemmablock}{Symmetric subspace}
\begin{itemize}
	\item The dimension of the symmetric subspace $P_{\sym^\dorder} (\CC^d)^{\otimes \dorder}$ is 
	\ifex\begin{equation}\else\begin{equation}\label{eq:dimSymSpace}\fi
	\Tr[P_{\sym^\dorder}] = \binom{\dorder+d-1}{d-1} \, .
	\end{equation}

	\item The orthogonal projectors onto the symmetric and anti-symmetric subspace are
	\ifex\begin{equation}\else\begin{equation}\label{eq:psym_expansion}\fi
	\begin{split}
		P_{\sym^\dorder} &= \frac1{k!} \sum_{\sigma \in \Sym_\dorder} \pi_\dorder(\sigma) \\
	 	\text{and\quad} P_{\wedge^\dorder} &= \frac1{k!}\sum_{\sigma \in \Sym_\dorder} \sign(\sigma)\pi_\dorder(\sigma) \, , 
	 \end{split}
	 \end{equation}
	 respectively. 
\end{itemize}
\end{lemmablock}
\begin{proof}
The first statement is a combinatorial one. 
The trace of the symmetric projector is the number of ways to distribute $\dorder$ indistinguishable particles (bosons) into $d$ boxes (modes), i.e., the dimension of the corresponding subspace of the bosonic subspace, which is known to be given by the binomial coefficient. 

The second statement follows, e.g., for $P_{\sym^\dorder}$ by realizing that any symmetric vector in the range of $P_{\sym^\dorder}$ and that this operator is indeed a projector, i.e., that $P_{\sym^\dorder}$ is self-adjoint and $P_{\sym^\dorder}P_{\sym^\dorder} = P_{\sym^\dorder}$. 
\end{proof}
}

For the case of $\dorder=2$ the decomposition into these two subspaces is very familiar. 
It is easy to see that any matrix can be decomposed into a symmetric and an anti-symmetric part, which are orthogonal to each other. 
This implies that 
\begin{equation}\label{eq:Schur-Weyl_k2}
	(\CC^d)^{\otimes 2} 
	= 
	\H_{\sym^2} \oplus \H_{\wedge^2} \, .
\end{equation}
Note that due to Corollary~\ref{cor:irreps_abelian}, both the symmetric and the antisymmetric subspace are 
isomorphic to $\CC^{m_{\sym^2}}$ and 
$\CC^{m_{\wedge^2}}$, respectively;
here $m_{\sym^2}$ and $m_{\wedge^2}$ are the multiplicities of the two distinct one-dimensional irreps of $\Sym_2$. 

For $\dorder > 2$ there is a similar decomposition with more summands called \emph{Schur-Weyl} decomposition. 
The Schur-Weyl decomposition relies on a duality relation between the commuting representations $\Delta_d^\dorder$ and $\pi_\dorder$. 
The representations $\Delta_d^\dorder$ and $\pi_\dorder$ span each other's commutant as algebras. 

\begin{theoremblock}[label={thm:commutantOfDelta}]{Schur-Weyl duality \cite[Theorem 4.2.10]{GooWal09}}
	For the two commuting representations \eqref{eq:Ud-action} and \eqref{eq:Sk-action} it holds that 
	\begin{equation}
		\operatorname{comm}(\Delta_d^\dorder(\U(d))) = \operatorname{span}\{\pi_\dorder(\Sym_\dorder)\} \label{eq:commutant_as_perms}
	\end{equation}
	and
	\begin{equation}
		\operatorname{comm}(\pi_\dorder(\Sym_\dorder)) = \operatorname{span}\{\Delta_d^\dorder(\U(d))\}\, .
	\end{equation}
\end{theoremblock}
By Schur's lemma such a duality relation implies that the multiplicity spaces of the irreducible representation of one representation are irreducible representations of the dual representation and vice versa. 
In other words, $\CC^d$ decomposes into multiplicity-free representations of the combined action $\U(d) \times \Sym_k$. 
In order to state this composition, 
 we write $\lambda = (\lambda_1, \lambda_2 \ldots, \lambda_{l(\lambda)}) \vdash \dorder$ for a partition of $\dorder$ into $l(\lambda)$ non-increasing, positive integers with $\lambda_1\geq 1$ and 
 fulfilling 
\begin{equation}
	\dorder  = \sum_{i=1}^{l(\lambda)} \lambda_i\, .
\end{equation}
Such partitions of integers label the irreducible representations of the symmetric group and the diagonal representation. 
As a consequence of Schur-Weyl duality one can prove the following statement. 

\begin{theoremblock}[label={thm:Schur-Weyl}]{Schur-Weyl decomposition \cite[Theorem 9.1.2]{GooWal09}}
The action of $\U(d)\times \Sym_k$ on $(\CC^d)^{\otimes k}$ given by the commuting representations \eqref{eq:Sk-action} and \eqref{eq:Ud-action} is multiplicity-free and $(\CC^d)^{\otimes k}$ decomposes into irreducible components as 
\begin{equation}\label{eq:Schur-Weyl_decomp}
	(\CC^d)^{\otimes \dorder } 
	\cong
	\bigoplus_{\lambda \vdash \dorder, l(\lambda) \leq d}  W_\lambda \otimes S_\lambda \, ,
\end{equation}
where $\U(d)$ acts non-trivially only on $W_\lambda$ and $\Sym_k$ acts non-trivially only on $S_\lambda$. 

For any $\dorder\geq 2$, both $\H_{\sym^\dorder}$ and $\H_{\wedge^\dorder}$ occur as components in the direct sum~\eqref{eq:Schur-Weyl_decomp}. 
\end{theoremblock}

The spaces $W_\lambda$ are called \emph{Weyl modules} and $S_\lambda$ \emph{Specht modules}. 
Schur-Weyl duality implies that the Weyl modules are the multiplicity spaces of the irreps of $\Sym_k$ and, similarly, 
the Specht modules are the multiplicity spaces of the irreps of $\U(d)$. 

Schur-Weyl duality, Theorem~\ref{thm:commutantOfDelta}, and the resulting decomposition, Theorem~\ref{thm:Schur-Weyl}, give a simple characterization of the commutant of the diagonal action of the unitary group. 
The relation \eqref{eq:commutant_as_perms} allows one to derive an expression for the $\dorder$-moment operator $\mo^{(\dorder)}_{\mu_{\U(d)}}$ 
as the orthogonal projector onto the span of the symmetric group. 
But one has to be careful since $\{\pi^d_\dorder(\sigma)\}_{\sigma\in \Sym_\dorder}$ is not an orthonormal basis. 
Note that it only becomes an orthogonal set asymptotically for large $k$, which can be exploited in some applications, e.g.\ in local random quantum circuits \cite{brandao_local_2016}. 
A general expression in terms of so-called Weingarten functions \cite{weingarten_asymptotic_1978} was derived by Collins and Sniady \cite{SWDuality}, see also the Supplemental Material of Ref.~\cite{RotKueKim18}
for a convenient expression of their result and a summary of the derivation. 
For our purposes we only need to derive an expression $\mo^{(\dorder)}_{\mu_{\U(d)}}$ for certain special cases, 
namely, for $\dorder=2$ and when restricted to symmetric endomorphisms as its input. 

We begin with the second moment, $\dorder=2$. 
\begin{propositionblock}[label={prop:invariant_operators_k2}]{Second moment operator}
For an operator $A \in \L(\CC^d \otimes \CC^d)$, $d \geq 2$, it holds that
\begin{equation}\label{eq:secondMomentOperator}
	\mo^{(2)}_{\mu_{\U(d)}}(A) = c_{\sym^2} P_{\sym^2} + c_{\wedge^2} P_{\wedge^2}\, 
\end{equation}
with 
$
	c_{\sym^2} = \frac{2}{d(d+1)}\Tr[A P_{\sym^2}]
$
and $
c_{\wedge^2} = \frac{2}{d(d-1)}\Tr[A P_{\wedge^2}]
$. 
\end{propositionblock}

\begin{proof}
	From 
	Lemma~\ref{lem:momentoperatorIsProjector} and 
	Theorem~\ref{thm:commutantOfDelta} 
	we know that $\mo^{(2)}_{\mu_{\U(d)}}(A)$ is a linear combination of the identity $\1$ and the swap operator $\FF$ from \eqref{eq:SwapOp}. 
	For $\Sym_2$ the expansion of the projectors \eqref{eq:psym_expansion} onto the symmetric and anti-symmetric subspace can be inverted resulting in $\id = P_{\sym^2} + P_{\wedge^2}$ and $\FF = P_{\sym^2} - P_{\wedge^2}$. 
	This establishes the form of \eqref{eq:secondMomentOperator}. 
	Since $P_{\sym^2}$ and $P_{\wedge^2}$ are mutually orthogonal projectors and $\mo^{(2)}_{\mu_{\U(d)}}$ is an orthogonal projector
	the coefficients are given by $c_{\sym^2} = \Tr[A P_{\sym^2}] / \Tr[P_{\sym^2}] = \frac2{d(d+1)} \Tr[AP_{\sym^2}]$ and $c_{\wedge^2}$ analogously. 
\end{proof}

Second, we allow for arbitrary $\dorder$ but restrict the input of $\mo_{\mu_{\U(d)}}^{(\dorder)}$ to endomorphisms that are itself symmetric, i.e., of product form. 
In this case we also find an orthogonal decomposition as given by the following lemma.
\begin{lemmablock}[label={lem:sym_moment_operator}]{Moment operator on symmetric operators}
	For any operator $A \in \L(\CC^{d})$ it holds that
	\begin{equation}
		\mo_{\mu_{\U(d)}}^{(\dorder)} (A^{\otimes \dorder}) = \sum_{\lambda \vdash k, l(\lambda) \leq d} c_\lambda P_\lambda\, , 
	\end{equation}
	with $P_\lambda$ the orthogonal projector onto $W_\lambda \otimes S_\lambda$ and 
	$c_\lambda = \Tr(P_{\lambda} A^{\otimes k}) / \Tr(P_{\lambda})$.
	Furthermore, if the operator $A$ is of \emph{unit rank}, then 
	\begin{equation}
		\mo_{\mu_{\U(d)}}^{(\dorder)} (A^{\otimes \dorder}) = c P_{\sym^k}\, ,
	\end{equation}
	with $c = \Tr(P_{\sym^k} A^{\otimes k}) / \Tr(P_{\sym^k})$.
\end{lemmablock}
\begin{proof}
	We fix some $A \in \L(\CC^{d})$ and denote 
	$E \coloneqq \mo_{\mu_{\U(d)}}^{(\dorder)}(A^{\otimes k})$. 
	By the definition of the moment operator \eqref{eq:moment_op}, 
	$E 
	= 
	\int_{\U(d)} (UAU^\ad)^{\otimes \dorder} \rmd\mu_{\U(d)}(U)$ and it becomes apparent that 
	$E$ commutes with $\pi^d_\dorder (\sigma)$ for any $\sigma \in \Sym_\dorder$. 
	In other words, $E \in \operatorname{comm}{\Delta^\dorder_d(\U(d))} \cap \operatorname{comm}{\pi^d_\dorder(\Sym_\dorder)}$ by Lemma~\ref{lem:momentoperatorIsProjector}. 
	By Schur's lemma (Theorem~\ref{thm:Schurs_lemma}) and the Schur-Weyl decomposition \eqref{eq:Schur-Weyl_decomp}, we thus conclude that 
	$E$ acts proportionally to the identity on every Weyl module $W_\lambda$ and Specht module $S_\lambda$. 
	Denoting the orthogonal projector onto $W_\lambda \otimes S_\lambda$ as $P_\lambda$, the  operator
	$E$ permits the decomposition $E = \sum_{\lambda \vdash k, l(\lambda) \leq d} c_\lambda P_\lambda$ with $c_\lambda \in \CC$.
	Since the projectors are mutually orthogonal the coefficients are given by $c_\lambda = \Tr(A^{\otimes k} P_\lambda) / \Tr(P_\lambda)$. 
	This establishes the lemma's first assertion for $E$. 

	Finally, for unit rank $A$, i.e.\ $A = \ketbra\psi\phi$ with $\ket\psi, \ket\phi \in \CC^d$, we observe that $P_{\sym^k} A^{\otimes k} P_{\sym^k} = P_{\sym^k} {\ket\psi}^{\otimes n} {\bra\phi}^{\otimes n}  P_{\sym^k} = A^{\otimes k}$. Hence, $c_\lambda = 0$ for all $\lambda$ that do not correspond to the symmetric subspace. This leaves us with the lemma's second expression for $E$. 
\end{proof}

\subsubsection{Uniformly random state vectors}\label{sec:haar-random-states}
One can also define a uniform distributed on pure quantum states in multiple equivalent ways. 
First, one can draw randomly from the complex sphere $\sphereCd$, i.e.\ the set of normalized vectors in $\CC^d$. 
Indeed, there is a unique uniform probability measure $\mu_{\sphereCd}$ on $\sphereCd$ that is invariant under the canonical action of $\U(d)$ on $\CC^d$. 
By definition we see that a column $\ket\psi = U\ket 0$ of a Haar-randomly drawn unitary $U \sim \mu_{\U(d)}$ is distributed 
according to $\mu_{\sphereCd}$. 
Finally, we can switch to density matrices by factoring out a global phase. 
In more detail, the \emph{complex projective space} $\Cproj^{d-1} \coloneqq \sphereCd/\U(1)$ is the space of state vectors modulo a phase in $\U(1)$, which can be identified with the space of pure density matrices $\Cproj^{d-1} \subset \DM(\CC^d)$. 
It also has a uniform unitarily invariant probability distribution: a
uniformly random pure state $\ketbra \psi\psi$ can be obtained by drawing $\ket\psi\sim \mu_{\sphereCd}$. 

We can calculate the moments of polynomials that depend on states drawn uniformly from $\mu_{\sphereCd}$ using the 
moment operator $\mo^{(\dorder)}_{\mu_{\U(d)}}$. 
To this end, note that any polynomial $p_\dorder(\ket\psi, \bra\psi )$ of degree $\dorder$ in the component of each $\ket \psi$ and $\bra \psi$
 can be written as a contraction of ${\ketbra\psi\psi}^{\otimes\dorder}$ with some operator in $
 \L(\CC^{d^k})$. 
For this reason the following lemma summarizes everything we need. 

\begin{lemmablock}[{label={lem:moments_rand_states},fonttitle=\normalfont}]{Moment operator of random states}
	Let $K^{(\dorder)}_{\mu_{\sphereCd}}$ be the moment operator for $\ket\psi \sim \mu_{\sphereCd}$ explicitly defined by
	\begin{equation}
		K^{(\dorder)}_{\mu_{\sphereCd}} 
		\coloneqq 
		\int_{\sphereCd} (\ketbra \psi \psi)^{\otimes \dorder} \rmd \mu_{\sphereCd}(\psi) \, .
	\end{equation}
	It holds that 
	\begin{equation}
		K^{(\dorder)}_{\mu_{\sphereCd}} = \frac{\dorder!(d-1)!}{(\dorder+d-1)!} P_{\sym^k}\, ,
	\end{equation}
where $P_{\sym^k}$ is the projector \eqref{eq:psym_expansion} onto the symmetric subspace. 
\end{lemmablock}

\begin{proof}
	As $\mu_{\sphereCd }$ is $\U(d)$-invariant, 
	we find $K_\dorder = \mo_{\mu_{\U(d)}}^{(\dorder)}((\ketbra\psi\psi)^{\otimes\dorder})$. Lemma~\ref{lem:sym_moment_operator} thus implies that $K_\dorder = c P_{\sym^k}$ with 
	\begin{equation*}
		c = \frac{\Tr(P_{\sym^k} (\ketbra\psi\psi)^{\otimes k})}{\Tr(P_{\sym^k})}. 
	\end{equation*}
	Since $P_{\sym^k}$ acts trivially on $\ket\psi$ and it is normalized, the enumerator evaluates to $1$. 
	The denominator is the dimension of $P_{\sym^\dorder}$ given by \eqref{eq:dimSymSpace}.
\end{proof}

\subsubsection{Unitary, spherical and complex-projective \texorpdfstring{$k$}{k}-designs}
With our excursion to representation theory we derived expressions to calculate the moments of random variables on uniformly random states and unitaries. 
The very same results can also be used for certain other interesting probability distributions. 
To this end, note that if we want to control only the first $t$ moments of a random variable that is a polynomial of degrees $\ell$ in a random state or unitary, then our calculation will only involve the moment operators $\mo^{(\dorder)}_{\mu_{\U(d)}}$ for $k \leq t\ell$. 
In many applications it is sufficient to control the expectation value and the variance of low-degree polynomials. 
In these cases, any probability distribution that reproduces the first couple of moments of the uniform distributions 
can be used without changing the mathematical expressions. 
This idea is formalized by the definition of $\dorder$-designs. 

\begin{definitionblock}[label={def:unitary_design}]{Unitary $\dorder$-design}
A distribution $\mu$ on the unitary group $\U(d)$ is a \emph{unitary $\dorder$-design} if its $\dorder$-th moment operator \eqref{eq:moment_op} coincides with the one of the Haar measure,  
\begin{equation}
	\mo^{(\dorder)}_{\mu} 
	= 
	\mo_{\mu_{\U(d)}}^{(\dorder)}.
\end{equation} 
Furthermore, a subset $\{U_1, \dots, U_\ng\}\subset \U(d)$ is called a \emph{unitary $\dorder$-design} if its uniform distribution is one. 
\end{definitionblock}
\vspace{\baselineskip}

\ex{
\begin{exerciseblock}{($\dorder-1$)-designs}
Prove that a unitary $\dorder$-design is also a unitary ($\dorder-1$)-design for $\dorder\geq 2$. 
\end{exerciseblock}
\vspace{\baselineskip}
}
\exreplace{
Note that by definition, any unitary $\dorder$-design is also a unitary ($\dorder-1$)-design for $\dorder\geq 2$. 
}

A famous example of a unitary design in the context of quantum computing is the Clifford group. 
\begin{block}{The Clifford group}
The $n$-qubit \emph{Clifford group} $\Cl_n\subset \U(2^n)$ is the normalizer of the Pauli group $\mcP_n$ (see Section~\ref{eq:STAB_as_sum}),
\begin{equation}
	\Cl_n \coloneqq \{U \in \U(2^n;\QQ): 
	\
	U \mcP_n U^\ad \subset \mcP_n \} \, ,
\end{equation}
where it is common to restrict to unitary matrices with complex rational entries, here denoted by 
$\U(d;\QQ)\coloneqq \U(d)\cap (\QQ^{d\times d}+\i \QQ^{d\times d})$, so that $\Cl_n$ becomes a finite group. 
This group is generated by the 
single qubit Hadamard gate $\operatorname{H}$ and the phase gate $\operatorname{S}$ given by
(see, e.g.~\cite[Theorem~10.6]{NieChu10})
\begin{equation}
\begin{split}
	\operatorname H = \frac1{\sqrt{2}}
		\begin{pmatrix} 1 & 1 \\ 1 & -1 \end{pmatrix}
	\quad\text{and}\quad
	\operatorname S = \begin{pmatrix} 1 &  \\  & \i \end{pmatrix}
\end{split}
\end{equation}
together with the two-qubit \CNOT\ gate 
\begin{equation}
	\CNOT = 
	\ketbra 00 \otimes \1 + \ketbra 11 \otimes \sigma_x\, ,
\end{equation}
all acting locally on any qubit. 

Together with the $\operatorname{T} = \sqrt{\operatorname S}$ gate the Clifford group is 
a universal gate set (see, e.g.~\cite[Section~4.5.3]{NieChu10}). 

The Clifford group is a unitary $3$-design but not a unitary $4$-design \cite{Web15,Zhu15,ZhuKueGra16}. 
Being a subgroup of the unitary group the commutant of the 
diagonal action of the Clifford group for $k > 3$ is, thus, a strictly larger space than the span of the permutation group. 
A classification of the `missing generators' of the commutant was done by Gross \emph{et al.}~\cite{gross2017schur}. 
\end{block}

Analogously to unitary designs, we can define spherical $k$-designs. 
For a distribution $\mu$ on the complex sphere $\sphereCd$ we define the \emph{$\dorder$-th moment operator} as
\begin{equation}
	K^{(\dorder)}_{\mu} 
	\coloneqq 
	\int_{\sphereCd } (\ketbra \psi \psi)^{\otimes \dorder} \rmd \mu(\psi) \, .
\end{equation}

\begin{definitionblock}[label={def:t-design}]{Complex spherical/proj\-ec\-tive $k$-design}
A distribution $\mu$ on $\sphereCd $ is a \emph{spherical $k$-design} if 
\begin{equation}
	K^{(k)}_{\mu} = K_{\mu_{\U(d)}}^{(k)}.
\end{equation} 
Furthermore, a subset $\sphereCd $ is called a \emph{spherical $k$-design} if its uniform distribution is $1$. 
The corresponding distribution of $\ketbra \psi \psi$ is called a \emph{complex projective $\dorder$-design}. 
\end{definitionblock}

See also Refs.~\cite{AmbEme07,RoySco07} for related definitions. 

Analogously to the relation of the uniform measure on $\U(d)$ and $\sphereCd$, 
a rather obvious but important example of a spherical $\dorder$-designs it given by the orbits of a unitary $\dorder$-design. 
If $\mu$ is a unitary $\dorder$-design for $\U(d)$ and $\ket\psi \in \CC^d$ then the induced distribution $\tilde\mu$ given by $U \ket\psi$ with $U \sim \mu$, is a complex spherical $\dorder$-design. 

One can use this relation to see that the Clifford group being a unitary $3$-design implies the analogous statement for stabilizer states. 

\begin{block}{Stabilizer states are $3$-designs\label{STABs2}} 
The set of all stabilizer states \eqref{eq:STAB_as_sum} is known to be a $2$-design \cite{GroAudEis07,DanCleEme09}, actually even a $3$-design but not a $4$-design \cite{Zhu15,Web15,KueGro15}. 
\end{block}
Other examples for spherical designs that play important roles 
in quantum system characterization are \emph{\acl{MUBs}} and \emph{\acl{SIC} \acp{POVM}}. 

\begin{block}{\Ac{MUBs}}
\ac{MUBs} are sets of bases with minimal overlaps. 
More explicitly, two orthonormal bases $\{\ket{\psi_i}\}_{i \in [d]}\subset \CC^d$ and $\{\ket{\phi_i}\}_{i \in [d]}\subset \CC^d$ are said to be \emph{mutually unbiased} if $|\braket{\psi_i}{\phi_j}|^2 = \frac 1d$ for all $i,j\in [d]$. 
For instance, if $U\in \U(d)$ is the discrete Fourier transform then the bases $\{\ket{i}\}_{i \in [d]}\subset \CC^d$ and $\{U\ket{i}\}_{i \in [d]}\subset \CC^d$ are mutually unbiased. 
The number of \ac{MUBs} in $\CC^d$ is upper bounded by $d+1$ and, in prime power dimensions (e.g., for qubits), there are exactly $d+1$ \ac{MUBs} \cite{Ivonovic:1981:GeometricalDescriptionOf,Wootters:1989:OptimalState-Determination}. 
However, it is a well-known open problem to exactly obtain this number for all $d$. 
Klappenecker and Roettler \cite{KlaRoe05} showed that maximal sets of \ac{MUBs} are complex spherical $2$-designs.
\end{block}

\begin{block}{\acs{SIC} \acp{POVM}}
A \emph{\acf{SIC} \ac{POVM}} is given by a set of $d^2$ normalized vectors $\{\ket{\psi_j}\}_{j \in [d^2]}\subset \sphereCd  \subset \CC^d$ satisfying 
\begin{equation}
	|\braket{\psi_i}{\psi_j}|^2 = \frac{1}{d+1} \qquad \forall i\neq j \, .
\end{equation}
``Symmetric'' refers to the inner products being all equal. 
Zauner \cite{zauner_quantendesigns_1999} has investigated \ac{SIC} \acp{POVM} systematically. 
Renes \emph{et al.}~\cite{RenBluSco04} have shown that \ac{SIC} \acp{POVM} are indeed $2$-designs. 
Both works provide explicit constructions for small dimensions. 
\end{block}

\subsection{\protchaptitle{Shadow fidelity estimation}}\label{sec:shadow_fidelity_estimation}
Another recently proposed approach to fidelity estimation makes use of estimating so-called classical shadows \cite{Huang2019PredictingFeatures, Huang2020Predicting}.
The principle idea of shadow estimation is to calculate the least-square estimator of a quantum state from recorded classical measurement outcomes with measurement setting drawn from a certain \emph{measurement frame}. 
As we see in this section such a \ac{POVM} that allows for a quite explicit analysis is given by a complex projective $3$-design. 

From the state's least-square estimator one can construct estimators of multiple target functions of the state, which are linear functions or even higher degree polynomials. 
The sampling complexity of the derived estimators can be captured by a so-called \emph{shadow norm} that is defined in terms of the measurement frame. 
The classical post-processing complexity is determined by the complexity of constructing the state estimator and evaluating the target functions. 
Operationally, the analyzed \ac{POVM} measurement is assumed to be implementable by 
random unitaries from a suitable ensemble and a consecutive basis measurement. 
While shadow estimation is a rather broad and flexible framework, 
we focus on the estimation of \emph{fidelities with pure target states} using unitaries that form a unitary $3$-design, e.g., multi-qubit Clifford gates or suitable subgroups thereof. 
Besides being an instructive example for shadow fidelity estimation, the $3$-design setting can be equipped 
with a performance guarantees that features a sampling complexity $\LandauO(\epsilon^{-2})$ that does not scale with the Hilbert space dimension.  
This system-size-independent scaling is not achievable in general for other measurement frames. 

The complete \acf{SFE} protocol is the following. 

\begin{protocolblock}[label={protocol:SFE}]{\Ac{SFE}}
Let $\mu$ be a distribution on $\U(d)$, 
$\{\ket b: b \in [d]\}\subset \CC^d$ an orthonormal basis and $\rhot\in \DM(\CC^d)$ be a target state.

The protocol consists of the following steps applied to state preparations $\{\rhop_i\}_{i=1}^\np \subset \DM(\CC^d)$. 
For each $\rhop_i$ perform the following steps:
\begin{compactenum}[(i)]
\item Draw $U_i \sim \mu$. 
\item\label{sfe:step:exp} Perform the following experiment: 
	\begin{compactenum}[I)]
		\item Prepare $\rhop_i$.
		\item Apply the gate $\rhop_i \mapsto U_i\rhop_i U_i^\ad$.
		\item Perform the basis measurement $\mathcal B$ and record the outcome $b_i \in \{0, 1\}^n$.
	\end{compactenum}
\item Calculate 
\begin{equation}\label{eq:sfe:single_estimators}
	\hat f_i = (d + 1) \sandwich {b_i} {U_i\rhot\, U^\ad_i} {b_i} - 1\, .
\end{equation}
\end{compactenum}
Output the median of means estimator \eqref{eq:MMestimator} of $\{\hat f_i\}_{i=1}^{n_\rhop}$. 
\end{protocolblock}

We present the protocol as iterations over combined experimental and classical pre- and post-processing steps. 
Note, however that one can complete the three stages separately:
first, one can classically generate the complete sequence of $\np$ random unitaries. 
Then, one can subsequently perform the quantum experiment, i.e.\ all repetitions of step~\eqref{sfe:step:exp}. 
Importantly, at this stage not even the knowledge of the target state $\rhot$ is required. 
Storage of the experimental outcomes, $\np$ bit strings, requires $\np\log n$ bits. 
These bit strings together with a prescription of the random sequence of unitaries are then taken as the input of the post-processing algorithm that calculates the median of means estimator. 
The complexity the classical post-processing depends on the complexity of calculating the overlap of Eq.\ \eqref{eq:sfe:single_estimators}. 
For an arbitrary target state $\rhot$ the effort of performing this task can scale exponentially in the number of qubits. 
In contrast, for stabilizer states and Clifford group unitaries the Gottesman-Knill theorem, see e.g.\ the book by Nielsen and Chuang \cite{NieChu10}, allows for an efficient computation of this expression. 

Shadow fidelity estimation comes along with the following guarantee. 

\begin{theoremblock}[label={thm:SFE}]{Guarantee for \ac{SFE}}
Consider Protocol~\ref{protocol:SFE} with $\mu$ being a unitary $3$-design and $\rhot$ a \emph{pure} target state. 
Choose $\delta \in (0,1)$, $\epsilon > 0$ 
and a number 
\begin{equation}
	\np \geq 160\, \frac1{\epsilon^2} \ln\frac1\delta
\end{equation}
such that it is a multiple of $k = \lceil 8\ln(1/\delta) \rceil$. 
Then, the median of means estimator of the protocol 
is an $\epsilon$-accurate unbiased estimator of $\fidelity(\rhot, \rhop)$ with confidence $1-\delta$ for $\np$ iid.\ state preparations;
the median is taken over $l = \np / k$ means, each of which is an empirical mean of $k$ realizations of $\hat f_i$. 
\end{theoremblock}

Theorem~\ref{thm:SFE} shows that \ac{SFE} requires a number of state copies that for arbitrary pure target states does not dependent on the Hilbert space dimension. 

With the \ac{DFE} protocol of Section~\ref{sec:fidelity_estimation} we already encountered another fidelity estimation protocol. 
In contrast to \ac{SFE}, recall that \ac{DFE} features a sampling complexity independent of the Hilbert space dimension only for the class of well-conditioned states, cmp.\ Theorem~\ref{thm:DFE_well-cond}. 
Keep in mind, however, that in order to additionally ensure an efficient classical post-processing 
also \ac{SFE} requires further structure such as provided by stabilizer states. 
Finally, note that \ac{SFE} and \ac{DFE}, as presented here, make use of different type of measurement data. While \ac{SFE} uses basis measurement randomly selected from a large set of bases, \ac{DFE} uses the expectation values of observables. 
Correspondingly, they differ in their requirements for experimental implementations. 

The proof of the performance guarantee, Theorem~\ref{thm:SFE}, proceeds in three steps: 
first, we have to establish that the \ac{SFE} estimator actually estimates the fidelity for pure target states. 
To derive the sampling complexity of the estimator a 
natural attempt would be to employ Hoeffding's inequality. 
Unfortunately, the random variables $\hat f_i$ defined in Eq.~\eqref{eq:sfe:single_estimators} only have bounds scaling as $\LandauO(d)$. 
This becomes exponentially large in the number of qubits and does not yield the desired scaling. 
The main insight underlying the efficiency of shadow fidelity estimation 
is that due to the structure of the unitary $3$-design the variance 
of $\hat f_i$ is still bounded in $\LandauO(1)$. 
Thus, as a second step we derive the bound for the variance. 
Finally, by combining both results we arrive at the sampling complexity using the tail bound for the median of mean estimator introduced in Theorem~\ref{thm:MME}. 
Using the median of mean estimator allows us to derive a sampling complexity in $\LandauO(\ln \delta^{-1})$ in the confidence $1-\delta$. 
Note that simply using an empirical mean estimator in the \ac{SFE} protocol can also be equipped with a guarantee with sampling complexity in $\LandauO(\delta^{-1})$ using {Chebyshev}'s inequality, Theorem~\ref{thm:Chebyshev}. 
A mean estimator might in a practical parameter regime even be more precise compared to the median of mean estimator. 

\begin{lemmablock}[fonttitle=\normalfont]{Unbiasedness of \ac{SFE} estimator}
Consider Protocol~\ref{protocol:SFE} with $\mu$ being a unitary $2$-design and $\rhot$  a pure target state. 
Let $\hat f_i$ be a random variable \eqref{eq:sfe:single_estimators} w.r.t.\ a state preparation $\rhop$. 
	Then
\begin{equation}\label{eq:SFE:Efi}
	\EE[\hat f_i] 
	= 
	\fidelity(\rhot, \rhop),
\end{equation}
where the expectation value is taken over both, $U \sim \mu$ and the subsequent random measurement outcome. 
\end{lemmablock}

\begin{proof}
	For convenience we suppress writing the index $i$. 
	Born's rule for the probability of the measurement outcomes yields
	\begin{equation}\label{eq:SFE:p}
		p(b) = \sandwich{b}{U\rhop\, U^\ad}{b}.
	\end{equation}
	Thus, the expectation value over $U$ and the measurement reads
	\begin{equation}\begin{split}
		\EE [\hat f] 
			&= \EE_{U \sim \mu} \Biggl[\sum_{b=1}^d
			\sandwich{b}{U\rhop\, U^\ad}{b} 
			\left[(d + 1)  \sandwich{b}{U\rhot\, U^\ad}{b} - 1 \right] \biggr] .
	\end{split}	\end{equation}
	The second term can be directly evaluated using the fact that we sum over a basis,
	\begin{equation} \label{eq:one-design-property}
	\begin{split}
		\sum_{b=1}^d \EE_{U\sim \mu} 
			\sandwich b {U \rhop \,U^\ad} b 
		&= \EE_{U\sim \mu}\bigl[ \Tr[U \rhop\, U^\ad] \bigr] 
		\\
		&= 
		\Tr[\rhop]\, . 
	\end{split}
	\end{equation}
	The first term can be calculated using the $3$-design property of $\mu$. 
	More precisely, at this point we need only $\mu$ to be a $2$-design. 
	Recall that if $U \sim \mu$ is a unitary $k$-design then for any state $\ket \tau$ its orbit $\ket \phi = U\ket \tau$ with the induced measure $\tilde \mu$ is a state $k$-design.
	Thus, using the swap-trick \eqref{eq:swap-trick} and Lemma~\ref{lem:moments_rand_states} we calculate that 
	\begin{align}\nonumber
		\EE_{U \sim \mu}&\biggl[(d+1) \sum_{b=1}^d
			\sandwich{b}{U\rhop\, U^\ad}{b} 
			\sandwich{b}{U\rhot\, U^\ad}{b}\biggr] 
			\\\nonumber
			&=  
				(d+1) \sum_{b=1}^d \Tr\myleft[
					\EE_{U\sim \mu}\Bigl[\left(U^\ad\ketbra bb U\right)^{\otimes 2}\Bigr]
					(\rhop \otimes \rhot) \FF
				\myright] \\\nonumber
			&= 
				(d+1) \sum_{b=1}^d \Tr\myleft[
					\EE_{\ket\phi\sim \tilde\mu}\Bigl[\left(\ketbra \phi \phi\right)^{\otimes 2}\Bigr]
					(\rhop \otimes \rhot) \FF
				\myright] \\\nonumber
			&= \Tr\left[ (\1 + \FF)(\rhop \otimes \rhot) \FF \right]  \\
			&= \Tr[\rhot\rhop] + \Tr[\rhop]\, \Tr[\rhot] \, .
	\end{align}
	Combining both terms again and using that $\Tr[\rhot]=1$, 
	we find that 
	\begin{equation}
	 	\EE[\hat f] = \Tr[\rhot\rhop] \, .
	 \end{equation} 
	Using that $\rhot$ was assumed to be a pure state establishes the statement \eqref{eq:SFE:Efi}. 
\end{proof}

Next we bound the variance. 
\begin{lemmablock}{Variance bound for \ac{SFE}}
Consider Protocol~\ref{protocol:SFE} with $\mu$ being a unitary $3$-design and $\rhot$  a pure target state. 
Let $\hat f_i$ be a random variable \eqref{eq:sfe:single_estimators} w.r.t.\ a state preparation $\rhop$. 
	Then
	\begin{equation}\label{eq:SFE:varbound}
		\Var[\hat f_i] < 5 \, , 
	\end{equation}
	where the variance is taken over both, $U \sim \mu$ and the subsequent random measurement outcome. 
\end{lemmablock}

\begin{proof}
	We again suppress the index $i$. 
	The variance is
	\begin{equation}\label{eq:SFE:var}
	\begin{split}
		\Var[\hat f] &= \EE \bigl[ \hat f^2 \bigr] - \EE[\hat f]^2 \, .
	\end{split}
	\end{equation}
	Using Born's rule \eqref{eq:SFE:p}, Eq.~\eqref{eq:one-design-property} and that $U \ket b$ is distributed as a complex spherical $3$-design $\tilde \mu$, 
	the second moment can be written as
	\begin{align}\nonumber
		\EE \bigl[ \hat f^2 \bigr] 
		= 
		d(d + 1)^2 &\EE_{\ket\phi \sim \tilde\mu}\bigl[   \sandwich \phi \rhop \phi \left| \Tr[\ketbra\phi\phi \rhot ]\right|^2 \bigr]
		\\ 
		&- 2 \EE[\hat f] + \Tr[\rhop]\, . 
		\label{eq:SFE:varFirstTerm}
	\end{align}
	The first term in this expression can be calculated using the $3$-design property of $\tilde\mu$ 
	and Lemma~\ref{lem:moments_rand_states}, 
	\begin{equation}
	\begin{split}
		\EE_{\ket\phi \sim \tilde\mu}&\bigl[   \sandwich \phi \rhop \phi \left| \Tr[\ketbra\phi\phi \rhot ]\right|^2 \bigr]
		\\
		&=
		\EE_{\ket\phi \sim \tilde\mu} \bigl[  
			 \Tr[\rhop\ketbra\phi\phi] \, \Tr[\rhot\ketbra\phi\phi]^2
			 \bigr]
		\\
		&=
		  \Tr\bigl[(\rhop\otimes \rhot\otimes \rhot) \EE_{\ket\phi \sim \tilde\mu}[\ketbra\phi\phi^{\otimes 3}]\bigr] 
		\\
		&=
		\frac{6}{d(d+1)(d+2)}\, 
		\Tr\bigl[(\rhop\otimes \rhot\otimes \rhot) P_{\sym^3}\bigr] .
	\end{split}
	\end{equation}
	We recall that the projector $P_{\sym^3}$ onto the symmetric representation of the symmetric group $\Sym_3$ is given by the sum of all six permutations in $\Sym_3$. Those are the identity, $3$ transpositions and the cyclic and anticyclic permutation. 
	Writing out this sum and tracking the resulting contractions (which can be most conveniently done using tensor network diagrams) yields 
	\begin{equation}\label{eq:SFE:varFirstTermExpr}
	\begin{split}
		\EE_{\ket\phi \sim \tilde\mu}&\bigl[   \sandwich \phi \rhop \phi \left| \Tr[\ketbra\phi\phi \rhot ]\right|^2 \bigr]
		\\
		&=
		\frac{1}{d(d+1)(d+2)}\, 
		\Bigl( 
			  \Tr[\rhop]\, \Tr[\rhot]^2 
			  \\ &\qquad \qquad 
			+ 2\Tr[\rhop\rhot]\, \Tr[\rhot]
			+ \Tr[\rhop] \Tr[\rhot^2] 
			\\ &\qquad \qquad \qquad 
			+2 \Tr[\rhop\rhot^2] 
		\Bigr) 
		\\
		&=
		\frac{2+ 4\fidelity(\rhot,\rhop)}{d(d+1)(d+2)}
	\end{split}
	\end{equation}
	where we use the normalization of the states and that $\rhot$ is pure in the last identity. 
	Combining \eqref{eq:SFE:var}, \eqref{eq:SFE:varFirstTerm} and using the expression \eqref{eq:SFE:Efi} from the previous lemma and \eqref{eq:SFE:varFirstTermExpr} we find the upper bound
	\begin{equation}
	\begin{split}
		\Var[\hat f] &= \frac{d+1}{d+2}(2+4F) - 2F + 1 - F^2 
		\\
		&<
		2(1+F) + 1-F^2\leq 5
	\end{split}	
	\end{equation}
	with $F \coloneqq \fidelity(\rhot,\rhop) \in [0,1]$.
\end{proof}

We have now the ingredients to simply invoke the median of means estimator (Theorem~\ref{thm:MME}) as the final step. 

\begin{proof}[Proof of the Theorem~\ref{thm:SFE}]
	By Theorem~\ref{thm:MME} and the assumptions of Theorem~\ref{thm:SFE} we have for the median of mean estimator $\hat \mu$ 
	with confidence $1-\delta$ 
	\begin{equation}
		|\hat \mu - \mu| \leq \sigma\sqrt{\frac{32 \ln(1/\delta)}\np}\, , 
	\end{equation}
	where $\mu = \EE[\hat f_i]$ and $\sigma^2 = \Var[\hat f_i]$. 
	Now, \eqref{eq:SFE:Efi} implies that $\mu = \fidelity(\rhot, \rhop)$ and by \eqref{eq:SFE:varbound} we have $\sigma < \sqrt 5$. 
	Requiring $|\hat \mu - \fidelity(\rhot, \rhop)| \leq \epsilon$ and solving the right-hand sides leads to the sufficient condition $\np \geq 160\, \epsilon^{-2} \ln (1 /\delta)$.
\end{proof}

\subsubsection*{Further reading}
Shadow fidelity estimation builds on the idea of extracting an incomplete description 
of a quantum state in order to subsequently estimate its properties. 
For such an incomplete description that correctly predicts the expectation of a set of observables 
Aaronson coined the term `shadow' in Ref.~\cite{Aaronson2018ShadowTomography}. 
The broader framework for shadow estimation developed by Huang \emph{et al.}~\cite{Huang2019PredictingFeatures, Huang2020Predicting} 
allows the sampling complexity of different measurement frames to be derived and is also not restricted to estimating 
fidelities. 
See also Paini and Kalev~\cite{PainiKalev:2019} for a parallel work analyzing the sampling complexity of estimating 
expectation values of observables from measurement frames that are generated using a group. 
Finally, we note that the linear cross-entropy benchmarking protocol \cite{BoiIsaSme16} presented in Section~\ref{sec:xeb} 
similarly to \ac{SFE} exploits a unitary $3$-design as the measurement frame to achieve a sampling complexity scaling independently of the system size, as explicitly worked out by Helsen~\emph{et al.} \cite{Helsen20AGeneralFramework}.

\section{Quantum processes}
\label{sec:processes}
In the first part of this tutorial we presented different approaches to certify quantum states. 
For the second part we now turn our attention to the certification of quantum processes, i.e.\ maps on quantum states. 

As quantum technologies typically involve processing quantum states, 
the task of their certification is omnipresent. 
For example in quantum computing, processes of interest might be individual quantum gates, entire algorithms or a noise process that accounts for the deviation from the ideal functioning of a device.

Many of the methods developed for quantum states can be employed to derive analogous results for quantum processes. 
In principle, we can always arrive at a certificate for a quantum process by certifying its output states on a suitably large set of input states. 
Similarly, maximally entangling the input of a quantum process with ancillary quantum systems allows one to operationally prepare a quantum state representing the quantum process via the so-called Choi-\Jamiolkowski{} isomorphism.

After reviewing the mathematical formalism for describing quantum processes and discussing several measures of quality, we briefly discuss examples of translating methods for direct state certification to quantum processes. 

These approaches come with potentially severe drawbacks concerning the feasibility of the measurements. 
The characterization of a quantum process always involves the preparation of input states and measurements on the output of the process. 
In this task so-called \ac{SPAM} errors can be a serious obstacle for a reliable characterization. 
This has motivated the development of quantum characterization and verification methods that are robust against such \ac{SPAM} errors to quite some extent. 
One way to achieve this robustness are \emph{self-consistent approaches} that aim at simultaneously characterizing quantum processes, the state-preparation and the measurement \cite{MerGamSmo13, blume2013robust, CerfontaineEtAl:2019}. 
These methods however require extensive effort in terms of the number of measurement settings, sampling complexity and classical post-processing, and deliver far more information than required for certification. 

An important class of certification methods in the context of digital quantum computing are \emph{randomized benchmarking} protocols \cite{EmeAliZyc05, LevLopEme07, DanCleEme09}. 
\Acf{RB} protocols extract performance measures for quantum gates by 
implementing random gate sequences of different lengths and measuring the error that accumulates with the sequence length. 
By studying the error dependence in the sequence length randomized benchmarking protocols 
are robust against \ac{SPAM} errors. 
We present two prototypical types of \ac{RB} protocols targeting performance measures of a gate set and of individual gates 
together with the theoretical analysis in the simplest setup 
 in Section~\ref{sec:RB}. 

Finally, in Section~\ref{sec:xeb} we turn our attention to a method that is used in order to 
certify the correct implementation of a quantum circuit in the context of demonstrating so-called quantum supremacy\footnote{We use the term `quantum (computational) supremacy' strictly in its established technical meaning \cite{preskill2013quantum}.}: 
\emph{cross-entropy benchmarking} \cite{BoiIsaSme16}.

\subsection{Quantum processes and measures of quality}\label{sec:quantum_processes_and_measures}
A \emph{quantum process} should model possible operations taking quantum states to quantum states. 
Mathematically, a \emph{quantum process} is, thus, a linear map taking density operators to density operators with suitable properties. 
Therefore, we start with introducing some notation related to linear maps between operator spaces. 

In the following, let $\H, \K$ be finite-dimensional Hilbert spaces. 
	The vector space of linear maps from $\L(\H)$ to $\L(\K)$ is denoted by 
	$\M(\H,\K) \coloneqq \L(\L(\H), \L(\K))$. 
	We set $\M(\H)\coloneqq \M(\H,\H)$ and denote the identity by $\id_\H \coloneqq \1_{\L(\H)} \in \M(\H)$. 
	Often we just write $\id$ when it is clear from the context what $\H$ is. 
	A map $\Phi\in \M(\H,\K)$ is called \emph{Hermicity-preserving} if 
	\begin{equation}
		\Phi(\Herm(\H))\subset \Herm(\K)\, ,
	\end{equation}
	\emph{positive} if 
	\begin{equation}
		\Phi(\PSD(\H))\subset \PSD(\K)\, ,
	\end{equation}
	and
	\emph{trace-preserving} if 
	\begin{equation}
		\Tr[\Phi(X)] = \Tr[X]
	\end{equation}
	for all $X \in \L(\H)$. 
	Note that positive maps are also Hermicity-preserving. 

	The map $\Phi$ is called \emph{\ac{CP}} if 
	$\Phi\otimes \id_{\H'}$ is positive for all Hilbert spaces $\H'$ with identity map $\id_{\H'}\in \M(\H')$. 
	The set of \ac{CP} maps is denoted by $\CP(\H,\K) \subset \M(\H,\K)$ and forms a convex cone. 
	We set $\CP(\H)\coloneqq \CP(\H,\H)$. 
	A \ac{CPT} map is also called a \emph{quantum channel} or just \emph{channel}. 
	The subset of \ac{CPT} maps is denoted by $\CPT(\H,\K) \subset \CP(\H,\K)$ and forms a convex set. 
	Again, we set $\CPT(\H)\coloneqq \CPT(\H,\H)$. 

	Lastly, a map $\Phi \in \M(\H, \K)$ is called \emph{unital} if $\Phi(\1_\H) = \1_\K$. 
	Note that $\Phi$ is trace-preserving if and only if its adjoint (w.r.t.\ the Hilbert-Schmidt inner product) $\Phi^\ad$ is unital. 

So, essentially, quantum channels are maps that take density matrices to density matrices even when applied to a part of a larger system. 
Usual unitary dynamics is of the following form. 

\begin{block}{Example (Unitary channels):}
We use calligraphic letters to denote the adjoint representation $\mcU \in \M(\H)$ of a unitary $U\in \U(\H)$ given by
\begin{equation}
	\mc U(X) \coloneqq U X U^\ad \, . 
\end{equation}
These maps are quantum channels and are called \emph{unitary (quantum) channels}. 
\end{block}

Unitary channels are invertible and the inverses are again unitary channels. 

\subsubsection{The Choi-\Jamiolkowski{} isomorphism}
The Choi-{\Jamiolkowski} isomorphism 
\cite{Jam72,Cho75} provides a duality between \ac{CP} maps and bipartite positive semidefinite operators and allows the identification of channels with certain states. 
It has many applications in quantum information theory and related fields. 
In particular, it gives a practical criterion to check whether a given map is a quantum channel. 
Furthermore, it allows us to derive certification methods for quantum processes from the already presented methods for quantum states. 

For any vector space $V$, recall that there is the canonical isomorphism 
\begin{equation}\label{eq:VVast}
	\L(V) = V\otimes V^\ast \, ,
\end{equation}
where $V^\ast \coloneqq \L(V,\C)$ is the dual space of $V$. 
Furthermore, if $V$ is equipped with an inner product $\braket \argdot\argdot$, we have 
the canonical isomorphism $v \mapsto (w \mapsto \braket vw)$ identifying $V \stackrel{\operatorname{hc}}{\cong} V^\ast$. 
For linear maps on linear maps $\M(\H,\K)$ this simple isomorphism induces a couple of identifications with other vector spaces. 

The \emph{Choi-Jamio{\l}kowski isomorphism} 
\begin{equation}
 	\choi: \M(\H,\K) \to \L(\K\otimes \H)
\end{equation} 
is one of these isomorphisms of vector spaces 
given by the following sequence of simple identifications:
\begin{equation}\label{eq:CJvia_isomorphisms}
\begin{aligned}
	\M(\H,\K) 
	&=	
	\L(\K)\otimes \L(\H)^\ast 
	=
	\K\otimes \K^\ast \otimes \H^\ast \otimes \H
	\\
	&\cong
	\K\otimes \H^\ast \otimes \K^\ast \otimes \H
	= 
	\L(\K\otimes \H^\ast)
	\\
	&\stackrel{\operatorname{hc}}{\cong} 
	\L(\K\otimes \H) \, ,
\end{aligned}
\end{equation}
where the natural isomorphism \eqref{eq:VVast} is denoted by ``$=$'', 
the isomorphism of changing the order of the vector spaces by ``$\cong$'', 
and identification marked by ``hc'' makes use of the Hilbert space isomorphism $\H \cong \H^\ast$. 

More explicitly, the Choi-Jamio{\l}kowski isomorphism can be written in the following way. 
Let $(\ket i)_{i \in [\dim(\H)]}$ be a basis of $\H$ and 
\begin{equation}
	\ket \1 = \sum_{i=1}^{\dim(\H)} \ket{i,i} \ \in \H\otimes \H
\end{equation}
the unnormalized maximally entangled state.
The \emph{Choi matrix} of $\X\in \M(\H,\K)$ is given as
\begin{equation}\label{eq:CJvia_max_ent_state}
	\choi(\X) = \X\otimes \id (\ketbra\1\1) \, .
\end{equation}

\ex{
\begin{exerciseblock}[{fonttitle=\normalfont}]{Choi-Jamio{\l}kowski iso\-mor\-phism}
Show that the characterizations of Choi-Jamio{\l}kowski isomorphism from \eqref{eq:CJvia_max_ent_state} and \eqref{eq:CJvia_isomorphisms} coincide. If you are familiar with tensor network diagrams, give a pictorial definition of the
Choi-Jamio{\l}kowski isomorphism. 
Moreover, show that 
\begin{equation}\label{eq:choi_expectation}
	\Tr[B \X(A)	] = \Tr[ (B \otimes A^\T) \choi(\X)]
\end{equation}
for all $\X\in \M(\H,\K)$, $A \in \L(\H)$ and $B \in \L(\K)$. 
\end{exerciseblock}
}%
\exreplace{
This characterization of $\choi(\X)$ implies 
\begin{equation}\ifex\else\label{eq:choi_expectation}\fi
	\Tr[B \X(A)	] = \Tr[ (B \otimes A^\T) \choi(\X)]
\end{equation}
for all $\X\in \M(\H,\K)$, $A \in \L(\H)$ and $B \in \L(\K)$, 
as can be seen by direct calculations with basis elements or tensor network diagrams. 
}

Now we can connect the Choi-\Jamiolkowski{} isomorphism to the properties of quantum channels. 

\begin{theoremblock}[label={thm:CPT_conditions}]{\Ac{CPT} conditions}
For any map $\X \in \M(\H,\K)$ the following equivalences hold:
\begin{enumerate}[label=(\roman*)]
	\item $\X$ is trace-preserving if and only if $\Tr_\K[\choi(\X)] = \1$. 
	
	\item $\X$ is Hermicity-preserving if and only if $\choi(\X)$ is Hermitian. 

	\item\label{item:CPTconditions:choi} $\X$ is completely positive if and only if $\choi(\X)$ is positive semidefinite. 
\end{enumerate}
\end{theoremblock}

\begin{proof}
As an exercise or see, e.g., \cite[Chapter~2.2]{Wat18}. 
\end{proof}

For completeness, we remark that another important consequence of the complete positivity of a map is the existence of so-called Kraus operators. 
This gives another item that could be added to Theorem~\ref{thm:CPT_conditions}:
$\X$ is a \ac{CP} map if and only if there are (Kraus) operators $K_1, \dots, K_{r} \in \L(\H,\K)$, 
where $r = \rank(\choi(\X))$ so that 
\begin{equation}\label{eq:Kraus}
  \X(A) = \sum_{i=1}^r K_i A K_i^\ad
\end{equation}
for all $A\in \L(\H)$. 
Moreover, $\X$ is a \ac{CPT} map if and only if \eqref{eq:Kraus} holds with $\sum_{i=1}^r K_i^\ad K_i = \1$. 

In the context of quantum information theory, another normalization convention for the Choi-\Jamiolkowski{} isomorphism is useful. 
For $\X\in \M(\H, \K)$ we set 
\begin{equation}\label{eq:jam_choi}
	\jam(\X) \coloneqq \frac{1}{\dim(\H)} \choi(\X)  
\end{equation}
with Choi matrix \eqref{eq:CJvia_max_ent_state}. 
The theorem tells us that $\X$ is a quantum channel if and only if $\jam(\X)$ is a density matrix with the reduction to $\H$ (obtained by tracing over $\K$) being a maximally mixed state. 
The so-called \emph{Choi state} of a channel $\X$ is 
\begin{equation}\label{eq:jam_def}
	\jam(\X) = \X\otimes \id_\H(\phi^+) \ \in \DM(\K\otimes \H) \, ,
\end{equation}
where 
\begin{equation}\label{eq:Phi_plus}
	\phi^+ \coloneqq \frac{1}{\dim(\H)} \ketbra\1\1 \quad  \in \DM(\H\otimes\H)
\end{equation}
is a \emph{maximally entangled state}, i.e., 
has the strongest bipartite quantum correlations possible in a precise sense. 
In particular, the Choi state can be prepared by applying the channel to this state. 

Note that not every bipartite state corresponds to a channel. 
Indeed, the Choi-\Jamiolkowski{} isomorphism is an isomorphism of convex cones, $\choi: \CP(\H,\K) \to \PSD(\K\otimes \H)$ but $\CPT(\H,\K)$ is mapped to a proper subset of $\DM(\K\otimes \H)$. 
The reason is that the trace-preservation constraint of channels corresponds to $\dim(\H)^2$ many equalities whereas the trace constraint of states is just one equality. 

An important quantum channel and frequent model for noise processes appearing in quantum technologies is the 
 depolarizing channel.
The \emph{(quantum) depolarizing channel} $\depol : \L(\CC^d) \to \L(\CC^d)$ with parameter $p \in [0,1]$ is the linear map defined by 
\begin{equation} \label{eq:def:depolChannel}
	\depol(X) \coloneqq p X + (1-p) \Tr[X] \frac{\1}{d}  \, .
\end{equation}

\ex{
\begin{exerciseblock}{Depolarizing channel}
From the definition of the depolarizing channel \eqref{eq:def:depolChannel} it is clear that $\depol \in \M(\CC^d)$. 
Show that $\depol \in \CPT(\CC^d)$ if and only if
\begin{equation*}
	-\frac{1}{d+1} \leq p \leq 1 \, .
\end{equation*}
For which of those values of $p$ is $\depol$ also invertible and when is the inverse also a channel? 
\end{exerciseblock}
}

\subsubsection{Inner products of superoperators and fidelity measures}
\label{sec:qp_fidelity}
The vector space of linear maps $\M(\H, \K)$ is also equipped with a canonical inner product 
(the Hilbert-Schmidt inner product for superoperators) 
given by
\begin{equation}
	\langle \X, \Y \rangle 
	= 
	\Tr[\X^\ad \Y]
\end{equation}
for any $\X,\Y\in \M(\H, \K)$, where the trace can be calculated using an orthonormal basis $\{E_0, E_1, \ldots, E_{d^2 -1}\}$ of $\L(\H)$ as 
\begin{equation}\label{eq:trace_channels}
	\Tr[\X] 
	= 
	\sum_{i = 0}^{d^2-1} \langle E_i , \X(E_i) \rangle 
	=
	\sum_{i = 0}^{d^2-1} \Tr[ E_i^\ad \X(E_i) ]\, .
\end{equation}
The Hilbert-Schmidt inner product on $\M(\H, \K)$ coincides with the inner product of the corresponding Choi matrices, i.e., for any $\X,\Y\in \M(\H, \K)$
\begin{equation}
	\langle \X, \Y \rangle 
	= 
	\langle \choi(\X), \choi(\Y) \rangle.
\end{equation}

We now consider the case where $\Y$ is a quantum channel and $\X$ a unitary quantum channel. 
 Then, as we see above, $\jam(\Y)$ and $\jam(\X)$ are quantum states (density matrices). 
 Moreover, $\jam(\X)$ is a pure state. 
In this case, the above Hilbert-Schmidt inner product with the proper normalization is the fidelity measure induced by the state fidelity \eqref{eq:pure_state_fidelity} via the Choi-\Jamiolkowski{} isomorphism \eqref{eq:CJvia_max_ent_state}, 
\begin{equation}\label{eq:def:EntanglementGateFidelity}
	\efidelity(\X, \Y) \coloneqq \fidelity(\jam(\X),\jam(\Y)) =\frac{1}{\dim(\H)^2} \langle \X, \Y \rangle \, ;
\end{equation}
it is referred to as the \emph{entanglement (gate) fidelity}. 

In the context of digital quantum computing, another very prominent fidelity measure for quantum processes is following. 
The \emph{average gate fidelity} (AGF) between maps $\X,\Y\in \M(\H, \K)$ is defined as
\begin{equation}\label{eq:def:AGV:HSdef}
\agf(\X, \Y) 
	\coloneqq 
	\int_{\sphereCd} \langle \X(\ketbra\psi\psi), \Y(\ketbra\psi\psi) \rangle\,  \rmd \mu_{\sphereCd}(\psi)
	\, , 
\end{equation}
where the integral is taken according to the uniform Haar-invariant probability measure on state vectors of 
Section~\ref{sec:haar-random-states}. 
Note that the inner product here is the Hilbert-Schmidt inner product of $\L(\K)$ not $\M(\H, \K)$. 
From the definition we see that the average gate fidelity $\agf(\X, \Y)$ is a measure of closeness of $\X$ and $\Y$ that compares the action of $\X$ and $\Y$ on pure input states on average. 
Intuitively, if $\X$ and $\Y$ deviate only in their action on a low-dimensional subspace of $\H$ they can still have an average gate fidelity close to $1$. 

	For any $\X,\Y\in \M(\H,\K)$ 
	\begin{equation}\label{eq:agf:covariance}
		\agf(\X, \Y)  = \agf(\id, \X^\ad \circ \Y) \, .
	\end{equation}
	This motivates the definition 
	$
	\agf(\X) \coloneqq \agf(\id,\X)
	$ for $\X\in \M(\H)$. 

The average gate fidelity is intricately related to the Hilbert-Schmidt inner product on $\M(\H, \K)$ \cite{HorHorHor99,Niel02} (see also Ref.~\cite{KueLonFla15}). 

\begin{propositionblock}[label={prop:AGFasOverlap}]{Inner product and $\agf$}
	For $\X, \Y \in \M(\H, \K)$ with $d = \dim(\H)$ it holds that 
	\begin{equation}\label{eq:AGF_as_inner_product}
		\langle \X , \Y \rangle
		= d(d+1) \agf(\X,\Y) - \langle \X(\1), \Y(\1) \rangle \, .
\end{equation}
\end{propositionblock}
\begin{proof}
	By the virtue of \eqref{eq:agf:covariance} which also holds for the inner products appearing in \eqref{eq:AGF_as_inner_product} it suffices to prove the statement for $\X = \id$. 
	Using \eqref{eq:choi_expectation} and denoting the transposition map as $T: \L(\H) \to \L(\H)$, $A \mapsto A^\T$, we can rewrite the average gate fidelity as
	\begin{equation}
	\begin{split}
		\agf&(\id, \Y) = \int_{\sphereCd} \langle \ketbra\psi\psi, \Y(\ketbra\psi\psi) \rangle\,  
			\rmd \mu_{\sphereCd}(\psi) \\
		&= \int_{\sphere^{d-1}} \Tr\left[ 
			\ketbra\psi\psi  \Y(\ketbra\psi\psi) 
		\right]
		\rmd \mu_{\sphereCd}(\psi) \\
		&= \int_{\sphereCd} 
			\Tr\left[ 
				\id\otimes T\left({\ketbra\psi\psi}^{\otimes 2}\right) \choi(\Y)
			\right]
		\rmd \mu_{\sphereCd}(\psi).
	\end{split}
	\end{equation}
	Due to linearity, we can recast this expression with the moment operator $K^{(k)}_{\mu_{\sphereCd}}$ of random states 
	and use the expression we derive in Lemma~\ref{lem:moments_rand_states}. Then, 
	\begin{equation}\label{eq:process:innerproductFavg}
	\begin{split}
		\agf&(\id, \Y) 
		= 
			\Tr\left[ 
				\id\otimes T \left(K^{(2)}_{\mu_{\sphereCd}}\right) \choi(\Y)
			\right] \\
		&= \frac{2}{d(d + 1)} \Tr\left[ \id\otimes T(P_{\sym^2}) \choi(\Y)\right] \\
		&= \frac{1}{d(d + 1)}\left(\Tr\left[ \1 \choi(\Y) \right] + \Tr\left[\ketbra \1\1 \choi(\Y)\right] \right),
	\end{split}
	\end{equation}
	where the last step follows from $P_{\sym^2} = \tfrac 12(\1+\FF)$ with the swap operator $\FF$ from \eqref{eq:SwapOp} 
	and $\id \otimes T (\FF) = \ketbra\1\1$. 
	Using \eqref{eq:choi_expectation} this time the other way around,  we see that the first summand of \eqref{eq:process:innerproductFavg} 
	is $\Tr[ \1 \choi(\Y) ]  = \Tr[ \1 \otimes \1 \choi(\Y) ] = \Tr[ \Y(\1)] = \langle \id(\1), \Y(\1) \rangle$. 
	From \eqref{eq:CJvia_max_ent_state} it directly follows that $\choi(\id) = \ketbra\1\1$. 
	Hence, the second term of \eqref{eq:process:innerproductFavg} is 
	$
	\Tr\left[\ketbra \1\1 \choi(\Y)\right]%
	= \Tr\left[\choi(\id)\choi(\Y)\right]%
	= \langle \choi(\id), \choi(\Y) \rangle%
	= \langle \id, \Y \rangle
	$.
	Plugging these two expressions into \eqref{eq:process:innerproductFavg} and solving for $\langle \id, \Y \rangle$ yields the assertion of the proposition. 
\end{proof}

Proposition~\ref{prop:AGFasOverlap} implies that the average gate fidelity is an inner product, i.e., a conjugate symmetric non-degenerate form that is linear in its second argument. 
	For Hermicity-preserving $\X$ and $\Y$ the average gate fidelity is real, $\agf(\X,\Y) \in \RR$.  
	Thus, on Hermicity-preserving maps it is \emph{symmetric},
	\begin{equation}
		\agf(\X, \Y) = \agf(\Y, \X).
	\end{equation}

	Associate to the average gate fidelity is
	the \emph{average error rate} or \emph{average infidelity},
	\begin{equation}
		\aer(\X,\Y) 
		\coloneqq 
		1- \agf(\X,\Y)\,  
	\end{equation}
	that is also real-valued for Hermicity-preserving maps.
	We set $\aer(\X) \coloneqq 1-\agf(\X)$. 
	For unital, completely positive $\X$, the average infidelity can be regarded as a distance to other quantum channels in the following sense:

\begin{lemmablock}{Infidelity as distance measure}
	Let $\X \in \CP(\H, \K)$ be unital. For all $\Y \in \CPT(\H, \K)$ it holds that 
	$\aer(\X, \Y) \geq 0$ and, $\aer(\X, \Y) = 0$ if and only if $\X = \Y$. 
\end{lemmablock}

\begin{proof}
	Using Proposition~\ref{prop:AGFasOverlap}, we have 
	$\agf(\Y) = \frac1{d(d+1)}\,\langle \id , \Y \rangle + \frac1{d+1}$. 
	The overlap of the two \ac{CP} maps can be bounded via the Cauchy-Schwarz inequality as 
	$\langle \id , \Y \rangle \leq \fnorm{\id} \fnorm{\Y}$ with equality if and only if $\Y = \id$. 
	For $\Y\in \CPT(\H)$ it holds that $\fnorm{\Y}^2 \leq d^2$ and 
	$\fnorm{\id}^2 = d^2$. This can be seen, e.g., from the basis expansion \eqref{eq:trace_channels} by choosing a unit-rank basis and applying the Hölder inequality \eqref{eq:matrix_Hoelder}. 
	Therefore, $\langle \id , \Y \rangle \leq d^2$.
	We conclude that $\agf(\Y) \leq 1$ again with equality if and only if $\Y = \id$ which implies the assertion.
\end{proof}

If $\X^\ad\Y$ is trace-preserving, \eqref{eq:AGF_as_inner_product} simplifies to
\begin{equation} \label{eq:tr(agf)}
	\langle \X, \Y \rangle 
	= 
	 d(d+1) \agf(\X, \Y) - d
	\, ,
\end{equation}
or, equivalently, 
\begin{equation} \label{eq:agf(tr)}
	\agf(\X, \Y)
	= 
	\frac{\langle \X, \Y \rangle + d}{d(d+1)}\,.
\end{equation}
We conclude that for trace-preserving and unital quantum channels the average gate fidelity and the Hilbert-Schmidt inner product are affinely related with a proportionality constant in $\LandauO(d^{-2})$. 
This is the same scaling as appearing for the entanglement fidelity in \eqref{eq:def:EntanglementGateFidelity}. 
More precisely, we find the affine relation between the two fidelities
\begin{equation}\label{eq:agfandFe}
		\agf(\X, \Y) = \frac{d \, \efidelity(\X,\Y) + 1}{d + 1}\, ,
\end{equation}
still assuming $\X^\ad\Y$ being trace-preserving and one of $\X$ and $\Y$ being a unitary channel.
For two unitary channels $\mc U, \mc V \in \CPT(\H)$ with $U, V\in \U(d)$ we can further simplify \eqref{eq:agf(tr)} to
\begin{equation}\label{eq:AGF_U}
	\agf(\mc V, \mc U)
	= 
	\frac{|\Tr[V^\ad U]\, |^2-d}{d(d+1)}\, .
\end{equation}
For $V = \1$ this equality reflects that the average gate fidelity measures how close $U$ is to $\1$ on average where the average is taken over its spectrum. 

Furthermore, the identity \eqref{eq:AGF_as_inner_product} also connects the average gate fidelity to the Frobenius norm.
This, in turn, shows that the Frobenius norm is an average case error measure as well. 

Lastly, beside the entanglement fidelity, the Hilbert-Schmidt inner-product, and the average gate fidelity, there is another 
affinely related measure of quality that is particularly convenient to work with in the analysis of randomized benchmarking:
the effective depolarizing parameter. 
Here, we define the effective depolarizing parameter only for trace-preserving maps via
its linear relation to the fidelity. 
If $\X$ is not trace-preserving one can more generally define it by explicitly first projecting on unital maps. 
Let $\X \in \M(\H, \K)$ be trace-preserving, its \emph{effective depolarizing parameter} is 
\begin{equation}\label{eq:effectiveP}
	p(\X) \coloneqq \frac{d \agf(\X) - 1}{d - 1}\, . 
\end{equation}

To justify its name let us have a look at the depolarizing channel $\depol[p]$, which is defined in \eqref{eq:def:depolChannel} as the convex combination of $\depol[1]=\id$ and $\depol[0]$. 
The average gate fidelity of these extremal channel can be quickly calculated to be $\agf(\id) = 1$ and $\agf(\depol[0]) = \frac1d$. 
Thus, $\agf(\depol[p]) = p + \frac{1-p}d$. Plugging this into the definition of the effective depolarizing parameter \eqref{eq:effectiveP} yields
\begin{equation}\label{eq:effective_depol_parameter}
	p(\depol[p]) = p.
\end{equation}

Another affinely related measure that is often used in this context is the $\chi_{0,0}$-entry of the so-called $\chi$-process matrix, see e.g.\ Ref.\ \cite{Carignan-Dugas2019BoundingTheAverage} for further details.

\subsubsection{The diamond norm}
\label{sec:diamond_norm}
The distance measures on quantum channels we encounter so far can be regarded as average error measures. 
A more pessimistic, worst-case error measure is induced by the trace-norm on operators, the so-called \emph{diamond norm}. 
It measures the operational distinguishability of quantum channels. 
Hence, it plays an important role in the certification of quantum processes. 
Indeed, also error-correction thresholds require worst-case guarantees without additional assumption on the error model, see e.g.\ the discussion Refs.~\cite{Sanders2015,KueLonFla15}. 
At the same time, certification schemes that directly deliver certificates in diamond norm are very resource intensive and typically practically infeasible. 
For this reason, the connection of the diamond norm to the already introduced average error measures is the focus of this section. 

We start with defining the \emph{($1\to 1$)-norm} on $\M(\H,\K)$ to be the operator norm induced by the trace norm,
\begin{equation}
	\norm{\X}_{1\to 1} 
	\coloneqq 
	\sup_{\tnorm{A} \leq 1} \tnorm{\X(A)} \, .
\end{equation}
Note that since the trace norm is a convex function, we have for any $\X \in \M(\H,\K)$ 
\begin{equation}\label{eq:dnorm_as_sup_unit_rank}
	\norm{\X}_{1\to 1} 
	= 
	\sup_{\substack{\lpnorm[2]{\ket\psi}=1,\\ \lpnorm[2]{\ket\phi}=1}} \bigl\{ \tnorm{\X (\ketbra\psi\phi)} \bigr\} \, ,
\end{equation}
i.e., the supremum is attained for rank-$1$ operators $\ketbra \psi\phi\in\L(\H)$. 

In order to operationally distinguish two quantum channels one can use ancillary systems and entangled states. 
This motivates the definition of the diamond norm as the so-called \emph{complete boundedness (CB)-completion} of the $(1\to 1)$-norm. 
We define the \emph{diamond norm} of $\X \in \M(\H)$ by 
\begin{equation}\label{eq:def:dnorm}
	\dnorm{\X} 
	\coloneqq 
	\norm{\X\otimes \id_{\H}}_{1\to 1}\, .
\end{equation}
Note that this norm inherits the property \eqref{eq:dnorm_as_sup_unit_rank} from the $(1\to 1)$-norm. 
For the relevant case where $\X \in \M(\H)$ is Hermitian-preserving we even have \cite[Theorem~3.51]{Wat18}
\begin{equation*}
	\dnorm{\X}
	= 
	\sup_{\lpnorm[2]{\ket\psi}=1} 
		\bigl\{ \tnorm{\X\otimes \id (\ketbra\psi\psi)} \bigr\} \, , 
\end{equation*}
i.e., the supremum is attained at a pure density operator $\ketbra \psi\psi\in\DM(\H\otimes\H)$. 
If $\X$ is additionally trace-preserving then its output 
$\X\otimes \id (\ketbra\psi\psi)$ 
is also a density operator. 
Hence, quantum channels are normalized in diamond norm: 
\begin{equation}
	\dnorm{\X} = 1 \qquad \forall \X \in \CPT(\H,\K) \, . 
\end{equation}

Moreover, the following theorem guarantees some further basic properties of the diamond norm. 
It has good stability properties concerning composition of Hilbert spaces, is multiplicative under tensor products and submultiplicative under products. 

\begin{theoremblock}[label={thm:dnorm}]{Complete boundedness and (sub)multiplicativity}
For any $\X \in \M(\H,\K)$ 
\begin{equation}\label{eq:dnorm:CB}
	\dnorm{\X} 
	= 
	\sup_{\H'} \norm{\X\otimes \id_{\H'}}_{1\to 1}\, , 
\end{equation}
where the supremum is taken over all finite-dimensional Hilbert spaces $\H'$. 
Moreover, 
\begin{align}
\dnorm{\X \otimes \Y} &= \dnorm{\X}\dnorm{\Y} 
\\	
\dnorm{\X\mc Z} &\leq \dnorm{\X} \dnorm{\mc Z}
\end{align}
for all $\X \in \M(\H,\K)$, $\Y \in \M(\H',\K')$ and $\mc Z \in \M(\H',\H)$. 
\end{theoremblock} 

\begin{proof}
For the proof we refer, e.g., to \cite[Chapter~3.3]{Wat18} or recommend proving it as an exercise. 
\end{proof}

Theorem~\ref{thm:dnorm} tells us that the diamond norm 
precisely captures the maximum distinguishability of quantum channels 
$\X,\Y \in \CPT(\H,\K)$
in the following sense. 
One can prepare copies of a state $\rho \in \DM(\H\otimes \H')$ and apply either $\X$ or $\Y$ to the parts on $\H$ to obtain states on $\K\otimes \H'$. 
Then Proposition~\ref{prop:tnorm_operational} tells us that 
$\frac 12 \tnorm{\Phi\otimes \id_{\H'}(\rho)}$ is the distinguishability of the output states. 
Taking the supremum over all (pure) states $\rho$ yields the distinguishability of $\X$ and $\Y$, which is given by the \emph{diamond distance} 
$\frac 12 \dnorm{\X-\Y}$. 
In particular, the theorem tells us that optimal distinguishability can be obtained by choosing $\H'=\H$ in a similar sense as it can be detected when a map is not \ac{CP} just using $\H'=\H$, cp.\ Theorem \ref{thm:CPT_conditions}\ref{item:CPTconditions:choi}. 

Another way to distinguish quantum processes is to prepare their Choi states and distinguish them, as characterized by Proposition~\ref{prop:tnorm_operational} via the trace norm. 
The following statements provide a relation of the two notions of distinguishability of quantum channels. 

\begin{propositionblock}[{label={prop:dnorm_tnorm},fonttitle=\normalfont}]{Diamond norm and trace norm}
For any map $\X \in \M(\H,\K)$ 
\begin{equation}
	\tnorm{\jam(\X)} 
	\leq
	\dnorm{\X}
	\leq 
	\dim(\H) \tnorm{\jam(\X)} \, , 
\end{equation}
where $\jam$ denotes the Choi-\Jamiolkowski{} isomorphism \eqref{eq:jam_def}. 
\end{propositionblock}

The upper bound can be improved. 
For a Hermitian-preserving map $\X\in \M(\H,\K)$ the improved bound implies \cite[Corollary~2]{NecPucPaw18}
\begin{equation}
	\dnorm{\X}\leq \dim(\H) \pnorm[\infty]{\Tr_2[|\jam(\X)|]} \, .
\end{equation}

\begin{proof}[Proof of Proposition~\ref{prop:dnorm_tnorm}]
	We prove the proposition in terms of $\choi(\X) = \dim(\H) \jam(\X)$. 
	Denoting the Frobenius norm again by $\fnorm{\argdot}$, 
	it holds that 
	\begin{equation}
		\dnorm{\X} 
		= 
		\sup_{\substack{A, B \in \L(\H) \\  \fnorm{A} = \fnorm{B}=1}} \bigl\{ \tnorm{(\1\otimes A) \choi(\X) (\1\otimes B)}\bigr\}
		\, ,
	\end{equation}
	as can be seen from \eqref{eq:dnorm_as_sup_unit_rank} and rearranging the contractions. 
	Choosing $A = B = \1/\sqrt{\dim(\H)}$ (corresponding to the maximally entangled state \eqref{eq:Phi_plus}) establishes the lower bound. 
	The upper bound follows using H\"older's inequality \eqref{eq:matrix_Hoelder}, 
	\begin{equation}
	\begin{aligned}
		&\tnorm{(\1\otimes A) \choi(\X) (\1\otimes B)} \\
		&\qquad\leq 
		\snorm{\1 \otimes A} \tnorm{\choi(\X)} \snorm{\1\otimes B}
		\\
		&\qquad=\snorm{\1} \snorm{A} \tnorm{\choi(\X)} \snorm{\1}\snorm{B}
		\\
		&\qquad\leq 
		\fnorm{A} \fnorm{B} \tnorm{\choi(\X)} \, .
	\end{aligned}
	\end{equation}
\end{proof}

\ex{
\begin{exerciseblock}{The diamond norm/trace norm inequalities are tight}
Show that the bounds in Proposition~\ref{prop:dnorm_tnorm} are tight, i.e., that there are $\X,\Y\in \M(\H,\K)$ so that $\tnorm{\jam(\X)} = \dnorm{\X}$ and 
$\dnorm{\Y} = \dim(\H) \tnorm{\jam(\Y)}$. 
\end{exerciseblock}
}
\exreplace{
It is not difficult to see that the bounds in Proposition~\ref{prop:dnorm_tnorm} are tight, i.e., that there are $\X,\Y\in \M(\H,\K)$ so that $\tnorm{\jam(\X)} = \dnorm{\X}$ and 
$\dnorm{\Y} = \dim(\H) \tnorm{\jam(\Y)}$. 
}
These results tell us that distinguishing quantum channels via their Choi states is in general not optimal. 

It is non-obvious how the diamond norm can actually be computed in practice. 
Watrous has shown that the diamond norm can be computed efficiently (in the dimension) via a semi-definite program \cite{Watrous2013}. 
However, for the highly relevant special case where the map is a difference of two unitary channels the computation is much simpler. 

\begin{propositionblock}[{label={prop:diamond_dist_unitaries}}]{Diamond norm distance of unitary channels}
For any $U,V \in \U(d)$ the diamond norm distance of the corresponding unitary channels is
\begin{equation}
	\frac 12 \dnorm{\mcU-\mc V} 
	=
	\sqrt{1-\dist\myleft( 0, \conv\{\lambda_i\}_{i\in [d]} \myright)^2} \, ,
\end{equation}
where $\lambda_i$ are the eigenvalues of $U^\ad V$, 
$\dist(\argdot, \argdot)$ denotes the Euclidean distance and $\conv(\argdot)$ the convex hull, both in the complex plane. 
\end{propositionblock}

This proposition reflects that the diamond distance is a worst-case quantity, where the worst-case optimization is done over the spectrum of the ``unitary difference'' $U^\ad V$. 
The geometric interpretation of this result is reviewed and visualized in Ref.~\cite{Ji2009NonIdentityCheck}. 

In order to prove the proposition we write the matrices $U$ and $V$ as vectors. 
In general, \emph{(column) vectorization} is a map $\vect \argdot: \CC^{n_1\times n_2} \to \CC^{n_1 n_2}$ that stacks the columns of a matrix $A \in \CC^{n_1\times n_2}$ on top of each other. 
For all matrices $A$, $B$ and $C$ with fitting dimensions it holds that 
\begin{equation}\label{eq:matrix_product_vectorization}
	\vect{ABC} = C^\T \otimes A \vect{B}, 
\end{equation} 
where $X\otimes Y \cong (X_{i,j}Y)_{i,j}$ (defined by a block matrix) denotes the \emph{Kronecker product} of matrices $X$ and $Y$. 

\begin{proof}[Proof of Proposition~\ref{prop:diamond_dist_unitaries}]
Starting with \eqref{eq:dnorm_as_sup_unit_rank} and using the Choi-\Jamiolkowski{} isomorphism \eqref{eq:choi_expectation} and the vectorization rules for matrix products \eqref{eq:matrix_product_vectorization},
we can write the diamond norm of the channel difference as  
\begin{align}\nonumber
&\dnorm{\mc U - \mc V } 
\\\nonumber
&\quad=
\sup_{\substack{A \in \L(\CC^d)\\ \pnorm[2]{A} = 1}}
	\myleft\{ \tnorm{ (\1\otimes A) (\ketbra UU - \ketbra VV) (\1\otimes A)}\myright\}
\\\nonumber
&\quad=
 \sup_{\pnorm[2]{A} = 1}\myleft\{ \tnorm{ \ketbra{AU}{AU} - \ketbra{AV}{AV}}\myright\}
\\
&\quad=
\sup_{\pnorm[2]{A} = 1}\myleft\{ \tnorm{ \ketbra{A}{A} - \ketbran{AU^\ad V}{AU^\ad V}} \myright\}\, .
\end{align}

Using \eqref{eq:pnorm_rank1} relating the trace-norm difference of two trace-normalized, hermitian, unit-rank matrices to their overlap yields
\begin{align}\nonumber
\frac 12\dnorm{\mc U - \mc V }
&= 
\sup_{\pnorm[2]{A} = 1 }\myleft\{ \sqrt{1- \left| \braket{A}{AU^\ad V} \right|^2} \myright\}
\\ \nonumber
&= 
\sup_{\pnorm[2]{A} = 1 }\myleft\{ \sqrt{1- \left|\Tr[A^2 U^\ad V]\right|^2}\myright\}
\\ \nonumber
&= 
\sup_{\rho \in \DM(\CC^d)} \myleft\{ \sqrt{1- \left|\Tr[\rho\, U^\ad V]\right|^2} \myright\}
\\ \nonumber
&=
\sqrt{1- \min_{\rho \in \DM(\CC^d)} \left|\Tr[\rho\, U^\ad V]\right|^2}
\\
&=
\sqrt{1- \min_{\rho \in \DM(\CC^d)} \left|\Tr[\rho \diag(\lambda)]\right|^2}\, , 
\label{eq:dnormUV}
\end{align}
where $\diag(\lambda)$ is a diagonal matrix with the same eigenvalues as $U^\ad V$. 
By writing $\rho = W\diag(q)W^\ad$ with a unitary matrix $W \in \U(d)$ and probability vector $q\in [0,1]^d$ we expand the trace in terms of matrix elements as
\begin{equation*}
	\Tr[\rho \diag(\lambda)] 
	= 
	\sum_{i,j=1}^d \lambda_i \left|W_{i,j}\right|^2 q_j \, .
\end{equation*}
Note that $T \coloneqq \bigl( \left|W_{i,j}\right|^2\bigr)_{i,j\in[d]}$ is a doubly stochastic matrix, i.e.\ its sums and columns are all probability vectors. 
Hence, $p\coloneqq Tq$ is a probability vector as well. 
Therefore, the minimization \eqref{eq:dnormUV} is equivalent to a minimization of $|\langle p, \lambda \rangle|^2$ over probability vectors $p$, i.e.,  
\begin{equation*}
\begin{aligned}
\frac 12\dnorm{\mc U - \mc V }
&=
\sqrt{1- \min_{\substack{p \in [0,1]^d \\ \sum_i p_i=1}} \left\{ \Bigl| \sum_i p_i \lambda_i \Bigr|^2  \right\} }
\\
&=
\sqrt{ 1 - \dist\myleft(0,\conv\{\lambda_i\}\myright) } \, .
\end{aligned}
\end{equation*}

\end{proof}

Practical certification schemes for quantum processes will typically certify w.r.t.\ the Hilbert-Schmidt overlap, average gate fidelity or an equivalent quantity. 
In terms of the infidelity $r(\X) = 1 - \agf(\X)$, the diamond norm and the average gate fidelity are in general related by the following inequalities.

\begin{propositionblock}[label={prop:infidelityDiamond}]{Infidelity and diamond norm \cite[Proposition~9]{WalFla14}}
For any $\X \in \CPT(\CC^d)$ it holds that
\begin{equation}
\begin{split}
	\frac{d+1}{d}\, r(\X) 
	\leq 
	\frac 12 \dnorm{\id-\X} 
	\leq 
	\sqrt{d(d+1) r(\X)} \, .
\end{split}
\end{equation}
\end{propositionblock}

\begin{proof}
The proof 
combines Proposition~\ref{prop:dnorm_tnorm} 
with the Fuchs-van-de-Graaf inequality \eqref{eq:Fuchs-van-de-Graaf}. 
Latter yields
\begin{equation}\label{eq:choiFuchsVanDeGraaf}
\begin{split}
	&1 - \fidelity( \jam (\id), \jam(\X)) \\
	&\qquad\leq \frac12\| \jam(\id) - \jam(\X) \|_1 \leq \sqrt{ 1 - \fidelity( \jam (\id), \jam(\X))} \, ,
\end{split}
\end{equation}
where we already drop a square-root on the lower bound.

Since $\jam(\id) = \frac1d \ketbra\1\1$ is of unit rank and Hermitian, it holds that 
$\fidelity(\jam (\id), \jam(\X)) = \langle \jam(\id), \jam(\X)\rangle = \efidelity(\id, \X)$. 
We can cast this in terms of the average gate fidelity via \eqref{eq:agfandFe}, 
\begin{equation}\label{eq:fidelityFavg}
	\fidelity(\jam (\id), \jam(\X)) =
	\frac{d+1}d \agf(\X) - \frac1d \,.
\end{equation}
Plugging \eqref{eq:fidelityFavg} into \eqref{eq:choiFuchsVanDeGraaf} yields 
\begin{equation}\label{eq:fidjambound}
\begin{split}
	&\frac{d+1}d(1 - \agf(\X)) \\
	&\qquad\leq \frac12\| \jam(\id) - \jam(\X) \|_1 \leq \sqrt{\frac{d+1}d} \sqrt{ 1 - \agf( \X)}.
\end{split}
\end{equation}
Finally, from Proposition~\ref{prop:dnorm_tnorm} 
the proposition's assertion follows.
\end{proof}

Proposition~\ref{prop:infidelityDiamond} leaves us with unsatisfactory state of affairs in two regards: 
first, the upper bound of the diamond norm introduces a dimensional factor $\LandauO(d)$. 
In the context of quantum computing, this leaves us with a potentially large factor scaling exponentially $\LandauO(2^n)$ with the number 
of qubits $n$.
Second, the upper bound scales with the square-root of the infidelity. 
For unitary quantum channels one can in fact tighten the lower-bound to $\sqrt{r(\X)}$ \cite{KueLonFla15}. 
The lower-bound for unitary quantum channels indicates that the square-root scaling is unavoidable in general.
Practically, this means that to certify in diamond norm requires a certificate in infidelity that is orders of magnitude smaller.
Particularly, for small system sizes this can be a key obstacle for the certification of the worst-case performance of quantum processes. 

Fortunately, if a quantum process is highly incoherent, i.e.\ far away from being unitary, one can derive a linear scaling of 
the diamond-norm distance in the infidelity. 
The incoherence can be controlled by the so-called unitarity introduced by Wallman \emph{et al.}\ \cite{WalGraHar15}. 
For $\X \in \M(\H)$ the \emph{unitarity} is defined as
\begin{equation}
	u(\X) = \frac d{d-1} \agf(\X', \X'),
\end{equation}
where $d = \dim \H$ and $\X' \in \M(\H)$ is defined by
\begin{equation}
	\X'(A) = \X(A) - \Tr[\X(A)] \1/\sqrt{d}\, .
\end{equation} 
One can straightforwardly check that $u(\mc U) = 1$ for every unitary channel $\mc U$. 
On the other hand, in Refs.~\cite{WalGraHar15, KueLonFla15} a lower-bound on $u$ in terms of the infidelity $r$ was derived 
for trace-decreasing maps. 
For $\X \in \M(\H)$, $d \coloneqq \dim(\H)$, and $\Tr(\X(\1)) \leq \Tr(\1)$ it holds that 
\begin{equation}\label{eq:def:umin}
	u(\X) \geq u_\text{min} =\left(1 - \frac{d}{d-1}\, r(\X) \right)^2 \, . 
\end{equation}

Kueng \emph{et al.}~\cite{KueLonFla15} established that quantum channels saturating this lower bound exhibit a 
linear scaling of the diamond norm distance in terms of the infidelity. 

\begin{theoremblock}[label={thm:RiKbound}]{Worst-case bound for incoherent channels \cite[Theorem 3]{KueLonFla15}}
Let $\X \in \CPT(\H)$ be unital. 
Then 
$\dnorm{\id - \X} \in \LandauO(r(\X))$ if $u(\X) = u_\text{min} + \LandauO(r^2(\X))$ with $u_\text{min}$ defined in \eqref{eq:def:umin}. 
\end{theoremblock}
The result implies that the infidelity is indeed particularly sensitive to depolarizing noise. 
We leave it with this qualitative statement and refer to Ref.~\cite[Proposition 3]{KueLonFla15} for a quantitative 
statement. See also Ref.~\cite{Wal15}. 

\subsection{\protchaptitle{Direct quantum process certification}}
\label{sec:DQPC}
We saw in Section~\ref{sec:StateCertification}, that
quantum states can be certified with measurement strategies resembling the optimal POVM $P^+$ for distinguishing quantum states of Proposition~\ref{prop:tnorm_operational}. 
By means of the Choi-\Jamiolkowski{} isomorphism strategies for quantum states can be lifted to quantum processes: 
operationally, one prepares the Choi state \eqref{eq:CJvia_max_ent_state} by applying the process to a state that is maximally entangled with an ancillary system. Then one certifies the Choi state using a protocol for quantum states. 
The resulting process certification protocols certifies with respect to the entanglement gate fidelity \eqref{eq:def:EntanglementGateFidelity}, which coincides with the state fidelity of the Choi states.  
Refs.~\cite{Liu2020EfficientVerificationOf,Zhu2020EfficientVerificationOf,Zeng2020QuantumGateVerification} use the direct state certification method of Section~\ref{sec:StateCertification} 
\cite{PalLinMon18,ZhuHay18} in 
this way.

Moreover, for certain measurement strategies the protocol can be performed without 
using entanglement with ancillary systems. 
These, \emph{prepare-and-measure versions} use an effective measurement strategy $\Omega$ of the form \cite{Liu2020EfficientVerificationOf}
\begin{equation}
	\Omega = \sum_{i} p_i N_i\otimes \rho_i^\T\, .
\end{equation}
For this measurement strategy the expectation value in the Choi state is
\begin{equation}
	\Tr[\Omega \jam(\tilde{\mcU})]
	=
	\sum_{i} p_i \Tr[(N_i\otimes \rho_i^\T) \jam(\tilde{\mcU})] 
\end{equation}
and can be recast, thanks to Eq.~\eqref{eq:choi_expectation}, as
\begin{equation}\label{eq:PAM}
	\Tr[(N_i\otimes \rho_i^\T) \jam(\tilde{\mcU})] 
	=
	\Tr[N_i\tilde{\mcU}(\rho_i)] \, . 
\end{equation}
While the dichotomic \ac{POVM} defined by $N_i\otimes \rho_i^\T$ for each $i$
originally acts on the Choi state $\jam(\tilde{\mcU})$,
the form \eqref{eq:PAM} suggests a simpler, straightforward experimental implementation of the dichotomic \ac{POVM}:
one prepares the state $\rho_i$, applies the channel $\mcU$ under scrutiny, and measures the dichotomic \ac{POVM} 
given by $N_i$ on the state $\tilde{\mcU}(\rho_i)$. 
Thus, effective measurement strategies of the form \cite{Liu2020EfficientVerificationOf} can indeed be implemented 
by simple prepare-and-measure schemes. 

For Clifford unitaries this method yields a simple direct certification test. 
The Choi state of a Clifford unitary channel is a stabilizer state and can hence be verified with the methods of Ref.~\cite{PalLinMon18} discussed in Section~\ref{sec:DSCofSTABs}. 
The following proposition gives a theoretical guarantee for this protocol. It can be derived as a corollary of the results of Section~\ref{sec:DSCofSTABs}. 

\begin{propositionblock}{Direct certification of Clifford operations, \cite[Proposition~3]{Liu2020EfficientVerificationOf}}
Let $\mc C$ be an $n$-qubit Clifford operation. 
We consider the state certification of Protocol~\ref{protocol:QSC} applied to its Choi state $\jam(\mc C)$, which is a stabilizer state.  
This yields an $\epsilon$-certification test of $\jam(\mc C)$ w.r.t.\ infidelity from $\np$ independent such state preparations for 
\begin{equation}
	\np \geq 2\, \frac{\ln(1/\delta)}{\epsilon} 
\end{equation}
with confidence $1-\delta$. 
Moreover, the target $\jam(\mc C)$ is accepted with probability $1$. 

This test corresponds to a similar certification test of $\mc C$ w.r.t.\ entanglement gate infidelity $1-\efidelity$ and can be implemented as a prepare-and-measure scheme via \eqref{eq:PAM}. 
\end{propositionblock}

\subsubsection*{Further reading}
The three works of Refs.~\cite{Liu2020EfficientVerificationOf,Zhu2020EfficientVerificationOf,Zeng2020QuantumGateVerification} all follow the presented certification strategy based on direct state certification. Moreover, they discuss several additional aspects:
Liu \emph{et al.}\ \cite{Liu2020EfficientVerificationOf} study non-trace-preserving processes and measurements, 
Zhu and Zhang \cite{Zhu2020EfficientVerificationOf} analyze the general multi-qudit case and strategies based on projective $2$-designs, and 
Zeng \emph{et al.}\ \cite{Zeng2020QuantumGateVerification} discuss entanglement property detection.  

Similar to direct state certification also fidelity estimation protocols can be lifted to quantum processes. 
To this end, one applies the state fidelity estimation to the output of the process applied to randomly chosen input states. 
The original \ac{DFE} proposal by Flammia and Liu \cite{FlaLiu11} already includes the application to quantum channels by sampling from the eigenstates of multi-qubit Pauli operators as the input states. Furthermore, simplifications arising for Clifford gates are discussed. 
See also the parallel work by da Silva \emph{et al.}~\cite{SilLanPou11}. 
A strategy to estimate the average gate fidelity by inputting states drawn at random from complex projective $2$-designs was studied by Bendersky \emph{et al.}~\cite{BenderskyPastawskiPaz:2008:Selective}. 
Reich \emph{et al.}~\cite{ReichGualdiKoch:2013:Minimum} determined the minimal number of required input states for the fidelity estimation of quantum processes. 
See also the related work by Hofmann~\cite{Hofmann:2005:Complementary}. 
Reich \emph{et al.}\ also provide a quantitative comparative overview over all the before-mentioned approaches in Ref.~\cite{Reich2013OptimalStrategiesFor}.

\subsection{\protchaptitle{Randomized benchmarking}}
\label{sec:RB}
The schemes presented in the previous section fail in the presence of sizeable \ac{SPAM} errors. 
In the context of digital quantum computing, this sensibility to \ac{SPAM} errors is dramatically reduced by so-called \emph{\ac{RB} protocols} \cite{EmeAliZyc05,LevLopEme07,KniLeiRei08,DanCleEme09,MagGamEme11}. 
These protocols can extract certain  quantitative measures of a quantum process 
associated to a \emph{quantum gate set}.
The process can be, for example, a certain gate, an error channel or an error map associated to the deviation of a quantum gate set from its ideal implementation.
While still concerned with the physical layer of a quantum device, randomized benchmarking protocols already make explicit use of a gate layer, the abstraction at the heart of digital quantum computing. 

\Acl{RB} comprises a large zoo of different protocols. 
Therefore, we begin with a fairly general description.
The principle idea to achieve the \ac{SPAM}-(error) robustness is the following:
after preparing an input state, one applies the quantum process under scrutiny multiple times in sequences of different lengths before performing a measurement.
Thereby, the effect of the process on the measurement is attenuated with increasing sequences length.
At the same time, errors in the state preparation and measurements enter the measured quantities only linearly and are independent of the sequence length. 
In this way, fitting the attained signals for different sequence lengths with functions depending on the length reveals properties of the quantum process disentangled from the \ac{SPAM} errors. 

A prototypical \ac{RB} protocol implements this rough idea for 
 a digital quantum computer as follows. 
Let $\gset \subset \U(d)$ be a subgroup of unitary operations and $\phi: \gset \to \M(\CC^d)$ be their implementation on a quantum computer. 
In simple \ac{RB} protocols $\phi(g)$ just models the faulty implementation of $\mcG$ on the actual device. 
More generally, the targeted implementation of the protocol can also include, e.g., a non-uniform sampling over the group or the implementation of another fixed gate after $\mcG$. 
Also in these cases $\phi$ is the faulty version of the targeted implementation.

Note that the assumption of the existence of such a map $\phi$ already encodes assumptions on the quantum device and its noise process: 
the map $\phi$ might model the compilation into elementary gates, effects and imperfections of the physical control, and noise. 
All these steps are not allowed to depend on the gate sequence the gate is part of, the overall time that evolves during the protocol, or other external variables. 
This noise can in particular be described as \emph{context-independent} and \emph{Markovian}. 

With these ingredients we can state a prototypical \ac{RB} protocol, 
see Figure~\ref{fig:RB} for an illustration. 

\begin{figure}
\centering
\includegraphics[width=\linewidth]{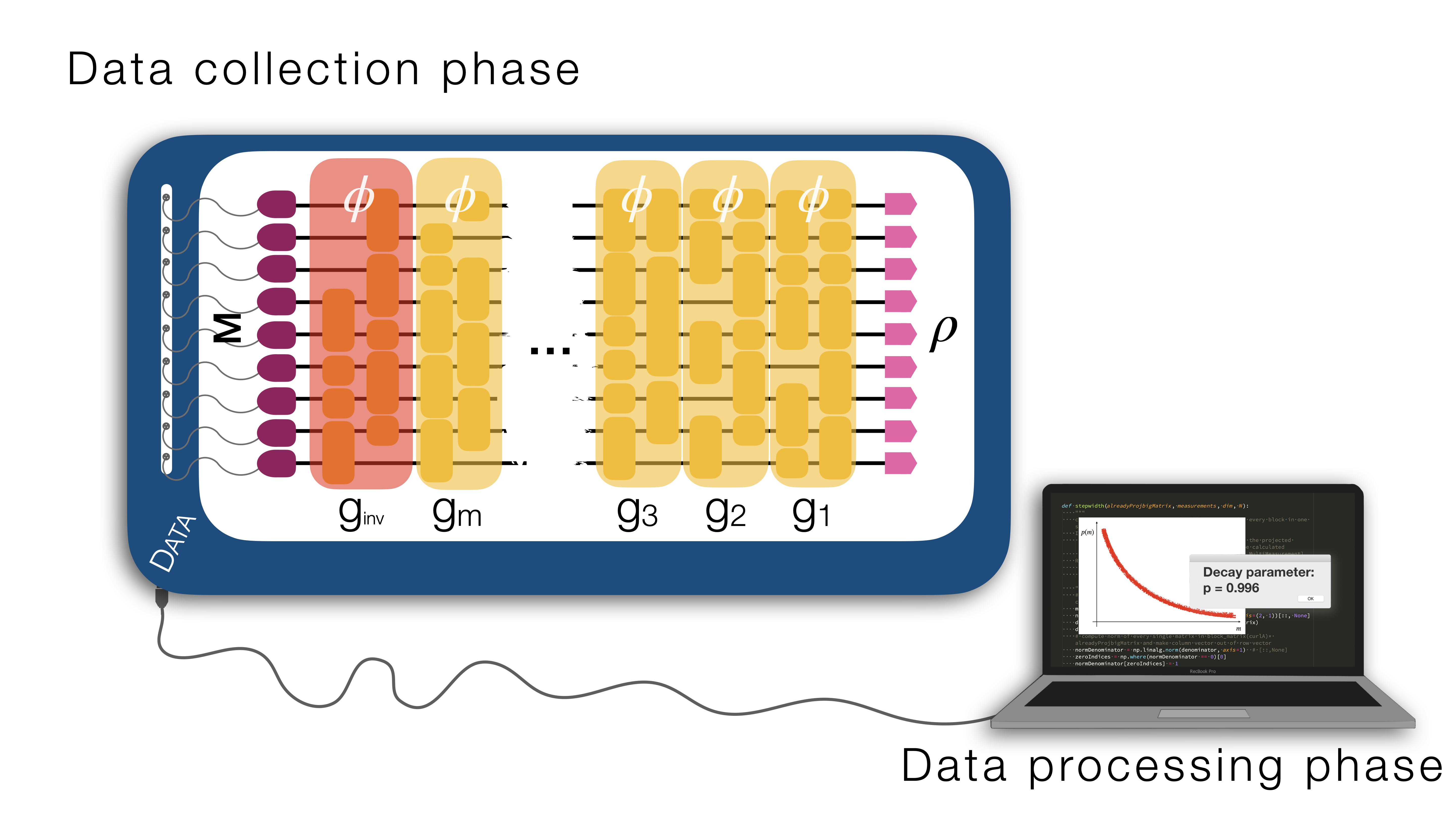}
  \caption{%
  Illustration of a prototypical \ac{RB} protocol. 
  After the preparation of an initial state, one applies a random sequence of unitaries 
  $g = (g_1, \ldots , g_m)$ succeeded by an inversion gate and final measurement of $M$. 
  This experiment is repeated for different sequences and different sequence lengths $m$. 
  In the classical postprocessing, the decay parameter of resulting empirical estimates for different sequence lengths m are extracted and reported as the \ac{RB} parameters.
  \label{fig:RB}%
  }
\end{figure}

\begin{protocolblock}[label={protocol:standardRB}]{Prototypical \ac{RB}}
Let $\gset  \subset \U(d)$ be a subgroup, 
$\rho \in \DM(\CC^d)$ an initial state, and
$\mset = \{\mop, \1-\mop\}\subset \PSD(\CC^d)$ a measurement. 
Furthermore, let $\mathfrak{M} \subset \mathbb N$ be 
a set of sequence lengths. 

For every sequence length $m \in \mathfrak{M}$, we perform the following procedure multiple times. 

Draw a 
sequence $g = (g_1, \ldots, g_m)$ of $m$ group elements i.i.d.\ uniformly at random. 
Calculate the inverse elements $g_\text{inv} =  g_1^{-1}g_2^{-1} \cdots g_m^{-1}$ of the sequence.

For each sequence preform the following experiment: 
\begin{itemize}
	\item Prepare $\rho$
	\item Apply %
	$S_g = \phi(g_\text{inv})\phi(g_m) \ldots \phi(g_2) \phi(g_1)$, i.e.\ the
	sequence of  implementations of $g$ followed by the implementation of $g_\text{inv}$, to $\rho$.
	\item Perform the measurement $\mset$.
\end{itemize}

Multiple repetitions of the experiment yield an estimator $\hat p_g$ for
the probabilities
\begin{equation}\label{eq:RB:p_g}
	p_g(m) = \Tr\left[M S_g \rho\right]
\end{equation}

Repeating these steps for different random sequences, we can calculate an estimator $\hat p(m)$ for
\begin{equation}
	p(m) = \EE_{g_1}\EE_{g_2} \cdots \EE_{g_m} p_{(g_1, g_2, \ldots g_m)}(m).
\end{equation}

\emph{Post-processing:} extract the decay parameters of the data $\mathfrak M \to [0,1]$, $m \mapsto \hat p(m)$ and report as the 
RB parameters.  
\end{protocolblock}

More generally, \ac{RB} protocols might go beyond Protocol~\ref{protocol:standardRB} in various ways: for example, by calculating the inverse of a sequence only up to specific gates, using a different measure than the uniform measure for drawing the group elements of the sequence, or performing a measurement POVM with multiple outputs or measurements adapted to the sequence. 
In addition, the post-processing might combine different \ac{RB} data series in order to get simpler decay signatures. 

The first step in the theoretical analysis of \ac{RB} protocols is to establish the fitting model of the \ac{RB} data $p(m)$. 
Ideally, $p(m)$ is well-approximated by a single exponential decay. 
Subsequently, the \ac{RB} decay parameters can in certain settings be connected to the average gate fidelity of a noise process effecting the implementation map, as we now discuss. 

The data model of most \ac{RB} protocols can be understood as estimating the $m$-fold self-convolution of the implementation map \cite{Merkel18}. 
More precisely, for $\phi, \psi: \gset \to \M(\CC^d)$ we can define a convolution operation as
\begin{equation}\label{eq:RB:convolution}
	\phi \ast \psi(g) = \EE_{\tilde g} \phi(g {\tilde g}^{-1})\psi(\tilde g).
\end{equation}
Note that this definition naturally generalizes, e.g., the discrete circular convolution on vectors in $\CC^n$, which can be seen as an operation on functions on the finite group $(\ZZ_n, +) \to \mathbb \CC$.
With the convolution \eqref{eq:RB:convolution}, we can rewrite the averages of the \ac{RB} sequences as
\begin{align}
	\EE_{g} S_g &= \EE_{g_1, g_2, \ldots, g_m}
	\phi(g_1^{-1}g_2^{-1} \cdots g_m^{-1})\phi(g_m) \cdots \phi(g_2)\phi(g_1) 
	\nonumber\\ \nonumber
	&= \EE_{h_1, h_2, \ldots, h_m} 
		\phi(h_m^{-1})\phi(h_{m} h_{m-1}^{-1})  \cdots\phi(h_2h_1^{-1})\phi(h_1) \\
	&= \phi^{\ast (m+1)}(\id),
\end{align}
where the replacements $h_1 = g_1$ and $h_j = g_j h_{j-1}$ for $j\in \{2, \dots, m\}$ have been made the second equality, 
$\id$ denotes the identity element of $\gset{}$ and $\phi^{\ast k}$ denotes the $k$-fold convolution of $\phi$ with itself.
In expectation the \ac{RB} data $p(m)$ is thus a contraction defined by $M$ and $\rho$ of the $(m+1)$-fold self-convolution of $\phi$ evaluated at the identity element. 

In the simplest instance of an \ac{RB} protocol one can directly calculate this expression: 
namely, when $\gset{}$ is a unitary $2$-design, the targeted implementation is simply the action of $\gset{}$ as quantum gates, and the noise in $\phi$ can be modeled by a single gate-independent quantum channel $\Lambda \in \CPT(\CC^d)$. 
Denoting by $\mathcal G$ the (adjoint) action of $g$ as the unitary channel $X \mapsto \mathcal G (X) = g X g^\ad$,
 we have the noise model
\begin{equation}\label{eq:gate_indep_noise}
	\phi(g) = \Lambda \circ \mcG \,.
\end{equation}
With this ansatz for $\phi$ we can calculate that
\begin{equation}\label{eq:RB:convolution2Design}
	\EE_{g \in {\gset^m}} S_g = \phi^{\ast(m + 1)}(\id) = \Lambda \left[\EE_{g \in \gset} \mcG^\dagger \Lambda \mcG \right]^m.
\end{equation}

The operator $\tw_\mu: \M(\mathbb C^d) \to \M(\mathbb C^d)$, 
$\mathcal X \mapsto \EE_{U \sim \mu}[ \mc U \mc X \mc U^\ad ]$ appearing in \eqref{eq:RB:convolution2Design} is the so-called \emph{(channel) twirling map} and appears 
in different contexts in quantum information. 
If we write out the twirling map with the individual unitaries it reads
\begin{equation}
	\tw_\mu(\X) = ( \rho \mapsto \EE_{U \sim \mu}[ U \X(U^\ad ( \rho)\, U)U^\ad ]\, ).
\end{equation}
It becomes apparent that $\tw_\mu$ is related to second moment operator $\mo^{(2)}_\mu$ from Eq.~\eqref{eq:moment_op} by 
a simple vector space isomorphisms. 
Recall that for a unitary $2$-design $\mu$ Proposition~\ref{prop:invariant_operators_k2} gives us an explicit description of $\mo^{(2)}_\mu$.  
We can track the isomorphism to derive the following convenient expression. 

\begin{theoremblock}[label={thm:channel-twirling}]{Twirling of channels \cite{Niel02,EmeAliZyc05}}
Let $\X \in \M(\CC^d)$ be trace-preserving and $\mu$ be a unitary $2$-design. 
Then 
\begin{equation}\label{eq:Xtw_depol}
	\tw_\mu(\X) 
	=
	\depol[p(\X)] \, ,
\end{equation} 
where $\depol[p]$ is the depolarizing channel \eqref{eq:def:depolChannel} and $p(\X)$ is the effective depolarizing parameter defined in Eq.~\eqref{eq:effectiveP}. 
\end{theoremblock}

\begin{proof}
First we note that any map $\X \in \M(\CC^d)$ is uniquely determined by 
$(\mc X \otimes \id)(\FF)$, which is a similar construction as the Choi-\Jamiolkowski\ isomorphism. 
This isomorphism is given by $\Tr_{2}[(\mc X \otimes \id)(\FF) (\1\otimes A)] = \X(A)$ 
but its explicit form is not needed. 
Hence, we can make the isomorphisms between the twirling map $\tw_\mu$ and the second moment operator $\mo^{(2)}_\mu$ from \eqref{eq:moment_op} explicit by writing
	\begin{align}
		&(\tw_\mu(\mc X) \otimes \id)(\FF) 
		\nonumber
		\\
		&\ = 
		\EE_{U \sim \mu}
		\left[ 
			(U \otimes \1) 
				\X \otimes \id 
				\left(
					(U^\ad \otimes \1) \FF (U \otimes \1)
				\right)
			(U^\ad \otimes \1)
		\right] 
		\nonumber\\
		&\ = 
		\EE_{U \sim \mu}
		\left[ 
			(U \otimes \1) 
				\X \otimes \id 
				\left(
					(\1 \otimes U) \FF (\1 \otimes U^\ad)
				\right)
			(U^\ad \otimes \1)
		\right] 
		\nonumber\\
		&\ = 
		\EE_{U \sim \mu}
		\left[ 
			(U \otimes U) 
				\X \otimes \id 
				\left(
					\FF
				\right)
			(U^\ad \otimes U^\ad)
		\right] \nonumber\\
		&\ = \mo_\mu^{(2)}(\X \otimes \id (\FF)) \, .
	\end{align}
For a unitary $2$-design $\mu$, $\mo_\mu^{(2)}$ coincides with the second moment operator of the Haar measure. 
Schur-Weyl duality (Theorem~\ref{thm:commutantOfDelta}) tells us that
\begin{equation}
	\mo_\mu^{(2)}(\X \otimes \id (\FF)) \in \operatorname{span}\{ \1,  \FF\}\, . 
\end{equation}
Observing that $\depol[0]\otimes \id(\FF) = \1 / d$ and trivially $\depol[1] \otimes \id (\FF) = \FF$, we 
conclude that 
\begin{equation}
	\tw_\mu(\X) \in \operatorname{span}\{ \depol[0], \depol[1]\}\, . 
\end{equation}
Furthermore, one quickly checks that if $\X$ is trace-preserving so is $\tw_\mu(\X)$. 
Hence, $\tw_\mu(\X)$ is an affine combination of $\depol[0]$ and $\depol[1]$. 
Thus, 
$\tw_\mu(\X) = \depol[p]$ holds for some $p \in \CC$ and it remains to determine $p$. 
One way forward is a straight-forward calculation using the expressions for the coefficients provided by Proposition~\ref{lem:sym_moment_operator}. 
A shortcut is to calculate the effective depolarization of both sides. 
Due to the unitary invariance of $\mu_{\sphereCd}$, it follows from 
\eqref{eq:def:AGV:HSdef} that
$\agf(\X) = \agf(\tw(\X))$
and correspondingly for the affinely related effective depolarization parameter that $p(\X) = p(\tw(\X))$. 
Combined with $p(\depol[p]) = p$ from Eq.~\eqref{eq:effective_depol_parameter} yields the theorem's assertion. 
\end{proof}

Theorem~\ref{thm:channel-twirling} allows us to explicitly calculate the \ac{RB} data model from Eq.~\eqref{eq:RB:convolution2Design}. 
To this end, a short calculation reveals that $\depol^m = \depol[p^m]$.
With this we find  
the \ac{RB} data model to be 
\begin{equation}\label{eq:RBdata_derivation}
\begin{aligned}
	p(m) 
	&=
	\Tr[\tilde M\Lambda \depol[p(\Lambda)^m] (\tilde \rho)]
	\\
	&=
	p(\Lambda)^m \Tr[\tilde M\Lambda(\tilde \rho)] + (1-p(\Lambda)^m) \Tr[\tilde M\Lambda(\1/d)] 
	\\
	&=
	p(\Lambda)^m \Tr[\tilde M\Lambda(\tilde \rho-\1/d)] + \Tr[\tilde M\Lambda(\1/d)] \, ,
\end{aligned}
\end{equation}
with $\tilde M$ and $\tilde \rho$ denoting the potentially faulty implementation of the measurement $M$ and initial state $\rho$.
In terms of the so-called \emph{\ac{SPAM} constants}
\begin{equation}
\begin{aligned}
	A &\coloneqq \Tr[\tilde M\Lambda(\tilde \rho-\1/d)] \, , 
	\\ 
	B &\coloneqq \Tr[\tilde M\Lambda(\1/d)] \, ,
\end{aligned}
\end{equation}
we obtain the simple \ac{RB} fitting model 
\begin{equation}\label{eq:RB_fitting}
		p(m) = A\, p^m + B\, . 
\end{equation}
Thus, fitting a single exponential decay to the estimator $\hat p(m)$ yields estimates $\hat p$, $\hat A$ and $\hat B$ for the model parameters $p$, $A$ and $B$.
In particular, 
the estimated \ac{RB} decay parameter $\hat p$ is an estimator for the effective depolarizing parameter $p(\Lambda)$ of the error channel $\Lambda$. 
Recall that the effective depolarizing parameter is affinely related to the average gate fidelity \eqref{eq:def:AGV:HSdef} via Eq.~\eqref{eq:effectiveP}. 
From the \ac{RB} decay parameter, we thus equivalently obtain an estimate for the average gate fidelity of the noise channel $\Lambda$ as 
\begin{equation} \label{eq:RB:AGFestimation}
		\hatagf
		= \left(1 - \frac1d\right) \hat p + \frac1d.
\end{equation}
Note that the resulting estimate of the average gate fidelity \eqref{eq:def:AGV:HSdef} is indeed robust against \ac{SPAM} errors, which only enter the \ac{SPAM} constants $A$ and $B$. 

Deriving rigorous performance guarantees for the estimator \ac{RB} estimator $\hat p$ is involved: it requires the analysis of confidence regions of the estimator $\hat p_g(m)$ of the probability \eqref{eq:RB:p_g} that is a random variable of the quantum measurement statistics and $\hat p(m)$ obtained by the subsampling of the sequences $g$. 
Furthermore, the error of these estimators for each $m$ enters the errors of the fidelity estimator via the exponential fitting procedure.
This step depends on the choice of algorithm and the estimated sequence lengths. 

Using the fact that $\hat p(m)$ is the mean estimator of a bounded random variable, one can use Hoeffding's inequality (Theorem~\ref{thm:Hoeffdings}) 
to derive confidence intervals for an overall sampling complexity that is independent of the number of qubits in the regime of high fidelity. 
Such bounds however are prohibitively large for practical implementations. 
A refined analysis by Wallman and Flammia \cite{WalFla14} derived tighter  bounds for short sequences and small number of qubits. 
However, bounds that are practical and scalable in the number of qubits require a careful analysis of the variance of the estimator $\hat p_g(m)$ over the choice of the random sequences. 
For $\gset{}$ being the Clifford group, Helsen \emph{et al.} \cite{HelsenEtAl:2019:mqbRB} work out explicit variance bounds for the estimator $\hat p_g(m)$ and derived sampling complexities for $\hat p(m)$ that are practical, independent of the number of qubits and scale favorable with the sequence length. 
To this end, 
they
employed a refined representation theoretical analysis of the commutant of the $4$-th order diagonal action of the Clifford group \cite{HelWalWeh16, ZhuKueGra16} in order to calculate the corresponding moment operator; an endeavor that is complicated by the fact that the Clifford group itself is not a unitary $4$-design. 

A rigorous analysis of a simplified fitting procedure was derived in Ref.~\cite{HarHinFer19}. 
Therein (again using trivial bounds on the variance) the authors show that a ratio estimator for the infidelity $r = 1 - p$ that employs the estimates of $p(m)$ for two different sequence length has multiplicative error using an efficient number of samples again in the regime of high fidelity. 

All of these performance guarantees indicate that in principle \ac{RB} protocols can be efficiently scalable in the number of qubits. 
To ensure also an efficient classical pre-processing of the prototypical \ac{RB} protocol it is 
important to have an efficiently tractable group structure so that the inverse of the gate sequence can be computed. 

For the important example of the Clifford group, the Gottesman-Knill theorem, see e.g.\ Ref.~\cite{NieChu10}, allows the efficient computation the inverse of a sequence $g_m \cdots g_2 g_1$ in polynomial time (w.r.t.\ the number of qubits). 
Furthermore, since the Clifford group is a unitary $3$-design \cite{Web15,Zhu15}, 
it meets the requirement of Theorem~\ref{thm:channel-twirling}. 
For this reason the presented analysis applies to the Clifford group under the assumption of gate-independent noise. 

It is natural to ask of additional examples 
of groups that constitute a unitary $2$-design and are covered by the presented analysis without modifications. 
But it has been established that these two requirements are already surprisingly restrictive. A complete classification of so-called $2$-groups ($2$-design groups) is summarized in Ref.~\cite{BannaiEtAl:2020:tgroups}. 
In fact, if one requires a family of $2$-groups that can be constructed for an arbitrary number of qubits, one is left with subgroups of the Clifford group or $\SU(d)$ itself as the only examples \cite{BannaiEtAl:2020:tgroups, SawickiKarnas:2017, HaferkampEtAl:2020}. 

We provide more details how the analysis of the prototypical \ac{RB} protocol can be generalized in the further-reading paragraph at the end of the section. 
Now, we want to discuss another variant of \ac{RB} that is particularly important as tool for certifying quantum gates. 

\subsubsection*{\protchaptitle{Interleaved randomized benchmarking}}
The prototypical \ac{RB} protocol yields estimates of the effective depolarizing parameter or the average gate fidelity of the average error channel of a gate \emph{set}. 
In contrast, \emph{interleaved \ac{RB}} protocols  \cite{MagGamJoh12} allow one to extract the effective depolarizing parameter of \emph{individual} gates from a group with respect to their ideal implementation provided the noise is sufficiently incoherent.  

In an interleaved \ac{RB} protocol one performs in addition to 
the standard \ac{RB} protocol a modified version, where the random sequences 
are interleaved with the specific target gate. 
The second experiment yields estimates for the effective depolarization parameter of the error channel associated to the group concatenated 
with the error channel of the individual target gate. 
Under certain assumptions the effective depolarization parameter of the implementation of the target gate can 
be estimated from the decay parameters of both \ac{RB} protocols.

\begin{protocolblock}[label={protocol:modifiedRB}]{Interleaved \ac{RB}} 
For $\gset\subset \U(d)$ and a target gate $\gt\in \gset$ 
\begin{enumerate}
	\item follow Protocol~\ref{protocol:standardRB}, 
	\item follow Protocol~\ref{protocol:standardRB} but modify the sequences to be
	\begin{equation}
		g = (g_1, \gt, g_2, \gt, g_3, \ldots, \gt , g_m),
	\end{equation}
	where $\gt$ is the target gate and $g_i \in \gset$ for $i \in [m]$ are drawn uniformly at random.
	The inverse $g_\text{inv}$ is also calculated w.r.t.\ the modified sequence $g$.
\end{enumerate}
The output of the protocol are the decay parameters of both experiments.
\end{protocolblock}

For the analysis we again consider a `mostly' gate-independent noise model and assume that $\gset$ is a unitary $2$-design. 
In the noise model we assume that the same noise channel 
$\Lambda \in \CPT(\H)$ 
follows the ideal implementation of all gates but the target gate, i.e., 
\begin{equation}\label{eq:tildeG}
	\phi(g) = \Lambda \circ \mathcal G
\end{equation}
for all $ g \in \gset \setminus \{\gt\}$.

The first step of the protocol is the unmodified \ac{RB} protocol. 
If we neglect that $\phi$ deviates from the form Eq.~\eqref{eq:tildeG} on $\gt$, 
we can apply the analysis of the previous section for gate-independent noise and 
conclude that the protocol outputs and estimator for the effective depolarizing constant $p(\Lambda)$. 
For example, for a large group it is plausible to neglect the contribution of the noise associated to the $\gt$ gate to the group average. 

It remains to analyze the second protocol. 
In analogy to Eq.~\eqref{eq:RB:convolution} we can in general rewrite
\begin{align*}
	\EE_{g_1, \ldots, g_m} &S_g \\
	= \EE_{g_1, \ldots, g_m} &\phi(g_1^{-1}\gt^{-1}g_2^{-1}\gt^{-1} \ldots g_m^{-1}) \\
							&\times \phi(g_m) \phi(\gt) \ldots \phi(g_2) \phi(\gt) \phi(g_1) \\
	= \EE_{g_1, \ldots, g_m} &\phi(g_m^{-1})\ldots \\
							& \times \phi(g_3g_2^{-1}\gt^{-1})\phi(\gt)\phi(g_2g_1^{-1}\gt^{-1})\phi(\gt)\phi(g_1),
\end{align*}
by substituting $g_i$ with $g_i g_{i-1}^{-1} \gt^{-1}$ for all $i > 1$. 

Inserting the noise model~\eqref{eq:tildeG} yields
\begin{equation}\begin{split}
	\EE_{g_1, \ldots, g_m} S_g %
	= \Lambda \left[%
					\EE_{g \in \gset{}}\,%
						\mcG^\ad \Gt^\ad \phi(\gt)\Lambda\mcG%
				\right]^m%
\end{split}\end{equation}

This is the same expression as Eq.~\eqref{eq:RB:convolution2Design} with $\Lambda$ replaced by $\Gt^\ad \phi(\gt) \Lambda$. 
Hence, applying the same arguments as in the analysis of the standard \ac{RB} protocol for unitary $2$-designs yields 
a single-exponential fitting model with 
decay parameter estimating the effective depolarizing parameter $p(\Gt^\ad\phi(\gt) \Lambda)$. 
The second part of the interleaved \ac{RB} protocol, thus, yields an estimate of the effective depolarizing parameter or equivalently, via Eq.~\eqref{eq:RB:AGFestimation}, of the fidelity of the error map $\Gt^\ad\phi(\gt)$ of the target gate $\Gt$ concatenated with the error channel $\Lambda$. 

From $p(\Lambda)$ and $p(\Gt^\ad\phi(\gt)\Lambda)$ it is indeed possible to infer $p(\Gt^\ad\phi(\gt))$.  
In meaningful practical regimes this however requires additional control the unitarity of $\Lambda$ \cite{Carignan-Dugas2019BoundingTheAverage}: 
for sequences of unitary channels the infidelity of their composition can scale quadratically in the sequence length in leading order. 
In contrast, highly non-unitary channels will feature a close to linear scaling in the sequence length. 
Thus, using the unitary one can derive bounds for fidelity measures of 
composite channels that exploit the linear scaling. 
We simply state the required bound without proof for interleaved \ac{RB}: 
\begin{theoremblock}[{label={thm:comp-channel-bound},fonttitle=\normalfont}]{Composite channel bound \cite{Carignan-Dugas2019BoundingTheAverage}}
	For any two quantum channels $\mc X, \mc Y$ it holds that 
	\begin{equation}\label{eq:comp-channel-bound}
		\left| p(\mc X) 
				- \frac
					{p(\mc X\mc Y)p(\mc Y)}
					{u(\mc Y)} 
		\right| 
		\leq 
		\sqrt{ 
			1 - \frac
				{p(\mc Y)^2}
				{u(\mc Y)}
		} 
		\sqrt{
			1 - \frac
				{p(\mc X \mc Y)^2}
				{u(\mc Y)}
		}
	\end{equation}
\end{theoremblock}

With an estimate for the unitarity $\hat u(\Lambda)$, Theorem~\ref{thm:comp-channel-bound} allows the estimation of the effective depolarizing constant and thus the average gate fidelity of the target gate by 
\begin{equation}
	\hatagf(\phi(\gt), \Gt) 
	= 
	\frac{d - 1}d \, \frac{\hat p(\Gt^\ad \phi(\gt))\hat p(\Lambda)}{\hat u^{\Lambda}} + \frac1d 
\end{equation}
up to a systematic error that is given by evaluating the right-hand side of Eq.~\eqref{eq:comp-channel-bound}. 
The systematic error is small in the regime where $u(\Lambda) \approx p(\Lambda)^2$ which is the case if $\Lambda$ is decoherent. 
The unitarity of $\Lambda$ can be estimated using variants of the \ac{RB} protocol itself developed in Refs.~\cite{WalGraHar15, dirkse2019efficient}. 

Alternatively, one can just assume that the error is sufficiently incoherent, i.e.\ that $| 1 - p(\Lambda)^2 / u(\Lambda)| \leq \epsilon$. 
Conditioned on this external belief, one obtains the simpler estimator 
\begin{equation}
	\hatagf(\phi(\gt), \Gt) 
	= 
	\frac{d - 1}d \, \frac{\hat p(\Gt^\ad \phi(\gt))}{\hat p(\Lambda)} + \frac1d
\end{equation}
that comes with a systematic error that is controlled in $\epsilon$. 
Thereby, interleaved \ac{RB} can be used to arrive at average-performance certificates of individual quantum gates. 

We have already seen that for interleaved \ac{RB} controlling the unitarity is helpful in deriving tighter error bounds. 
In addition, estimating the unitarity can also yield relevant worst-case performance bounds in terms of the average gate fidelities using Theorem~\ref{thm:RiKbound}.

\subsubsection*{Further reading} \label{sec:rb:furtherreading}
Randomized benchmarking was originally developed in a series of work focusing on the unitary group and Clifford gates \cite{EmeAliZyc05,LevLopEme07,KniLeiRei08,DanCleEme09,MagGamEme11}. 

The early analyses used the gate-independent noise model \eqref{eq:gate_indep_noise}, which we also assume here. 
In many applications this is however a questionable assumption. 
After first perturbative approaches to derive the \ac{RB} signal model under gate-dependent noise by Magesan \emph{et al.}~\cite{MagGamEme11,Magesan2012} and Proctor \emph{et al.}~\cite{Proctor2019DirectRandomizedBenchmarking}, Wallman rigorously derived the fitting model for unitary $2$-designs in Ref.~\cite{wallman2018randomized}. 

Using the elegant description of the \ac{RB} data as the $m$-fold convolution of the implementation map, recently proposed by Merkel \emph{et al.}~\cite{Merkel18}, one can abstractly understand the result as follows: as the standard discrete circular convolution, the convolution operator of maps on a group can be turned into a (matrix) multiplication using a Fourier transform. 
This abstract Fourier transform for functions on the group is defined to be a function on the irreducible representations of the group. 
In the case of \ac{RB}, this function is matrix-valued, and we observe matrix powers of the Fourier transforms for every irreducible representation superimposed by a linear map. 
For every irreducible representation, for sufficiently large $m$, the matrix powers are proportional to the $m$-th power of the largest eigenvalue of the matrix-valued Fourier transformation. Contributions from other eigenvalues are suppressed. 
In this sense \ac{RB} is akin to the power method of numerical linear algebra but in Fourier space \cite{HelsenPC}. 
A rigorous analysis requires to perturbatively bound the contribution of the subleading eigenvalues. 
For unitary $2$-groups the adjoint representation decomposes into two irreducible representations, 
the trace representation and the unital part of the quantum channel. For close to trace-preserving maps the trace representation will only contribute a very slow decay, i.e.\ a constant contribution to the fitting model, and the \ac{RB} decay parameter is the dominant eigenvalue of the unital representation. 
Wallman \cite{wallman2018randomized} derived norm bounds for the contribution of subleading eigenvalues and showed that the contribution is exponentially suppressed with the sequence length.  
Furthermore, Wallman showed that there is a gauge choice of the gate set such that the decay parameter can be connected to the average gate fidelity of the average error channel over the gate set. 
For qubits this gauge was demonstrated to yield a physical gate set by Carignan-Dugas \emph{et al.}~\cite{Carignan-DugasEtAl:2018:From}. 
The physicality of this gauge is, however, in general not guaranteed and a counter example is given by Helsen \emph{et al.}~\cite{Helsen20AGeneralFramework}. 
As discussed by Proctor \emph{et al.}~\cite{proctor2017WhatRandomizedBenchmarking}, this complicates the interpretation of the \ac{RB} decay rates as related to average fidelities that have a clear physically interpretation. 

While the Clifford gates are definitely a prominent use case in the benchmarking of digital quantum computers, more flexible \ac{RB} protocols require analyzing groups that are not a unitary $2$-design.

Randomized benchmarking protocols for other groups were developed in Refs.~\cite{Gambetta2012CharacterizationOfAddressability,CarWalEme16,CroMagBis16, HashagenEtAl:2018,BrownEastin:2018,Franca2018ApproximateRB,Chasseur2015LeakageErrors,HelsenEtAl:2019:character}. 
These protocols, for example, allow inclusion of the $T$-gate in the gate set \cite{CarWalEme16} or characterization of leakage between qubit registers by using tensor copies of the Clifford group \cite{Gambetta2012CharacterizationOfAddressability}.
As the adjoint representation of other groups typically decomposes into multiple irreducible representation, \ac{RB} data is expected to feature multiple decays in general. 
For a description of a flexible post-processing scheme for general \ac{RB} type data and performance guarantees see Ref.~\cite{Helsen20AGeneralFramework}. 

In order to isolate the different decays, multiple \ac{RB} variants have been developed. 
These either rely on directly preparing a state that has high overlap with only one irreducible representation or cleverly combining data from different \ac{RB} experiments to achieve the same effect. 
Many of these techniques can be understood as variants of the character benchmarking protocol developed by Helsen \emph{et al.}~\cite{HelsenEtAl:2019:character}. 
Character benchmarking uses inversions of the \ac{RB} sequence not to the identity but randomly drawn gates from the group. 
In the classical post-processing data sequences of different end gates are linearly combined by weighting them according to the character formulas. 
Thereby, the data is projected onto the irreducible representation of the respective character and can be subsequently fitted by a single decay. 

\emph{Interleaved \ac{RB}} was proposed in Refs.~\cite{MagGamJoh12,Gaebler2012RandomizedBenchmarking} and demonstrated in practice. 
Already standard RB provides a trivial bound for individual gates of the group by simply 
attributing the average error to a single gate. 
In the original proposal of interleaved \ac{RB}, the analysis 
does not allow for rigorous certificates that go significantly beyond this trivial bound for 
few qubits \cite{Carignan-Dugas2019BoundingTheAverage}. 
A general bound by Kimmel \emph{et al.}~\cite{KimSilRya14}, was considerably refined using the unitarity by Carignan-Dugas \emph{et al.}~\cite{Carignan-Dugas2019BoundingTheAverage}.
Thereby it was established that if the error channel is sufficiently incoherent interleaved \ac{RB} yields 
rigorous certificates for individual gates with reasonable error bars. 
There exist multiple variants of the interleaved \ac{RB} scheme \cite{Erhard2019CharacterizingLarge-scale, sheldon2016characterizing, harper2017estimating, chasseur2017hybrid}. 
Another class of interleaved \ac{RB} was introduced in Ref.~\cite{onorati2019randomized}. Here, the average gate fidelity of individual gates is inferred from measurements of random sequences of gates that are drawn from the symmetry group of the gate. The individual gates are not part of the group itself and are also not included in the inversion of the sequence.

Another practically very interesting variation of \ac{RB} arises when one does not draw the gates from the uniform but another distribution over the group \cite{KniLeiRei08, Franca2018ApproximateRB, boone2019randomized, Proctor2019DirectRandomizedBenchmarking}.
For example, drawing the sequences randomly from the generating gates of the group, reduces the required sequence lengths \cite{Proctor2019DirectRandomizedBenchmarking}. 

Other quantities that can be measured by variants of the \ac{RB} protocols are the unitarity \cite{WalGraHar15, dirkse2019efficient}, measures for the losses, leakage, addressability and cross-talk \cite{Gambetta2012CharacterizationOfAddressability,WalBarEme15,WalBarEme16b}. Furthermore, \ac{RB} of operations on the logical level of an error correcting quantum architecture was proposed in Ref.~\cite{combes2017logical}. 

Combining different relative average gate fidelities obtained by interleaved \ac{RB} schemes can be used to acquire tomographic information about the error channel providing actionable advise to an experimentalist beyond a mere benchmarking and certification \cite{KimSilRya14}. 
Using \ac{SPAM}-robust data, these tomography schemes are in addition resource optimal for the unitary gates \cite{KimLiu16} and Clifford gates \cite{RotKueKim18}. 
For Pauli channels tomographic information can be efficiently obtained performing a character \ac{RB} protocol on multiple qubits simultaneously \cite{HarFlaWal19,Flammia2019EfficientEstimation,Harper2020FastEstimationOf,Franca2020EfficientBenchmarkingAnd}.

A general framework with few theorems that establishes the \ac{RB} fitting model of essentially all known \ac{RB} schemes under gate-dependent noise is developed in Ref.~\cite{Helsen20AGeneralFramework}. 
The central assumption employed therein to control contributions from subdominant eigenvalues of the Fourier transformation is a closeness condition to a reference representation in diamond norm averaged over all group elements. 
Moreover, a unifying review of \ac{RB} is provided.

\subsection{\protchaptitle{Cross-entropy benchmarking}}\label{sec:xeb}
The final protocol we discuss in this tutorial is \emph{\ac{XEB}} \cite{BoiIsaSme16}. 
XEB gained importance recently: 
it was used in order to experimentally collect evidence that a quantum computer can perform a task that basically no existing classical computer can solve in a reasonable amount of time \cite{Google2019QuantumSupremacy}. 

In Ref.~\cite{Google2019QuantumSupremacy} \ac{XEB} is performed in two distinct variants: 
one variant aims at extracting fidelity measures averaged over random sequences of individual gates. 
This protocol can be regarded as a special case of the character randomized benchmarking protocol \cite{HelsenEtAl:2019:character, Helsen20AGeneralFramework} 
that we have touched upon in Section~\ref{sec:rb:furtherreading}.
The second variant aims at certifying the correct sampling from the measurement output distribution of a single specific circuit. 
This second variant of \ac{XEB} is the focus of this section. 
It can be seen as an instance of a certification protocol on the application layer of a digital quantum computer. 
In consequence, it is commonly also referred to as a \emph{verification protocol} for sampling tasks. 
But the application, sampling from a distribution encoded in a quantum circuit, 
is deliberately chosen very close to the physical layer. 

\ac{XEB} was proposed as a protocol in the context of demonstrating quantum supremacy.
Experimentally demonstrating that a quantum computer can outperform current classical computers in some task is regarded as one of the 
mayor milestones in developing quantum computing technologies. 
The accuracy of the quantum operations and numbers of qubits of today's 
devices do not permit instances of interesting quantum algorithms 
that solve problems without known efficient algorithms, such as Shor's algorithm for integer factorization, 
at least not problem instances that come even close to being troublesome for a classical computer \cite{Pre18}. 
This motivated the proposal of demonstrating quantum supremacy in 
the task of generating samples from a probability distribution that is specified as the measurement distribution of a quantum circuit. 
This is a task that a quantum computer solves very naturally even though it might not be of any practical use \cite{BremnerJozsaShepherd:2011:Classical, BoiIsaSme16}. 
At the same time one can prove that certain random ensembles of quantum circuits yield probability distributions that can not be efficiently sampled from on a classical computer \cite{bouland_quantum_2018}. 

Besides establishing evidence for the hardness of solving the sampling task on a classical computer, 
a crucial ingredient in demonstrating 
quantum supremacy 
is a certification protocol that 
guarantees that one has implemented the correct distribution. 

The approach taken in Ref.~\cite{Google2019QuantumSupremacy} is to 
build trust in the correct functioning of the device for 
circuits that are still amenable to calculating a couple of outcome probabilities on a classical super-computer. 
To this end, the \ac{XEB} protocol was used. 
The measures that \ac{XEB} tries to estimate are the \emph{cross-entropy difference} and its variant the \emph{cross-entropy fidelity}. 

\subsubsection*{Cross-entropy and cross-entropy fidelity}
In the context of certifying a sampling task it is natural to directly consider 
measures of quality that compare two probability densities describing the measurement outcomes. 
While the measures we have studied in this tutorial so far are concerned with the physical layer, 
measures directly comparing two probabilities can be regarded as measures on the application layer. 

For a quantum circuit $U$ acting on $n$ qubits, we denote its measurement probability mass function in a basis $\{\ket x\}_{x \in [d]}$ after preparing a fixed initial state $\ket \psi$ by $p_U: [d] \to [0,1]$ with
\begin{equation}\label{eq:def:measdist}
	p_U(x) = \left|\sandwich x U \psi \right|^2.  
\end{equation} 

A well-known statistical measure \cite{CsiszarKoerner:2011:Information} to relate two 
 probability mass functions $q, p: [d] \to [0,1]$ is the \emph{cross-entropy} 
\begin{equation}
	H_X(q,p) = - \sum_{x \in [d]} q(x) \ln(p(x)).
\end{equation}
For $p=q$ we find that $H_X(q,q) = -\sum_{x} q(x) \ln( q(x)) \eqqcolon H(q)$ is the standard Shannon entropy. 
One can show that $H(q)$ is the minimal value of the cross-entropy $H_X(q,p)$, a relation known as \emph{Gibbs' inequality} \cite{MacKay:2003:Information}.

In the context of quantum supremacy demonstrations one expects the target probability distribution that one aims to implement to 
be of \emph{Porter-Thomas shape}. 
We say that a probability mass function $p: [d] \to [0,1]$ is of \emph{Porter-Thomas shape} if 
the tail distribution of $p(x)$ regarded as a random variable for $x$ drawn uniformly at random from $[d]$ 
is well-approximated by an exponential decay function, 
\begin{equation}\label{eq:xeb:porter-thomas}
	\PP_{x \sim p_\mathrm{uni}}[p(x) > p] \approx \e^{-d p}\, , 
\end{equation}
where $p_\mathrm{uni}$ denotes the uniform distribution. 
Note that while the left-hand side of Eq.~\eqref{eq:xeb:porter-thomas} is discontinuous, the right-hand side allows 
us to approximately think of the distribution of $p(x)$ as being described by the continuous probability density
$p_\mathrm{PT}(p) = d\e^{-d p}$ of the Porter-Thomas distribution \cite{porter_fluctuations_1956}. 
We use this description in our theoretical analysis multiple times. 
\ex{%
The motivation to study distributions of Porter-Thomas shape stems from considering Haar random unitaries in place of the quantum circuit $U$ and is further illuminated in the following exercise.

\begin{exerciseblock}{Densities of Porter-Thomas shape \cite{BoiIsaSme16}}
For $U \in \U(d)$ drawn from the Haar measure $\mu_{\U(d)}$
one can show that the squared absolute value $p = |U_{ij}|^2$ of its matrix entries $\{U_{ij}\}$ have the probability density function
$p_\mathrm{abs}(p) = 
(d - 1)(1-p)^{d-2}$. 
In the limit of $d \gg 1$, $p_\mathrm{abs}(p)$ is described by the \emph{Porter-Thomas distribution} \cite{porter_fluctuations_1956}
\begin{equation}\label{eq:p_PT}
	p_{\mathrm{PT}}(p) = d\exp(-dp) \, .
\end{equation}

Argue that for a fixed $U$ and again in the limit of large $d$ 
the probability mass function $p_U$ is of Porter-Thomas shape. 

Assuming that $p_U$ is of Porter-Thomas shape, show that 
\begin{align}\label{ex:porter-thomas:hxpp}
	H(p_U) &= \ln(d) + \gamma - 1, \\
	\label{ex:porter-thomas:hxup}
	H_X(p_\mathrm{uni}, p_U) &= \ln(d) + \gamma,
\end{align}
where $\gamma$ is the Euler–Mascheroni constant and $p_\mathrm{uni}(x) = 1 / d$ is the uniform probability mass function.  

\emph{Hint: Recall the definite integral formulas $\int_0^\infty p \ln(p)\, \e^{-p} \rmd p = 1 - \gamma$ and $\int_0^\infty \ln(p)\, \e^{-p} \rmd p = \gamma$.}
\end{exerciseblock}
}
\exreplace{
The motivation to study distributions of Porter-Thomas shape stems 
from considering Haar random unitaries in place of the quantum circuit $U$ and is further illuminated in the following box.

\begin{block}{Densities of Porter-Thomas shape}
For $U \in \U(d)$ drawn from the Haar measure $\mu_{\U(d)}$
one can show that the squared absolute values of its matrix entries have the probability density function
$p_\mathrm{abs}(p) = 
(d - 1)(1-p)^{d-2}$. 
In the limit of $d \gg 1$, $p_\mathrm{abs}(p)$ is described by the \emph{Porter-Thomas distribution} \cite{porter_fluctuations_1956}
\begin{equation}\ifex\else\label{eq:p_PT}\fi
	p_{\mathrm{PT}}(p) = d\exp(-dp)\, .
\end{equation}
Note the similarity with the density of the $\chi^2$-distribution with $2$ degrees of freedom. 
For fixed $U$ 
and again in the limit of large $d$, 
one can hence argue that 
the probability mass function $p_U$ is of Porter-Thomas shape \cite{BoiIsaSme16}.

Assuming that $p_U$ is of Porter-Thomas shape, Boxio \emph{et al.}~\cite{BoiIsaSme16} showed that 
a straightforward calculation reveals 
\begin{align}\ifex\else\label{ex:porter-thomas:hxpp}\fi
	H_X(p_U, p_U) = H(p_U) &= \ln(d) + \gamma - 1, \\
	\ifex\else\label{ex:porter-thomas:hxup}\fi
	H_X(p_\mathrm{uni}, p_U) &= \ln(d) + \gamma,
\end{align}
where $\gamma$ is the Euler–Mascheroni constant and $p_\mathrm{uni}(x) = 1 / d$ is the uniform probability mass function. 
\end{block}
}

The introduction of the so-called \emph{cross-entropy difference} 
as a performance measure in quantum supremacy sampling tasks 
brought the cross-entropy into focus. 

\begin{block}{Cross-entropy difference}
\label{block:dxe}
Ref.~\cite{BoiIsaSme16} introduced the \emph{cross-entropy difference} as a performance 
measure in sampling tasks
\begin{equation}\label{eq:dxe}
	\dxe(q, p) \coloneqq H_X(p_\mathrm{uni},p) - H_X(q, p)\, ,
\end{equation}
where $p_\mathrm{uni}$ is the uniform distribution. 
The cross-entropy difference, thus, measures the excess in cross-entropy that $q$ has with $p$ beyond the uniform distribution. 

In the previous box we argue that for Haar-random unitaries the corresponding measurement densities 
$p_U$ are generically of Porter-Thomas shape. 
The motivation of the cross-entropy difference is highly relying on this observation. 
By definition, we have that $\dxe(p_\mathrm{uni}, p) = 0$ for any $p$. 
If $p$ is of Porter-Thomas shape, Eqs.~\eqref{ex:porter-thomas:hxpp} and \eqref{ex:porter-thomas:hxup} show that $\dxe(p, p) = 1$. 
Note however that there still exist probability distributions that score even higher in cross-entropy difference than 
$p$ itself.
\end{block}

Another measure introduced in this context 
is the \emph{cross-entropy fidelity} \cite{Google2019QuantumSupremacy}
\begin{equation}\label{eq:XEF}
	F_X(q, p_U) = \sum_{x \in [d]} q(x) (d p_U(x) - 1 ). 
\end{equation}

Before discussing the \ac{XEB} protocol to estimate $H_X$ and $F_X$ let us illuminate the 
motivation of $F_X$ in the context of certifying sampling tasks. 

First, the cross-entropy fidelity can be regarded as a linear proxy to the cross-entropy difference 
and, as such, as a simpler version of it.
The shift of minus one in the definition of $F_X$ is chosen such that 
$F_X(p_\text{uni}, p) = 0$ for $p_\text{uni}$ the uniform density and any probability density $p$.  
If $p_U$ is assumed to be of Porter-Thomas shape 
one can calculate that $F_X(p_U, p_U) = 1$. 
This motivates the expectation that performing high in cross-entropy fidelity indicates successfully 
solving the sampling task for typical random circuits $U$. 

Note that if $U$ is drawn at random from a unitary $2$-design $\mu$, 
we can reproduce the Porter-Thomas value of $F_X(p_U, p_U)$ in expectation over $U$ using Lemma~\ref{lem:sym_moment_operator}: 
we first calculate 
\begin{align}\nonumber
	\EE_{U\sim \mu}\left[p^2_U(x) \right] %
	&= \EE_{U\sim \mu}\myleft[ |\sandwich x U \psi|^4\myright] 
	\\\nonumber
	&= {\bra x}^{\otimes 2} \,
		\EE_{U\sim \mu} 
			\myleft[ (U \ketbra\psi\psi U^\dagger)^{\otimes 2} \myright]
	  {\ket x}^{\otimes 2} 
	\\\nonumber
	&= \frac{{\bra x}^{\otimes 2} P_{\sym^2} {\ket x}^{\otimes 2}}{\Tr(P_{\sym^2})} \\
	&= \frac2{d(d+1)}
	\label{eq:xeb:fx:exp}
\end{align}
ande hence find that
\begin{equation}%
\begin{split}%
	\EE_{U\sim \mu} [F_X(p_U, p_U) ]
		&= \sum_{x \in [d]} d\, \EE_{U\sim \mu}\myleft[p^2_U(x)\myright] - 1 \\
		&= \frac{2 d}{d + 1} - 1 =\frac{d-1}{d+1} 
		\\
		&= 1 + \LandauO(1/d) \, .
\end{split}
\end{equation}
Thus, if $U$ is drawn from a distribution, where we have suitable control over higher moments we can hope to 
proof concentration around the expectation with high probability for large $d$. 
For Haar random unitaries Levy's lemma \cite{Ledoux:2001:concentration} directly yields a corresponding statement. 

For the moment, we leave this as a motivation for the estimating $F_X$ and $H_X$ and 
turn to the \ac{XEB} protocol. 

\subsubsection*{\protchaptitle{Cross-entropy benchmarking protocol}}
The crucial structural insight of \ac{XEB} is that $F_X$ and $H_X$ are both of the form
\begin{equation}\label{eq:xeb:fform}
	E_f = \sum_{x \in [d]} q(x) f(p_U(x))
\end{equation}
with $f(p) = f_F(p) = d p - 1$ for the cross-entropy fidelity and $f(p) = f_H(p) = -\ln(p)$ for the cross-entropy. 
This observation suggests a simple protocol, akin to importance sampling (Section~\ref{sec:ImportanceSampling}), for empirically estimating both quantities if we have access to samples of one of the distributions. 

\begin{protocolblock}[label={protocol:xeb}]{\Acf{XEB} \cite{BoiIsaSme16,Google2019QuantumSupremacy}}
	Let $U$ be a description of a quantum circuit, $\ket{\psi} \in \CC^d$ be an initial states and $\mathcal B = \{\ket x\}_{x \in [d]}$ an orthonormal basis of $\CC^d$.  
	\begin{enumerate}
		\item Prepare $U\ket{\psi}$ on a quantum computer and measure in the basis $\mathcal B$ a number of $m$ times to collect the measurement outcomes $\mathcal O = (\tilde x_1,\ldots, \tilde x_m) \in [d]^m$. 
		\item \label{prot:xeb:classical} Calculate on a classical computer for each $\tilde x \in \mathcal O$ the value of $p_U(\tilde x)$. 
		\item Return the estimator 
		\begin{equation}
			\hat E_f = \frac1{|\mathcal O|}\sum_{\tilde x \in \mathcal O} f(p_U(\tilde x)), 
		\end{equation}
		where $f$ is $f_F$ or $f_H$ for estimating the cross-entropy fidelity or cross-entropy, respectively. 
	\end{enumerate}
\end{protocolblock}

It is important to keep in mind that step~\ref{prot:xeb:classical} 
requires that a classical computer can compute individual probabilities of the circuit. 
For this reason, \ac{XEB} cannot be used directly for circuits that are not classically simulable. 
Instead, one can investigate the performance on restricted subclasses of circuits that are still tractable on a 
powerful classical computer and from these results extrapolate the performance in the regime where one expects quantum 
supremacy. 

If we assume that the target distribution $p_U$ is defined using a Haar-randomly drawn unitary $U$, we can derive a guarantee for Protocol~\ref{protocol:xeb} for 
the linear cross-entropy using the techniques that we presented in this tutorial. 
Such a guarantee was derived by Hangleiter~\cite{Hangleiter:2020:thesis}. 

\begin{theoremblock}[label={thm:XEBsampling}]{Linear \ac{XEB} sampling complexity \cite{Hangleiter:2020:thesis}}
	Let $U \in \U(d)$ be a Haar random unitary, $\ket{\psi}\in \CC^d$ and $\mathcal B = \{\ket x\}_{x=1}^d \subset \CC^d$ an orthonormal basis. Denote by $p_U$ the associated measurement probability mass function \eqref{eq:def:measdist} and by $\tilde p_{U}$ the implemented probability mass function. 
	Choose $\epsilon, \delta > 0$ and 
	\begin{equation}
		m \geq \frac{\e^2}{2\epsilon^2} \ln^2\left(\frac{2d}\delta\right)\, \ln\left(\frac2\delta\right) \, . 
	\end{equation}
	Then, Protocol~\ref{protocol:xeb} returns with confidence $1-\delta$
	an unbiased $\epsilon$-accurate estimator $\hat E_f$ for 
	$F_X(\tilde p_U, p_U)$. 
\end{theoremblock}

The proof of the theorem relies on bounding the range of the random variable $p_U(\tilde x)$ and applying 
the Hoeffding's inequality \eqref{eq:Hoeffdings}. 
We have already seen that for $U$ drawn from the Haar measure, $p_U$ is asymptotically of Porter-Thomas shape. 
In particular, large probabilities in $p_U$ are exponentially suppressed. 
For this reason, we expect that with high probability over the choice of $U$, 
$p_U(x)$ will be bounded for all $x$. 
The following lemma makes this expectation explicit. 

\begin{lemmablock}[label={lem:PUbound}]{$p_U$ is bounded w.h.p.}
	Let $U \in \U(d)$ be a Haar random unitary and $\{\ket x\}_{i=1}^d$ be 
	an orthonormal basis of $\CC^d$. 
	Then, 
	the measurement probability mass function $p_U: [d] \to [0,1]$, $p_U(x) = \left|\sandwich x U 0 \right|^2$, fulfills 
	$p_U(x) \leq b$ for all $x$ with probability of at least 
	$1 - d \e^{- d b / \e}$. 
\end{lemmablock}

One way to prove the lemma is via the Porter-Thomas density \eqref{eq:p_PT}.
We follow a more self-contained strategy by calculating the moments of $p_U$. 
Then, the bound on the moments can be translated to an exponential tail bound 
using the following consequence of Markov's inequality. 

\begin{theoremblock}[label={thm:subexponentialtailbound}]{Sub-exponential tail bound, e.g.\ \cite[Proposition 7.11]{FouRau13}}
	Let $X$ be a random variable satisfying
	\begin{equation}
		\EE[|X|^k]^{1/k} \leq \alpha \beta^{1/k} k
	\end{equation}
	for all $k \geq 2$. 
	Then, for all $t \geq 2$, 
	\begin{equation}
		\Pr[|X| \geq \e \alpha t ] \leq \beta \e^{-t}\, . 
	\end{equation}
\end{theoremblock}

\begin{proof}
	Applying Markov's inequality \eqref{eq:MarkovsIneq} and the theorem's assumption gives for $k \geq 2$
	\begin{equation}
	\begin{split}
		\Pr[|X| \geq \e \alpha t] &= \Pr[|X|^k \geq (\e \alpha t)^k]  \\
		&\leq \frac{\EE[|X|^k]}{(\e \alpha t)^k} \leq \beta \e^{-k} \left(\frac kt\right)^k \, .
	\end{split}
	\end{equation}
	Now choosing $k = t$ yields the claim. 
\end{proof}

\begin{proof}[Proof of the Lemma~\ref{lem:PUbound}]
	We start by calculating the moments of $p_U(x)$ as a random variable depending on $U \sim \mu_{\U(d)}$. 
	First note that by definition $p_U(x) = \left|\sandwich x U 0\right|^2 = \left|\braket x \psi\right|^2$ with $\psi$ drawn uniformly from the sphere $\sphereCd{}$. 
	Using the moment operator $K^{(k)}_{\mu_{\sphereCd}}$ for $\ket\psi \sim \mu_{\sphereCd}$, Lemma~\ref{lem:moments_rand_states} and \eqref{eq:psym_expansion}, we find that for all $x \in [d]$ 
	\begin{equation}
	\begin{split}
		&\!\!\!\!%
		\EE_{U \sim \mu_{\U(d)}}[ p_U(x)^{k}] \\
		&= \EE_{\ket\psi \sim \mu_{\sphereCd{}}}\left[ {\bra x}^{\otimes k} (\ketbra\psi\psi)^{\otimes k} {\ket x}^{\otimes k}  \right] \\
		&= {\bra x}^{\otimes k} K^{(k)}_{\mu_{\sphereCd}} {\ket x}^{\otimes k}   \\
		&= \frac{k! (d-1)!}{(k + d - 1)!} (\bra x)^{\otimes k} P_{\sym^k} {\ket x}^{\otimes k} \\
		&=  \frac{k! (d-1)!}{(k + d - 1)!} \norm{\ket x}^{k} =  
		\binom{k + d - 1}{k}^{-1} .
	\end{split}
	\end{equation}
	Due to the inequality $\binom{n}{k} \geq (n/k)^k$, it holds for $k \geq 1$ that 
	 $\binom{d + k - 1}k \geq \left( \frac{d + k - 1}k \right)^k \geq (d / k)^k$ and, 
	thus,  
	\begin{equation}
		\EE_{U \sim \mu_{\U(d)}}[ p_U(x)^k] = \binom{k + d -1}k^{-1} \leq k^k d^{-k}\, .
	\end{equation}
	By Theorem~\ref{thm:subexponentialtailbound}, this moment bound translates into the tail bound 
	\begin{equation}
		\Pr[p_U(x) \geq t] \leq \e^{-d t / \e}\, 
	\end{equation}
	for $t \geq 2\e / d $. 
	Finally, using the union bound we conclude that 
	\begin{equation}
		\Pr[p_U(x) \geq t\quad\forall x \in [d]] \leq d\e^{-d t / \e}\, ,
	\end{equation}
	which completes the proof.
\end{proof}

\begin{proof}[Proof of Theorem~\ref{thm:XEBsampling}]
	Let $d = 2^n$. The estimator $\hat E_f$ is the sum of $m$ i.i.d.\ random variables $f(p_U(\tilde x))$. 
	By the form \eqref{eq:xeb:fform} it is clear that $\hat E_f$ is an unbiased estimator for $E_f$. 
	The estimator $\hat E_f$ is the sum  of $m$ i.i.d.\ random variables $f(p_U(\tilde x))$. 
	Using Lemma~\ref{lem:PUbound} $p_U(x) \leq b \coloneqq \frac \e d \ln\myleft(\frac{2d}{\delta}\myright)$  with probability $1 - \delta / 2$. 
	Thus, with the same probability, the random variable $f(p_U(\tilde x))$, with $f = f_F$ as defined below \eqref{eq:xeb:fform}, is bounded by $db - 1 = \e \ln(2d /\delta)$.  
	Now Hoeffding's inequality \eqref{eq:HoeffdingsAbs}, with failure probability $\delta / 2$ yields the statement. 
\end{proof}

Following the same strategy, one can also derive a sampling complexity in 
$\LandauO\bigl(\epsilon^{-2}\ln^2(d)\ln(1/\delta)\bigr)$ 
for estimating the cross-entropy $H_X(\tilde p_U, p_U)$ by Protocol~\ref{protocol:xeb} \cite{Hangleiter:2020:thesis}. 
Since the cross-entropy $f(p_U(\tilde x))$ involves the logarithm, the 
upper bound on the range of $p_U$ of Lemma~\ref{lem:PUbound} is no longer sufficient to ensure boundedness of the random variables that enter the estimator. 
In addition, one needs a lower bound on the range of $p_U$. 
This is not possible with our bounds on the moments. 
Instead, one has to explicitly calculate the tail distribution \eqref{eq:p_PT}. 

From an estimate of the cross-entropy one can calculate an estimate of the cross-entropy difference by shifting with $H_X(p_\text{uni}, p_U)$. 
If the ideal circuit is sufficiently close to a Haar-random unitary, one can analytically calculate $H_X(p_\text{uni}, p_U)$. 
Alternatively, taking the average of the values calculated in step~\ref{prot:xeb:classical} provides a numerical estimate for $H_X(p_\mathrm{uni}, p_U)$.

Ultimately, theoretical results for the hardness of sampling tasks require closeness of the probability mass functions
in \emph{\ac{TV} distance} or \emph{TV norm}
\begin{equation}
	\norm{q - p}_\mathrm{TV} = \frac12 \sum_{x \in [d]} | q(x) - p(x) |.
\end{equation}
Without additional assumptions, it is not possible to derive a \ac{TV} norm bound from the cross-entropy. 
A counter example is discussed in Ref.~\cite{bouland_quantum_2018}. 
Therein, Bouland~\emph{et al.}~also hint at a possible bail out. 
An insightful presentation of the argument is also given in Ref.~\cite{Hangleiter:2020:thesis}. 
Very close to the desired bound is Pinsker's inequality \cite{CsiszarKoerner:2011:Information} 
\begin{equation}
	\norm{q - p}_\text{TV} \leq \sqrt{\frac{D_\mathrm{KL}(q, p)}2}
\end{equation}
that bounds the \ac{TV} norm in terms of the 
\emph{Kullback-Leibler divergence} $D_\text{KL}(q, p) \coloneqq H_X(q, p) - H(q)$. 
The Kullback-Leibler divergence $D_\text{KL}(q, p_U)$ is unfortunately not of the form \eqref{eq:xeb:fform} and cannot 
be directly estimated by a \ac{XEB} protocol. 
In addition to the estimate of the cross-entropy, the $D_\mathrm{KL}(q, p_U)$ requires an estimate of the entropy of the implemented mass function $q$. 
If we assume that the noise in our implementation only increases the entropy such that $H(q) \geq H(p)$, 
we can avoid this obstacle and swap $H(q)$ for $H(p)$, the entropy of the ideal probability mass function. 
Thus, instead of $D_\text{KL}(q,p)$ we consider $D_\mathrm{XE}(q,p) = H_X(q,p) - H(p)$. 
If $H(q) \geq H(p)$, then $D_\text{KL}(q, p) \leq D_\text{XE}(q,p)$ and a \ac{TV} norm bound is given 
in terms of $D_\mathrm{XE}(q,p)$ via Pinsker's inequality. 

Similar to the cross-entropy difference~\eqref{eq:dxe} $D_\mathrm{XE}(q,p)$ can be estimated by 
measuring $H_\text{XE}(q,p)$ with Protocol~\ref{protocol:xeb} and either estimating the shift 
$H(p)$ analytically or numerically from the computed values $p_U(\tilde x_i)$ of step~\ref{prot:xeb:classical}. 
If the ideal probability mass function is of Porter-Thomas shape then 
one can calculate that $D_\mathrm{XE}(q, p) = 1 - d_\mathrm{XE}(q,p)$ and the above discussion can be translated to the \emph{cross-entropy difference}.

\subsubsection*{Further reading}
The idea of demonstrating quantum supremacy in the task of sampling from certain probability distribution that naturally arise 
in quantum systems goes back to the proposal of boson sampling in a linear optics \cite{AarArk10, BremnerJozsaShepherd:2011:Classical}. 
Even earlier, Terhal and DiVicenzo derived evidence for the hardness of the sampling task associated with simulating 
restricted classes of quantum circuits \cite{terhal_adaptive_2004}. 
Besides random circuit sampling \cite{BoiIsaSme16} multiple supremacy proposals exist, e.g.\ for 
other restricted classes of quantum computations 
\cite{BremnerJozsaShepherd:2011:Classical, BoulandMancinskaZhang:2016, morimae_hardness_2017,Bremner2017AchievingQuantum, bouland_complexity_2018} 
or for processes arising in quantum simulation \cite{gao_quantum_2017, bermejo-vega_architectures_2018}, see also Ref.~\cite{FeffermanUmans:2015,Farhi2016QuantumSupremacy, MannBremner:2017}. 
A series of additional theoretical works collects evidence for the robust hardness of the resulting approximate sampling tasks, e.g.\ \cite{Speedups,hangleiter_anticoncentration_2018, bouland_quantum_2018, HaferkampEtAl:2019:ClosingGaps}, and more fine-grained statements about the sufficient scaling \cite{DalzellEtAl:2020,MorimaeTamaki:2019,MorimaeTamaki:2020}. 

It was realized early on that the verification of quantum supremacy is a daunting task~\cite{Gogolin2013BosonSamplingIn,AarArk14}. 
One might hope that it is possible to perform a non-interactive black-box verification. 
Such a verification certifies the sampling task solely from the samples itself. 
Unfortunately, the same features of a probability distribution that guarantee the classical hardness of the sampling task 
prohibit the efficient verification from samples on a classical computer \cite{HanKliEis19}. 
Optimal but non-efficient strategies for general verification problems were studied in Ref.~\cite{ValVal17}. 

We focus on cross-entropy estimation for the quantum supremacy verification \cite{BoiIsaSme16}. 
Another measure of the form \eqref{eq:xeb:fform} is employed in the \ac{HOG} test  which 
uses a heavy-side function as $f$ \cite{Aaronson2016ComplexityTheoreticFoundations}. A refined notion of the heavy outcome generation test is the \ac{BOG} test proposed in Ref.~\cite{bouland_quantum_2018}. 
Note that these protocols still require an efficient simulation of the quantum circuit on another computing device. 
One approach to overcome this bottleneck is to run the quantum circuit as part of a larger circuit that includes so-called traps, subcircuits that can be efficiently simulated \cite{Ferracin18AccreditingOutputsOf}. 
Naturally, approaches for quantum state and processes certification can also be used to verify a sampling task under a various assumptions. 
It is an ongoing endeavor to develop classical strategies for spoofing verification protocols for quantum supremacy with successes reported e.g.\ in Refs.~\cite{Kahanamoku:2019:Forging,BarakEtAl:2020:Spoofing} and for collecting evidence for the hardness of classical spoofing \cite{AaronsonGunn:2019:Classical}. 

An extensive, recent overview over verification and certification methods in the context of quantum supremacy can be found in Ref.~\cite{Hangleiter:2020:thesis}.

\section*{Acknowledgements}
Our present work builds on countless insightful discussions 
on the topic of quantum system characterization and beyond
with many friends and colleagues over the last years. 

Among those are a couple that we explicitly thank for valuable support while drafting this tutorial. 
We thank Raphael Brieger for comments on the presentation of the lecture notes preceding this tutorial. 
We are grateful to Dominik Hangleiter for countless exciting and enlightening discussions on 
various topics of the tutorial, especially on the anatomy of certification protocols, \ac{DFE} and cross-entropy benchmarking. 
We are grateful to him for graciously providing a draft of his thesis ---a valuable resource on the certification of quantum supremacy--- and helpful comments on the tutorial. 
We thank Richard Kueng for explanations and discussions on \ac{SFE} and quantum channels as well as valuable feedback on the tutorial. 
I.R.\ thanks Jonas Helsen, Emilio Onorati, Albert Werner, and Jens Eisert for valuable and insightful discussions on \ac{RB}, 
Jonas Haferkamp, Markus Heinrich, Felipe Montealegre-Mora and David Gross for discussions on 
Schur-Weyl duality, Yi-Kai Liu for discussions on \ac{DFE} and \ac{RB}. 
Furthermore, the section on the anatomy of certification protocols 
builds on many enjoyable conversations with Nathan Walk. 
We are grateful to Susane Calegari for contribution to the illustrations and valuable feedback on the draft. 
The work of I.R.\ has been funded by the DFG (EI 519/9-1, EI 519/14-1, and CRC 183). 
The work by M.K.\ is supported by 
the Deutsche Forschungsgemeinschaft (DFG, German Research Foundation) via the Emmy Noether grant 441423094.

\section*{Acronyms} 
\begin{acronym}[POVM]\itemsep.5\baselineskip
\acro{NISQ}{noisy and intermediate scale quantum}
\acro{POVM}{positive operator valued measure}
\acro{PVM}{projector-valued measure}
\acro{CP}{completely positive}
\acro{CPT}{completely positive and trace preserving}
\acro{DFE}{direct fidelity estimation} 
\acro{MUBs}{mutually unbiased bases} 
\acro{SIC}{symmetric, informationally complete}
\acro{SFE}{shadow fidelity estimation}
\acro{RB}{randomized benchmarking}
\acro{XEB}{cross-entropy benchmarking}
\acro{SPAM}{state preparation and measurement}
\acro{TV}{total variation}
\acro{HOG}{heavy outcome generation}
\acro{BOG}{binned outcome generation}
\end{acronym}

\bibliographystyle{./myapsrev4-2}
\bibliography{mk} 

\end{document}